\documentclass[numbib]{imamat}

\usepackage{graphicx}
\usepackage{bm}
\usepackage{subeqnarray}
\usepackage{changepage}
\usepackage{geometry}
\usepackage{bm}
\usepackage{amsmath}
\usepackage{subcaption}

\begin{document}

\title{Shape optimisation of stirring rods for mixing binary fluids}

\shorttitle{Optimised stirring rods for mixing binary fluids} 
\shortauthorlist{M.F. Eggl \& P.J. Schmid} 

\author{
\name{Maximilian F. Eggl}
\address{Department of Mathematics, Imperial College London, London SW7 2AZ, U.K. \email{$^*$Corresponding author: maximilian.eggl11@imperial.ac.uk}}
\name{Peter J. Schmid}
\address{Department of Mathematics, Imperial College London, London SW7 2AZ, U.K.}
}

\maketitle

\begin{abstract}
{Mixing is an omnipresent process in a wide-range of industrial
applications, which supports scientific efforts to devise techniques for
optimising mixing processes under time and energy constraints. In this endeavor, we present a computational framework based on nonlinear
direct-adjoint looping for the enhancement of mixing efficiency in a
binary fluid system. The governing equations consist of the nonlinear
Navier-Stokes equations, complemented by an evolution equation for a
passive scalar. Immersed and moving stirrers are treated by a
Brinkman-penalisation technique, and the full system of equations is
solved using a Fourier-based pseudospectral approach. The adjoint
equations provide gradient and sensitivity information which is in turn used to improve an initial mixing strategy, based on shape, rotational and path modifications. We utilise a Fourier-based approach for
parameterising and optimising the embedded stirrers and consider a
variety of geometries to achieve enhanced mixing efficiency. We consider
a restricted optimisation space by limiting the time for mixing and
the rotational velocities of all stirrers. In all cases, non-intuitive
shapes are found which produce significantly enhanced mixing efficiency.}
{Mixing, Optimization, Penalization, Adjoint method, Fourier-based Shape Parameterisation}
\\
2000 Math Subject Classification: 34K30, 35K57, 35Q80,  92D25
\end{abstract}

\section{Introduction}
The mixing of binary fluids is a problem of fundamental concern in
fluid dynamics, as its underlying mechanisms play an important role in a wide variety of industrial fields and in many fluid processes encountered
in daily life. This provides ample motivation to study the processes and characteristics that drive this natural phenomenon. 

The study of mixing can broadly be split into two areas, theoretical and application-driven. The theoretical aspects of mixing have been investigated in great detail, and there exists a vast body of literature pertaining to many aspects of mixing for binary fluids. Issues that have been studied include the formulation of emulsions and the stirring mechanics used to create them \citep{Taylor1934}, mixing in stratified flows \citep{long1978,peltier2003} as well as the optimisation of mixing \citep{marcotte2018}, among many other efforts. For example, Stokes mixing, i.e., the mixing of highly viscous fluids, has drawn great interest in the research community, exemplified by investigations \citep{spencer1951} that identified stretching, cutting and folding as the primary mechanisms of mixing. \cite{chien1986}, in a related effort,  found that chaotic mixing can be induced from laminar mixing by varying the oscillations of the walls, a configuration that was further studied by \cite{ottino1988} and \cite{mohr1957} who argued for a simplified analysis of viscous mixing by ignoring interfacial tension. Their simplification formed the basis for many subsequent developments in the mixing of viscous fluids.
Concurrently, at the other end of the parametric range, turbulent mixing, i.e., the mixing of binary fluids mostly by inertial effects, has been studied extensively, starting with the early articles by \cite{corrsin1957,corrsin1969} and \cite{batchelor1952}. More recently, mixing has been linked to chaotic processes and chaotic advection, yielding the accepted consensus that turbulence naturally leads to `better' mixing \citep{dimotakis1999}. This fundamental finding was subsequently the topic of many more studies and has been verified experimentally, e.g. in \cite{dimotakis2000}. As a consequence, it has also spawned a trend where efficient mixing is achieved by techniques that inject energy into the system so as to generate turbulence, which subsequently produces better mixing \citep{aamo2013,jalali2015}. Turbulence is also a key component in studies of mixing in stratified fluids, e.g., \cite{Linden1979}; these investigations suggest that turbulence exists for all values of the ratio of buoyancy to flow shear and argue for the existence of an overall stratification where mixing reaches an optimum. This general configuration has also been studied by \cite{Tang2009}, who put a value on this optimal stratification and confirmed Linden's supposition.

The above selection of articles, while certainly incomplete and not exhaustive, is intended to demonstrate the breadth and depth of scientific efforts that have been devoted to mixing. For readers who seek a more in-depth review of the different aspects of mixing theory, the monographs by \cite{uhl2012}, \cite{paul2004} and \cite{ottino1989} provide a more complete background on the rich aspects of fluid mixing. 

The above monographs and investigations focus on the more theoretical aspects of fluid mixing, but are less representative of studies concerned with the applications of mixing in natural or technological settings. Samples from the latter category examine the spreading and mixing of pollutants in the
ocean~\citep{Lekien2005}, the ventilation of 
buildings~\citep{Hunt1999,Linden1999}, the mixing of air and fuel for
subsequent combustion~\citep{Annaswamy1995} or the mixing in
microfluidic devices~\citep{Hessel2005, Nguyen2005}. Additional fields of application include the food
processing, pharmaceutical and consumer-product industry. Interest in these industrial application fields is fueled by the fact that even modest improvements of mixing efficiency would translate into
immediate and significant profits as well as into a more consistent quality of the respective final product.

Mixing in industrial applications is often accomplished by stirrers,
i.e., rotating bodies of a given shape embedded in a mixing vessel, whose
task it is to produce long filaments \citep{Aref1984} which
subsequently diffuse. This two-step shearing-diffusion process, which is principally at
the core of binary fluid mixing by stirrers, has been recognised as a
fundamental mechanism in inertia-driven mixing \citep{spencer1951} and has been studied extensively to gain further insight
and to guide control strategies. The present article follows in this path and considers stirrer-induced inertia-driven mixing coupled with a shape optimisation of the stirrers. 

From a computational point of view, several difficulties arise from the treatment of the embedded stirrers, in particular the treatment of the fluid-structure
interaction between these moving bodies and the binary
fluid. Various options exist and have been pursued by previous
studies, such as body-fitted meshes~\citep{Mattiussi2000} or the method
of fictitious domains~\citep{Glowinski1996}. These computational techniques are, however, often restricted to simple configurations or are excessively
costly due to the need to remesh at each time step (after the solid
body has moved). An attractive alternative to these
methods is the Brinkman-penalisation approach which will be adopted in this paper and will provide the necessary flexibility and efficiency to treat more complex geometries and flow configurations. Introduced by \cite{arquis1984}, solutions obtained
by the penalisation method have been rigorously shown to converge to
the corresponding solutions of the Navier-Stokes
equations~\citep{Angot1999} for the respective complex
domain. Supporting studies, including an asymptotic analysis of this convergence process, are
summarised in~\cite{Liu2007}, and applications to high Mach-number
flows, e.g. \cite{Boiron2009}, and turbulent flow past
cylinders, e.g. \cite{Kevlahan2001}, among many other examples, have
demonstrated the effectiveness and adaptability of this approach.
Brinkman-penalisation has also been employed by~\cite{Chantalat2009},
coupled with a level-set technique to express the geometry, and
applied by~\cite{Bruneau2013} to optimise the shape of stirrers. An additional appeal of this method lies in its simple derivation and
straightforward numerical implementation.

A second difficulty of the mixing problem lies in the nonlinear nature
of its underlying governing equations. Optimising stirring strategies
using a gradient-based approach
will have to cope with the solution of nonlinear equations and the
checkpointing problem for the dual/adjoint problem (see details
below). The complexity of the flow does not furnish equilibrium points
about which to linearise, rather the full nonlinear problem has to be tackled, and
nonlinear adjoint looping techniques~(\cite{Juniper}) have to be
employed for the stirrer geometry and/or stirrer paths.

A key challenge arises from the parameterisation for the stirrers' geometry. Even though the pointwise definition of the penalisation method would allow for an extensive and diverse number of solid shapes \citep{hejlesen2015}, the resulting excessive dimensionality of the control variable does not lend itself to realistic and applicable optimisation outcomes. On the other hand, any choice of reducing the dimensions of the design space will influence the final optimisation results, as more exotic, yet potentially optimal, shapes are excluded. Consequently, care must be taken in selecting a suitable parameterisation to represent the embedded stirrers. Examples of shape parameterisations include B\'ezier curves \citep{han2009} or Hermite splines \citep{deboor1987}. While these methods allow for an acceptable range of possible complex shapes, they quickly approach a break-even point between flexibility and computational cost. Instead, we turn to a Fourier-based shape parameterisation, also referred to as Fourier descriptors \citep{lin1987}. Relying on Fourier series, we are thus able to select the dimensionality of the control space -- without substantially compromising on the design space. The effectiveness of this approach will be demonstrated throughout this paper.

With these challenges identified, and respective algorithms chosen, we can then attempt to enhance mixing via stirrer-based strategies using a nonlinear direct-adjoint optimisation framework, coupled to a shape-optimisation approach. While various partial aspects of this undertaking have been attempted earlier, the synthesis of all components constitutes a novel pursuit.

Previously,
optimal control of
mixing via entropy maximisation of a flow governed by two orthogonal
shear flows has been attempted~\citep{DAlessandro1999}, as well as the  optimal mixing of binary fluids by optimising the mix-norm~\citep{Mathew2007, Lin2011,vermach2018}; a
concise review is presented in~\cite{Ottino1990} and~\cite{Thiffeault2012}. These studies, however, do not address shape optimisation in the context of mixing. 

Relevant studies that do attempt shape optimisation fall into two general categories: (i) Brinkman-penalisation-based shape optimisation of mixing coupled to algorithms that are not gradient-based or nonlinear, and (ii) gradient-based methods for shape optimisation that do not focus specifically on mixing. 
Among some samples from the former category, we mention the 
mixing optimisation of CO$_2$ ejectors using genetic algorithms~\citep{palacz2016}, which achieved modest gains in mixing efficiency, the optimisation of diffuser shapes by a derivative-free optimisation approach~\citep{madsen2000}, and the shape optimisation of a heating element for desired temperature uniformity~\citep{smolka2013}, again using genetic meta-heuristic algorithms. In the latter category, we mention the work by \cite{alexandersen2014,pingen2007,saglietti2017,saglietti2018} which employ adjoint (gradient-based) technology as well as Brinkman-penalisation, but consider natural convection rather than inertia-driven mixing of binary fluids.

The work presented in this article builds on the nonlinear direct-adjoint framework presented in~\cite{Eggl2018}, where a first attempt at shape optimisation, computing the optimal parameters of elliptical cross-sections, has been made. Building on this starting point, we generalise the stirrer geometry, while maintaining the circular shape of the mixing container (a commonplace configuration in industrial applications~\citep{Handbook6}). More specifically, we consider three test cases of increasing geometric complexity, which allows the optimisation framework to fully explore the design space and enhance mixing efficiency through the use of optimal shapes. 

The remainder of this article is organised as follows. Section \S\ref{Sec:GovEq} introduces the governing equations and the mathematical background related to the optimisation of mixing. This is followed by a brief discussion of Brinkman-style penalisation, applied to the governing equations; discretisation details are given as well. We then turn to presenting the direct-adjoint optimisation procedure, together with the Fourier-shape parameterisation, in section \S\ref{Sec:OptFrame}. Next, we present the results of the optimisation studies in section \S\ref{Sec:Results}; enhanced mixing is reported in all cases. Conclusions are offered in section \S\ref{Sec:conclusion}.

\section{Mathematical background}
\label{Sec:GovEq}
\subsection{Governing equations}

We will consider the active mixing of a binary fluid by stirring devices, for a parameter regime where inertial effects cannot be dismissed, whilst turbulent fluid motion is not yet present. The material behavior of the two fluid components is aptly described as a Newtonian fluid. This particular choice of regime and characteristics reflects the fact that  
these properties represent values commonly encountered in industrial settings; moreover, they allow a rich and interesting suite of physical mechanisms that can be exploited for the design of optimised stirring protocols.  

The underlying governing equations that describe the motion of our fluid are the incompressible Navier-Stokes equations, which in primitive form read 
\begin{align}
\partial_t \bm{u} + \bm{u}\cdot \nabla \bm{u} + \nabla p - Re^{-1}\nabla^2 \bm{u} & = 0 \\
\nabla \cdot \bm{u} & = 0  \label{Eq:GovEq}
\end{align}
with $\bm{u}$ as the velocity vector and $p$ denoting the pressure field. We have stated the equations in non-dimensional form, using an appropriate characteristic length $(L_0)$ and velocity $(u_0),$ based upon which the dependent and independent variables are rendered dimensionless. 
This latter process introduces the non-dimensional fluid viscosity as the  inverse Reynolds number with $Re = \frac{u_0 L_0}{\nu}$ and $\nu$ is the kinematic viscosity. 

To fully describe a mixing process, an additional equation is introduced that tracks the two fluids (and their mixing) in the form of a passive scalar $\theta.$ This passive scalar, which takes on the value of $\theta=0$ in fluid one, $\theta = 1$ in fluid two, and values in-between for a mixture of the two fluids, is governed by an advection-diffusion equation of the form  
\begin{align}
\partial_t \theta + \bm{u}\cdot \nabla\theta - Pe^{-1}\nabla^2\theta = 0. \label{Eq:ScalEq}
\end{align}
Another non-dimensional parameter, the P\'eclet number $Pe,$ is defined as $Pe = \frac{u_0 L_0}{\kappa}$ and represents the dimensionless form of the diffusivity $\kappa$ of the mixing process. The passive scalar will be used to track the composition of the binary mixture, to measure mixedness of our fluid and to subsequently design active stirring strategies to optimise this mixedness. 

\subsection{Measuring mixedness}

Numerous studies have considered effective measures for quantifying mixedness, among them \cite{liu2008,Thiffeault2012,Foures2014}, where the latter performed a comparative study of energy norm, variance and mix-norm. While all measures have been successful in generating efficient mixing strategies within an optimisation framework, the most effective has been found to be the mix-norm defined by
\begin{align}
||\theta||_{\rm{mix}} = \frac{1}{V_\Omega}\int_\Omega ||\nabla^k \theta(\bm{x},t)||\text{d}\Omega .
\label{eq:Variance}
\end{align}
We note that $ k=0 $ is equivalent to the variance of the passive scalar.
The mix-norm, which is a Sobolev norm of fractional and/or negative index $k,$ was introduced by \cite{Mathew2007} and has its mathematical origins in measure theory. The mix-norm, which requires $k$ to be negative \citep{Thiffeault2012}, focuses specifically on the amount of small scales in the passive scalar and the transfer between large and small scales during mixing. There are two mechanisms which drive mixing, advection and diffusion, which are active and passive mechanisms, respectively. As we are only able to directly control advective processes, the seemingly most efficient way to exploit the passive diffusion process is by creating small scales and elongated filaments, using advective processes. Once these filaments have been created, diffusion will be far more effective and rapid. This emphasis in scale-transfer is best represented by the mix-norm -- making the mix-norm a crucial ingredient of our optimisation attempt.
We wish to stress, however, that the choice of measure does not affect the optimisation framework: no conceptual (but certainly quantitative) changes  would occur if, for example, the variance were chosen instead.

\subsection{Fluid-solid interactions}
\begin{figure}
\centering
\includegraphics[width=1.5in]{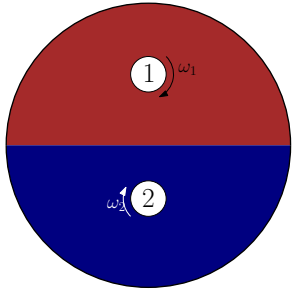}
\caption{Representative geometric setup of the computational mixing examples.}
\label{Fig:Geometry}
\end{figure}
We accomplish our mixing with the help of embedded stirrers; this setup defines our attempt as a fluid-solid interaction problem. A representative geometry of our setup is displayed in figure \ref{Fig:Geometry}. We will use a penalisation method \citep{arquis1984} to reformulate the problem. In essence, this method recasts the governing equations by including additional penalisation terms, which model the solid stirrers as Brinkman-style porous media with a vanishing permeability $C_{\eta}.$ As the permeability tends to zero, it has been shown that the modified equations converge towards solutions of the Navier-Stokes equations for the complex geometry \citep{Angot1999}. The attractiveness of this method lies in its simple implementation, its flexibility in expressing boundary conditions and its numerical efficiency. Solids are simply implemented using an indicator function, $\chi$, which is also referred to as the mask function.
As we will treat a number of stirrers, we will introduce sub-masks which describe each individual solid
\begin{align}
\chi_i =
\begin{cases}
1, \text{	if } \bm{x}\in \Omega_{s_i} \\
0, \text{	else } 
\end{cases}.
\end{align}
Since we impose a non-overlapping geometry we can combine these sub-masks into one general mask $\chi$ according to
\begin{align}
\chi = \sum_{i=1}^n\chi_i
\end{align}
which allows us to distinguish between the solid and the fluid parts of the computational domain.
An additional benefit of this method is the rather straightforward treatment of time-dependent (moving) solids. The movement of solid bodies through the fluid is achieved by remapping the mask function according to the updated locations of the bodies, without any need for a complex regridding operation.
The mask function allows us to augment the Navier-Stokes equations (\ref{Eq:GovEq}) which then model our fluid-solid interactions as
\begin{align}
\partial_t \bm{u} + \bm{u}\cdot \nabla \bm{u} + \frac{\chi}{C_\eta}\bm{u} - \frac{\chi_i}{C_\eta}\bm{u}_{s,i} + \nabla p - Re^{-1}\nabla^2 \bm{u} & = 0, \\
\nabla \cdot \bm{u} & = 0.  \label{Eq:PenGovEq}
\end{align}
We note that for $\chi=0,$ we describe the fluid domain and recover the standard Navier-Stokes equations.
Previous studies have shown that the optimal value of $C_\eta,$ that provides the most accurate numerical result, is proportional to $(\Delta x)^2;$ see \cite{Kadoch2012}. 
This penalisation approach will analogously be applied to the passive scalar equation (\ref{Eq:ScalEq}), yielding the penalised version \citep{Engels2015}
\begin{align}
\partial_t\theta + (1-\chi)\bm{u}\cdot\nabla \theta + \chi_i(\bm{u}_{s_i}\cdot \theta) - \nabla \cdot \left(\left[Pe^{-1}(1-\chi)+\frac{\chi}{C_\eta}\right]\nabla \theta\right) = 0
\end{align}
The terms $(1-\chi)\bm{u}\cdot\nabla \theta$ and $\chi_i(\bm{u}_{s_i}\cdot \theta) - \nabla \cdot \left(\left[Pe^{-1}(1-\chi)+\chi/ C_\eta\right]\nabla \theta\right)$ prevent the scalar field $\theta$ from advecting or diffusing, respectively, into any of the solid bodies, while the term $\chi_i(\bm{u}_{s_i}\cdot \theta)$ imposes the transport of the passive scalar with the velocity of the solid.

\subsection{Numerical method}

As details of the numerical method can be found in \cite{Eggl2018}, we will only briefly touch upon implementation specifics; for more technicalities the interested reader is referred to the original paper. 
Starting point for the numerical implementation of the penalised governing equations is the open-source software FluSI \citep{Engels2015}.
 This {\tt Fortran 90} Fourier-based, pseudo-spectral code solves the nonlinear, three-dimensional, incompressible Navier-Stokes equations on an equidistant grid and incorporates adaptive time-stepping as well as periodically uncoupled domains. A formulation in spectral space and a pressure-projection approach is used. The fluid-solid interactions are treated by the Brinkmann-style penalisation method discussed above, and specific grid layers are used to impose outflow and open boundaries.
 
 Following the spectral discretisation strategy implemented in FluSI, we will replace the spatial derivatives with a multiplication of the discretised velocity, pressure and scalar fields by a Fourier discretisation matrix. To this end, we introduce the discrete differentiation analog
 \begin{align}
     \frac{\partial}{\partial \bm{x}_i} \rightarrow \mathsf{A}_i
 \end{align}
 where $\mathsf{A}_i$ is a $n\times n$ matrix with $n$ as the number of grid points in a single coordinate direction. To represent the associated gradient vector $\nabla$, we introduce the tensor $\mathbf{A}$, with $\mathsf{A}_i$ as its entries. State variables are similarly discretised, i.e. $\bm{u}$ and $\bm{u}_{s,i}$ become $3\times n$ vectors containing the fluid and solid velocity components, respectively. We note that $(\bm{u}_{s,i})_j$ is a $1\times n$ vector and refers to the $j^{\text{th}}$ velocity component of the $i^{\text{th}}$ solid. The variables $\chi,\theta$ and $p$ turn into $1\times n$ vectors that represent the mask, passive scalar field and pressure, respectively. With these new variables, and assuming Einstein summation convention, we can state the spatially discretised form of the governing equations as follows
 \begin{align}
  \partial_t \bm{u}+ \bm{u}_j \circ \left[ {\mathsf{A}}_j\bm{u}
    \right]+ \frac{\chi}{C_{\eta}} \circ \bm{u} -
  \frac{\chi_i}{C_{\eta}} \circ \bm{u}_{s,i} +
  \mathbf{A} p - Re^{-1} {\mathsf{A}}_i {\mathsf{A}}_i \bm{u} &=
  0, \label{FLUSI:Eq1}\\
  {\mathsf{A}}_i \bm{u}_i &= 0, \\
  \partial_t \theta - {\mathsf{A}}_i \left( \left[ Pe^{-1} \left(
    \bm{1}-\chi \right) + \kappa\chi \right] \circ {\mathsf{A}}_i
  \theta \right) + \left( \bm{1}-\chi \right) \circ \bm{u}_j \circ
  \left[ {\mathsf{A}}_j \theta \right]
  + \chi_i \circ \left( \bm{u}_{s,i} \right)_j \circ \left[
    {\mathsf{A}}_j \theta \right] &= 0 \label{FLUSI:Eq2}
\end{align}
where $\circ$ denotes the Hadamard (elementwise) product \citep{Horn2012}. To enforce the continuity equation, we utilise a operator-splitting approach, producing a pressure Poisson equation of the form
\begin{equation}
  {\mathsf{A}}_j {\mathsf{A}}_j p + {\mathsf{A}}_i \left( \bm{u}_j
  \circ \left[ {\mathsf{A}}_j \bm{u}_i \right] \right) +
        {\mathsf{A}}_i \left[ \frac{\chi}{C_{\eta}} \circ
          \bm{u}- \frac{\chi_i}{C_{\eta}} \circ
          \bm{u}_{s,i} \right] = 0\label{FLUSI:Eq3}
\end{equation}
which completes the discretised system of equations describing the motion of the fluid and passive scalar. 

\section{Optimisation framework}
\label{Sec:OptFrame}
\subsection{The cost functional}

In the previous study by \cite{Eggl2018}, the above system of equations has been used to optimise the velocities of rotating circular stirrers or the eccentricities of rotating elliptical stirrers.
The primary conclusion from this study was that (i) increasing the rotational velocity generally enhanced mixing efficiency, and (ii) the shape of the stirrers changed significantly, once vortex shedding from distorted geometries could be exploited. The fusing of these two effects leaves unclear the role of pure shape optimisation. For this reason, we will concentrate in this study only on topology changes and assess their isolated effect on the overall mixing efficiency. 
More specifically, the control variables of our optimisation are shape-describing parameters. As a consequence of this setup, there is no need for an otherwise required $L_2$-penalisation of $\bm{u}_s.$ We thus can state our cost functional solely in terms of the mixing measure as 

\begin{align}
\mathcal{J} = \frac{1}{V_\Omega}\int_\Omega ||\nabla^{k}\theta||\text{d} \Omega\biggr|_{T}. \label{Eq:ContCost}
\end{align}
For our study, we select the mix-norm with an exponent $k$ of $-2/3$, slightly deviating from the previous choices of $k=-1$ \citep{Lin2011} or $k=-1/2$ \citep{Foures2014}, but falling within this range. In line with the discretisation of the governing equations, we present the semi-discretised form of the cost functional $\mathcal{J}$ as
\begin{align}
\mathcal{J} = \frac{\sqrt{[\mathsf{A}^{-2/3}\theta]^H M [\mathsf{A}^{-2/3}\theta]}}{V_\Omega}\biggr|_{T}.
\end{align}
where $\mathsf{M}$ is a symmetric, positive weight matrix, taking into account the grid spacing and spatial weightings, and the superscript $^H$ refers to the conjugate transpose of the particular matrix or vector. As we are confined to an equidistant grid of $n^2$ mesh points, the weight matrix in our case is simply a diagonal matrix of the form,
\begin{align}
\mathsf{M} = \frac{1}{n^2} I,
\end{align}
with $I$ as the identity matrix. We restate that this type of norm encourages the formation of filamented structures in the passive scalar field, which are subsequently subjected to diffusion, once the optimisation window has passed.

So far, we have not imposed any constraints on the control variables. However, as the shapes need to remain physically realisable, as well as non-overlapping, some restrictions need to be enforced. However, these constraints are not included directly in the cost functional, but are instead applied during the update step of the control variables, once  the direct-adjoint optimisation is completed.
 
 \subsection{The adjoint system}

 The cost functional $\mathcal{J}$ depends explicitly only on $\theta.$ The dependence on our control variables as well as other state variables is implicit via their effect on $\theta.$ The minimisation of our cost functional $\mathcal{J}$ has to observe the full governing equations, which is accomplished by embedding the governing equations as PDE-constraints into a single augmented cost functional $\mathcal{L}$. This embedding is achieved using Lagrange multipliers or adjoint variables. The augmented cost functional then takes the following form
 \begin{align}
 \mathcal{L} = \mathcal{J} - &\int_0^T (\bm{u}^\dag)^H_kM\left\{\partial_t\bm{u} + \bm{u}_j\circ[\mathsf{A}_j\bm{u}] + \frac{\chi}{C_\eta}\circ\bm{u}-\frac{\chi_i}{C_\eta}\circ\bm{u_{s,i}}+
 \mathbf{A}p - Re^{-1}[\mathsf{A}_i\mathsf{A}_i\bm{u}]\right\}_k \nonumber\\
 &+ p^{\dag,H}M\left\{[\mathsf{A}_i\mathsf{A}_i]p + \mathsf{A}_i(u_j\circ[\mathsf{A}_ju_i]) + \mathsf{A}_i\left[\frac{\chi}{C_\eta}-\frac{\chi_i}{C_\eta}\circ\bm{u_{s,i}}\right]\right\} \nonumber \\
 & +\theta^{\dag,H}M\left\{\partial_t\theta + (1-\chi)\circ\bm{u}_j\circ[\mathsf{A}_j\theta]-\chi\circ(\bm{u}_{s,i})_j\circ[\mathsf{A}_j\theta]-\mathsf{A}_i([Pe^{-1}(1-\chi)+\kappa\chi]\circ \mathsf{A}_i\theta)\right\} \nonumber \\
 &+\chi_i^{\dag,H}\mathsf{M}[\chi_i-g_i(\bm{x},t)] \text{d} t .
 \label{Eq:AugLagrange}
 \end{align}
This expression demonstrates that all governing equations and boundary conditions have been enforced using Lagrange multipliers, denoted by $(\cdot)^\dag$; these multipliers are also referred to as adjoint variables and will subsequently supply our optimisation algorithm with sensitivity or gradient information. The key to deriving the optimality conditions is the minimisation of the augmented cost functional $\mathcal{L}$ with respect to all dependent variables, direct or adjoint. This is accomplished by taking the first variation of (\ref{Eq:AugLagrange}) and evaluating it to zero, i.e., enforcing $\delta \mathcal{L} = 0,$ breaking it into its various subcomponents. 
 We note that the variation with respect to the adjoint variables recovers the governing equations (momentum, continuity and passive scalar equations). The first variation with respect to the state variables involves more algebraic effort, but leads to an evolution equation/algebraic equation for the adjoint variables. The explicit derivation will not be shown here; details of it can be found in  the appendix of \cite{Eggl2018}. The final adjoint equations, governing $\bm{u}^{\dag}, p^{\dag}$ and $\theta^{\dag}$ reads as follows 

\begin{align}
  \partial_t\bm{u}^{\dag}_i - \Pi^{\dag}_k\circ
          [{\mathsf{A}}_i\bm{u}_k ] - {\mathsf{A}}_j^H[\bm{u}_j \circ
            \Pi^{\dag}_i] \label{Adjoint:EqStart} -
          \frac{\chi}{C_{\eta}}\circ\Pi^{\dag}_i + Re^{-1}{\mathsf{A}}_j^H
               {\mathsf{A}}_j^H \bm{u}_i^{\dag}
   -(\bm{1}-\chi)\circ\theta^{\dag}\circ[ {\mathsf{A}}_i\theta] &=&0,
  \\
  {\mathsf{A}}_j^H\Pi_j^\dag &=& 0, \\
  \partial_t\theta^{\dag} -{\mathsf{A}}_j^H[(\bm{1}-\chi)\circ\bm{u}_j
    \circ \theta^{\dag}] + {\mathsf{A}}_i^H([Pe^{ -1}(\bm{1}-\chi) +
    \kappa\chi] \circ {\mathsf{A}}_i^H\theta^{\dag})
  -{\mathsf{A}}_j^H[ \chi_i \circ (\bm{u}_{s,i})_j\ \circ
    \theta^{\dag}] &=&0,
  \label{Adjoint:EqEnd}
\end{align}
with initial conditions
\begin{align}
  \bm{u}^{\dag}(\bm{x},T) = 0, \qquad
  \theta^{\dag}(\bm{x},T) = \frac{2}{V_\Omega}(\mathsf{A}^{-2/3})^HM(\mathsf{A}^{-2/3}\theta).
\end{align}
The optimality condition arises from the first variation with respect to the control variable, i.e., the mask $\chi.$ We obtain 
\begin{align}
\chi^{\dag}_i = \left[\theta^{\dag}\circ[\mathsf{A}_j\theta] - \frac{\Pi_j^{\dag}}{C_\eta}\right]\circ(\bm{u}_j-(\bm{u}_{s,i})_j) + (\kappa - Pe^{-1}\mathsf{A}_j^{H}\theta^{\dag}\circ \mathsf{A}_j\theta \label{Eq:OptChi}
\end{align}
where $\Pi^{\dag}_i = \bm{u}_i^{\dag} + \mathsf{A}_i^{H}p^{\dag}$.
We recognise that the above system constitutes a set of simultaneous equations that has to be solved for the optimal solution. In practice, this solution is obtained progressively, by solving the direct and adjoint equations exactly and by iterating the optimality condition until a user-specified criterion is satisfied. 

Due to the fact that our governing equations are nonlinear and, as a consequence, the optimisation problem is non-convex, the iterative scheme outlined above can only yield a local minimum. A globally optimal solution cannot be guaranteed. For this reason, care must be exercised  when choosing an initial condition, as this choice may have a noticeable influence on the overall result.

The optimality equation (\ref{Eq:OptChi}), based on the mask $\chi_i,$ allows for a maximum of flexibility in choosing the cross-sectional shape of the stirrers. To avoid discontinuous geometries and to reduce the dimensionality of the control space, we will make further modifications to the geometric representation of the stirrers' shape (beyond the mask function) to make the optimisation problem computationally more tractable. 

\subsection{Complex shape parameterisation}

Previous studies on stirrer shape optimisation dealt with a very simplified geometry, the morphing from circular to elliptical cross-sections, intended to validate the computational framework \citep{Eggl2018} rather than accommodate complex shapes. However, 
with a view towards industrially applicable configurations that cannot be approached by intuition alone, a geometric parameterisation with a far higher degree of flexibility -- while still conforming to the optimisation framework of the previous section -- has to be adopted. In our choice, we are motivated by our spatial discretisation and hence introduce a Fourier-based shape parameterisation \citep{Crimmins1982} shown to allow for an wide array of logically circular shapes. We have
\begin{subeqnarray}
f_x(\alpha) &=& \frac{a_0}{2}+\sum_{k=1}^n a_k\cos\left(k \alpha \right) - b_k \sin\left(k \alpha \right), \\
f_y(\alpha) &=& \frac{c_0}{2}+\sum_{k=1}^n c_k\cos\left(k \alpha \right) - d_k \sin\left(k \alpha \right),
\label{Eq:FourierShape}
\end{subeqnarray}
where $\alpha$ is a vector representing the discretised interval $[0,\ 2\pi ]$. A parametric curve in Cartesian coordinates is then given by 
\begin{subeqnarray}
x &=& f_x(\alpha), \\
y &=& f_y(\alpha),
\end{subeqnarray}
or, in a more compact form,
\begin{align}
    f(a_i,b_i,c_i,d_i) = [f_x,f_y],
\end{align}
which forms the basis for the optimisation of the $i^{\text{th}}$ stirrer shape.
The parametric representation of the stirrer geometries introduces additional computational issues that need to be addressed. The relation of the parametric shape back to the mask $\chi$ requires an indicator function which specifies whether a point falls inside or outside a chosen geometry. For the previous cases of circular or elliptical cross-sections this condition could be formulated in a trivial manner. As our new parameterisation allows for shapes of significant concavity, we have to introduce a more sophisticated scheme to distinguish the interior from the exterior of our shape. Furthermore, additional complications arise  from highly distorted geometric shapes whose barycentre falls outside the solid structure. To resolve these issues, we introduce the winding number $w(\bm{x})$. This number is defined (within a discrete setting) as 
\begin{align}
w(\bm{x}_k) &= \sum_{i=1}^{n} \phi_{i,i-1},
\end{align}
where $\phi_{i,j}$ denotes the angle of the arc, which is centred at the test point $\bm{x}_k$ and runs along the parametric curve from $\bm{x}_i$ to $\bm{x}_j$. As the points $\bm{x}_i,\bm{x}_j$ and the angle $\phi$ sweep the full perimeter of the parametric curve (from index $i$ to $n$), the total angle travelling around the curve will be $2\pi$ if the point $\bm{x}_k$ is inside the curve, or zero otherwise.

Combining this concept with our previous definition of the mask $\chi,$ we obtain
\begin{equation}
\chi_i(\bm{x},t) =
\begin{cases}
 1, \hspace{12pt} &w_i(\bm{x})=2\pi, \\
 0, \hspace{12pt} &w(\bm{x}) = 0,\\
\end{cases}
\end{equation}
We note that this formulation is discontinuous and thus would generate spurious numerical oscillations as the stirrers move across the Cartesian background grid. For this reason, we introduce a smoothing layer to avoid these artefacts. The design of this layer is further complicated, as the smoothing layer cannot simply be defined as a distance from the centre (as in \cite{Engels2015} or \cite{Eggl2018}); instead we must rely on the minimum distance from our chosen point to our curve, i.e.,
\begin{align}
d = \min_{\alpha} \sqrt{(x-f_x)^2+(y-f_y)^2} \label{eq:minDist}.
\end{align}
The angle of $\alpha$ that assumes this minimum value will be defined as $\tilde{\alpha}$ and can be uniquely determined for each point $(x,y)$ inside the shape.
We then augment our mask functions $\chi_i$ with $d$ to obtain the smoothed mask
\begin{equation}
\chi_i(x,t) =
\begin{cases}
 1, \hspace{12pt} &w=2\pi \ \text{and} \ d > h, \\
  \frac{1}{2}\left(1+\cos\left(\frac{\pi (h-d)}{h}\right)\right), & w=2\pi \ \text{and} \ d < h, \\
 0, \hspace{12pt} &w = 0.\\
\end{cases}
\end{equation}
With this formulation we ensure a continuous and differentiable mask-based geometry.

A benefit of using the parameterisation above is its adaptability to the adjoint system. Without loss of generality, we will focus on optimising the Lagrangian $\mathcal{L}$ with respect to one representative coefficient $a_k$, with the remaining Fourier coefficients following the same procedure. Since $\mathcal{L}$ does not explicitly depend on $a_k$, but is linked implicitly through the geometric mask, we have to express the derivative with respect to $a_k$ in terms of its partial components. We have 
\begin{align}
\frac{\partial \mathcal{L}}{\partial a_k} &= \frac{\partial \mathcal{L}}{\partial \chi_i}\frac{\partial \chi_i}{\partial d}\frac{\partial d}{\partial a_k}.
\label{eq:LagrangeFourier}
\end{align}
The first term, the gradient $\frac{\partial \mathcal{L}}{\partial \chi_i}$,  follows from equation (\ref{Eq:AugLagrange}), while the second term has been reported (in analogous form) in \cite{Eggl2018}. The last term is novel and will be further discussed below.  While equation (\ref{eq:minDist}) does include a minimisation, the derivative term in equation (\ref{eq:LagrangeFourier}) is independent of this minimum and does not affect the subsequent derivation. Given the form of the parameterisation (equation (\ref{Eq:FourierShape})), the derivative is in essence isolating the wavenumber within the minimum distance function, i.e.,
\begin{align}
\frac{\partial d}{\partial a_k} = \frac{(f_x-x)\cos(k\alpha)}{d}\biggr|_{\tilde{\alpha}}.
\end{align}

Combining this expression with the terms above, we arrive at the complete formulation of the optimality condition with respect to the chosen Fourier-coefficient $a_k$. The composite nature of equation (\ref{eq:LagrangeFourier}) allows for an efficient implementation as the first two terms are independent of our choice of $a_k$ and can be determined ahead of the direct-adjoint iterations. Furthermore, the final term only changes minimally, either by including a $(f_y-y)$ term or by replacing a sine- with a cosine-term, depending on our choice of coefficient. In short, the calculation of the optimality condition is efficient and readily scalable, as the majority of the required terms can be calculated \emph{a priori}.

One further issue that needs attention is area conservation. The optimisation routine at this point is not constrained by the cross-sectional area of the stirrers and therefore can and will modify the stirrers accordingly. This may encourage shapes that achieve better mixing simply by changing the size of the stirrers. To counter this tendency, we need to define and conserve the area of our parametric curves. We start by deriving the area as follows

\begin{align}
A &= -\int_{\alpha=0}^{\alpha=2\pi} f_y\text{d} f_x \\
&= -\int_{0}^{2\pi} f_y\frac{\partial f_x}{\partial \alpha} \text{d} \alpha \\
&=  -\int_{0}^{2\pi}\left[\frac{c_0}{2}+\sum_{k=1}^n c_k\cos\left(k\alpha \right) - d_k \sin\left(k\alpha \right)\right]\left[\sum_{l=1}^n l(-a_l\sin\left(l \alpha \right) - b_l \cos\left(l\alpha \right))\right] \text{d} \alpha.
\end{align}
We will absorb the factor $\frac{1}{2}$ into $c_0$ to simplify the notation
\begin{align}
A &= \sum_{k=0}^n\sum_{l=1}^n l\left( \int_{0}^{2\pi}\left[ c_k\cos\left(k\alpha \right) - d_k \sin\left(k\alpha \right)\right]\left[ a_l\sin\left(l \alpha \right) + b_l \cos\left(l\alpha \right)\right] \text{d} \alpha\right) \\
&=\sum_{k=0}^n\sum_{l=1}^n l\Biggl( \int_{0}^{2\pi} c_kb_l\cos\left(k\alpha \right)\cos\left(l\alpha \right) + c_ka_l\cos\left(k\alpha \right)\sin\left(l\alpha \right) \nonumber \\
 & \hspace{48pt} -d_ka_l\sin\left(k \alpha \right)\sin\left(l\alpha \right) - d_kb_l \sin\left(k \alpha \right)\cos\left(l\alpha \right) \text{d} \alpha\Biggr) \label{eq:areaConserve}.
\end{align}

Observing orthogonality between the trigonometric functions over the interval $(0,2\pi)$, the only non-zero contributions are given by $k=l$ for the $\cos \times \cos$ and $\sin \times \sin$ terms. This results in 
\begin{align}
\sum_{k=0}^n\sum_{l=1}^n  \int_{0}^{2\pi} c_kb_l\cos\left(k\alpha \right)\cos\left(l\alpha \right) \text{d} \alpha &=\sum_{k=0}^n \pi c_kb_k,  \\
\sum_{k=0}^n\sum_{l=1}^n  \int_{0}^{2\pi} c_ka_l\cos\left(k\alpha \right)\sin\left(l\alpha \right) \text{d} \alpha &=  0, \\
\sum_{k=1}^n\sum_{l=1}^n  \int_{0}^{2\pi} d_kb_l\sin\left(k\alpha \right)\cos\left(l\alpha \right) \text{d} \alpha &=   0, \\
\sum_{k=1}^n\sum_{l=1}^n  \int_{0}^{2\pi} d_ka_l\sin\left(k \alpha \right)\sin\left(l \alpha \right) \text{d} \alpha &=\sum_{k=1}^n d_ka_k\pi.
\end{align}
Using the above and substituting into equation (\ref{eq:areaConserve}) we obtain
\begin{align}
A &=\sum_{k=1}^n \pi k(c_kb_k - d_ka_k).
\end{align}
The special cases for the area of a circle and ellipse can be recovered easily. We will use the general expression to normalise the coefficients and thus preserve the initial area. More specifically, for an initial area
\begin{align}
\tilde{A} &=\sum_{k=1}^n \pi k(\tilde{c}_k\tilde{b}_k - \tilde{d}_k\tilde{a}_k),
\end{align}
we multiply all Fourier coefficients for the new geometric shape (with area $A$) by
 $\sqrt{\frac{\tilde{A}}{A}}$ to obtain
\begin{align}
\sum_{k=1}^n \pi k(\sqrt{\frac{\tilde{A}}{A}}c_k\sqrt{\frac{\tilde{A}}{A}}b_k - \sqrt{\frac{\tilde{A}}{A}}d_k\sqrt{\frac{\tilde{A}}{A}}a_k) &= \frac{\tilde{A}}{A}\sum_{k=1}^n \pi k(c_kb_k - d_ka_k) \\
&=  \frac{\tilde{A}}{A}A \\
&= \tilde{A},
\end{align}
i.e., an area-normalised new shape.

We now have in place the key ingredients to successfully optimise any number of stirrers of arbitrary Fourier-based shapes. This geometric parameterisation has been found to be particularly simple to implement and robust to optimise.

\subsection{Shape constraints}

\begin{figure}
\centering
\includegraphics[trim = 1.5in 0.6in 1.3in 0.4in, clip,width=0.4\textwidth]{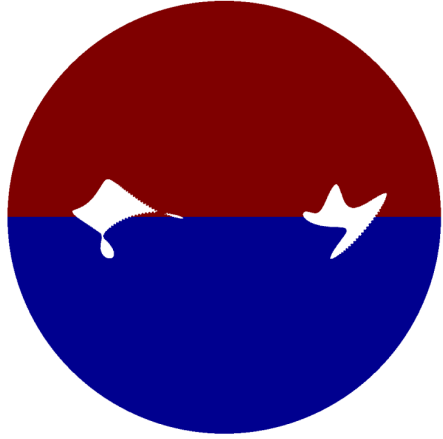}
\caption{Unconstrained shape optimisation of two rotating stirrers. We can see that the left shape has resulted in two pinched off pieces due to  the optimisation.}
\label{Fig:PinchOff}
\end{figure}

The possibilities within the Fourier-based approach 
require additional constraints that go beyond area conservation. These additional constraints arise from the desire to generate physically feasible shapes that are implementable in an industrial setting.

When generating geometric shapes with a multitude of Fourier coefficients, it is rather likely that the parametric curve self-intersects, an example of which can be seen in the left shape of figure \ref{Fig:PinchOff}. In essence, the direct-adjoint optimisation overrules the imposed number of prescribed stirrers by generating new pinched-off geometric components that subsequently enhance mixing in later optimisation steps, albeit with the same dynamic constraints. While this option may be beneficial to the mix-norm, the geometric shape is, of course, unrealisable in a realistic setting. It may appear sufficient to simply expand the distance between the points on the curve and therefore avoid such a pinching off. However, the full path along the parametric curve needs to be considered as well, as any pure thickening of the curve will inevitably leave some twisting due to our use of the winding number $w$. What is instead needed is an untwisting strategy. 

\begin{figure}
    \begin{tabular}{cc}
     \begin{subfigure}{0.4\textwidth}
         \centering
         \includegraphics[trim=5cm 2cm 3cm 6cm,width=1.5in]{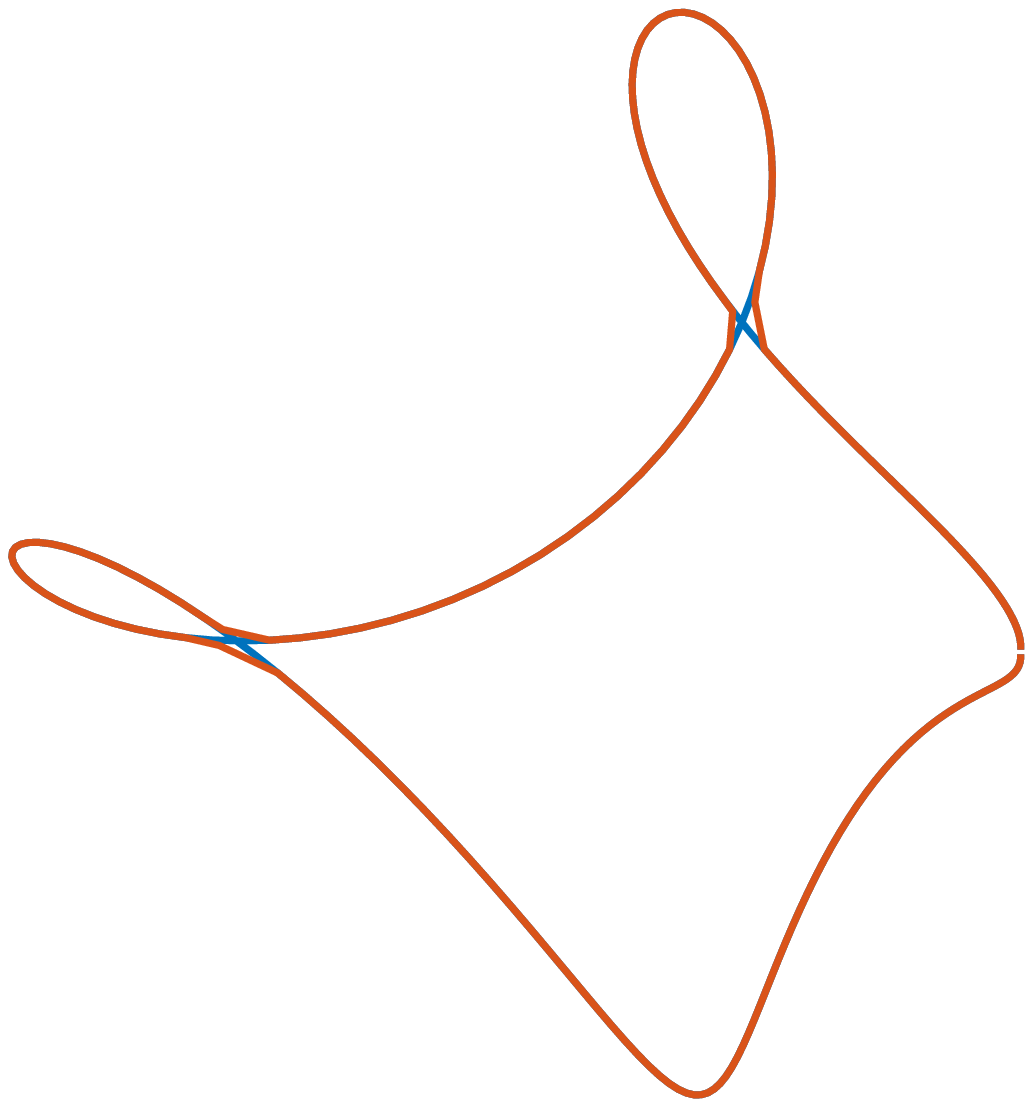}
         \caption{Untwisting algorithm applied to the left shape from figure \ref{Fig:PinchOff}. The new curve is displayed in red.}
         \label{Fig:UnTwist}
     \end{subfigure}
     &
     \begin{subfigure}{0.4\textwidth}
         \centering
         \includegraphics[trim=5cm 2cm 3cm 6cm,width=1.5in]{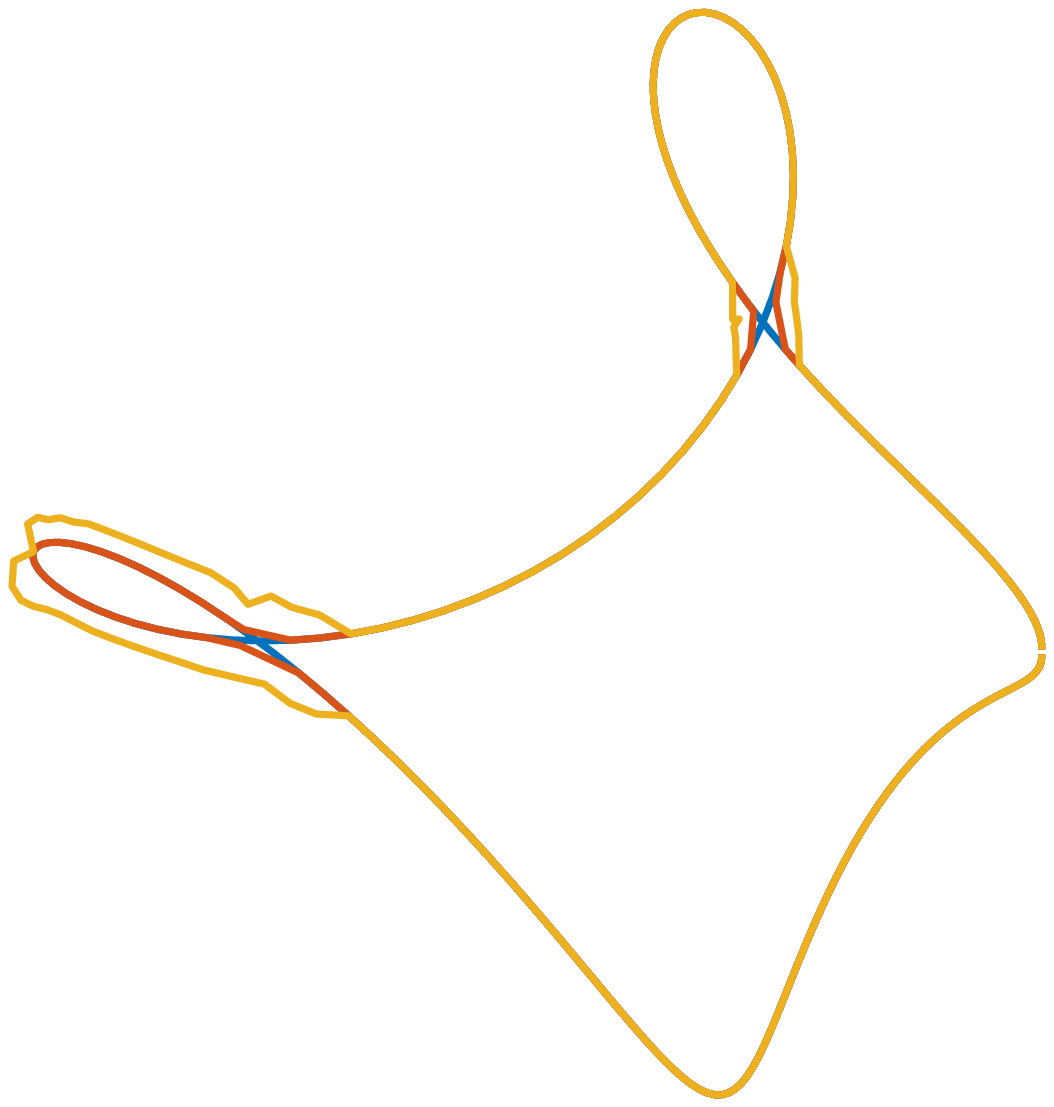}
         \caption{Application of thickening routine to untwisted shape.}
         \label{Fig:Thicken}
     \end{subfigure}
     \end{tabular}
\end{figure}

We consider two line segments $\Vec{AB}$ and $\Vec{CD}$ of arbitrary length and evaluate a criterion for their mutual intersection. If this criterion is met, without loss of generality, we exchange $B$ and $D$ and mirror the parameterisation of the part of the parametric curve from $A$ to $D$ to generate an untwisted, and simply connected, shape. An example of such an untwisting can be seen in figure \ref{Fig:UnTwist}, which contains the previously shown shape that had two pinched-off areas.
After this untwisting procedure, new Fourier coefficients can be calculated, supplied to the area calculation and passed on to further optimisation steps.

The above untwisting would be sufficient for a continuous setting. However, on a finite grid we can arrive at pinched-off shapes even after untwisting has been performed. This is due to small geometric features that fall below the resolution of the Cartesian background grid despite the fact that the shape is simply connected. To avoid this complication, we introduce a displacement strategy for points that fall below a chosen distance, so that their modified distances remain above the minimum grid spacing. This displacement is based on the following steps. Consider two points which are a distance $r$ apart. If this distance falls below the minimum distance constraint, i.e., $r<r_{min}$, we displace the two points along their respective outward pointing normals to the parametric curve. The amount of displacement, $\zeta$, is given by a Buckingham potential \citep{Buckingham1938}
\begin{align}
    \zeta = r_{min}\left[e^{-\frac{r}{r_{min}}}\right].
\end{align}
A demonstration of this procedure is displayed in figure \ref{Fig:Thicken}.

This strategy, in combination with the untwisting method, ensures a simply connected shape that is properly represented on our underlying grid.

\section{Optimisation results}
\label{Sec:Results}
\subsection{Geometric configuration}

\begin{figure}
  \centering
  \includegraphics[width=0.7\textwidth]{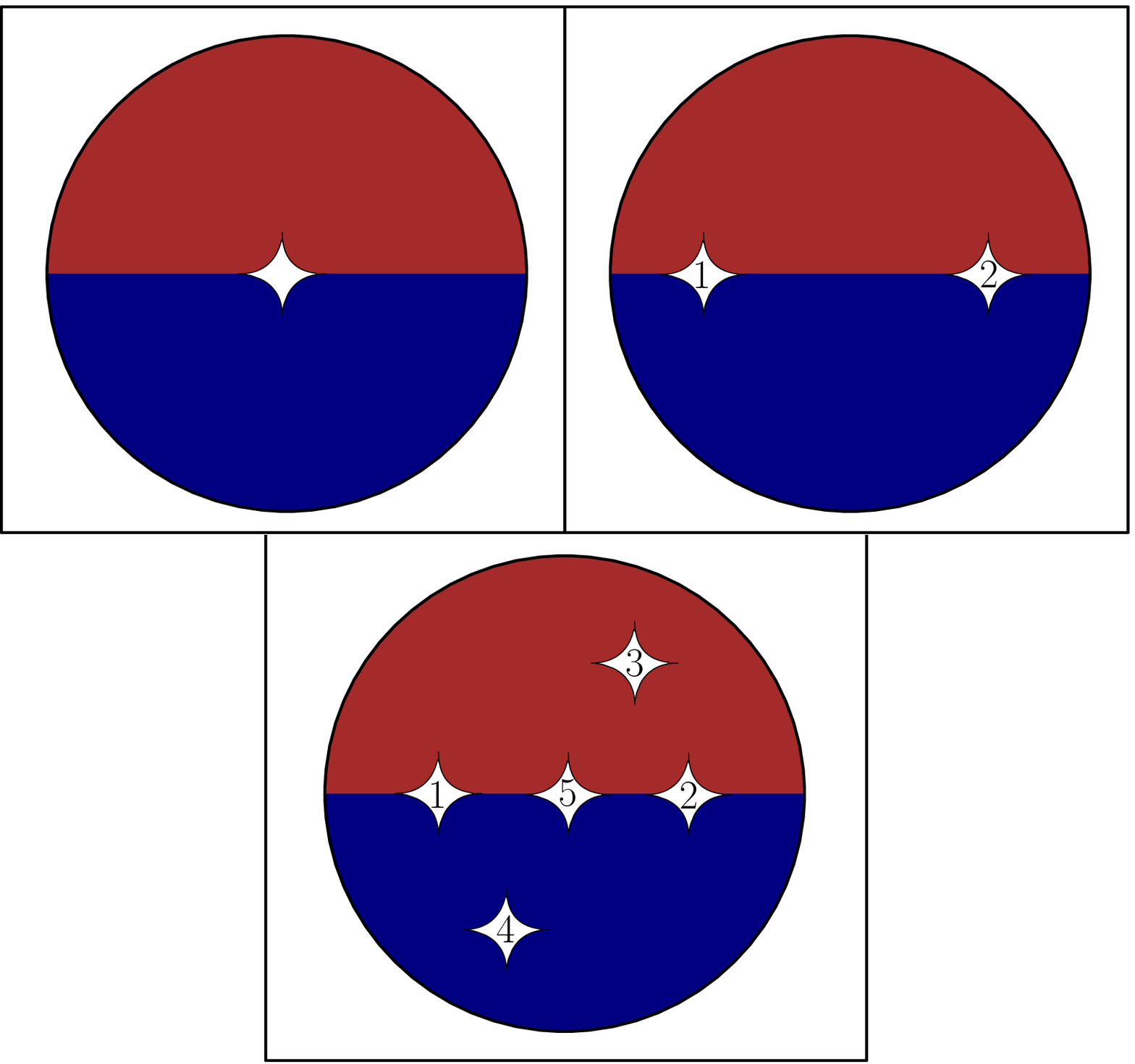}
  \caption{\label{Fig:ShapeGeometriesPic} Sketch of the initial configurations
    for the three shape optimisation cases. Case 1 (top left): one centred stirrer
    rotating about its centre. Case 2 (top right): two stirrers aligned on the
    horizontal axis, rotating in the clockwise directions.
    Case 3 (bottom centre): five rotating stirrers, with three cylinders
    aligned along the horizontal axis and two cylinders vertically displaced.}
\end{figure}

Our previous investigations of shape-optimisation by direct-adjoint looping \citep{Eggl2018} using a circular initial geometry for the stirrers revealed a severely impeded convergence towards an optimal design as the initial solid-body rotation did not induce vortex shedding and hence resulted in negligible gradient information for enhanced mixing. As a consequence, a substantial amount of computational effort was expended, and ultimately squandered, over many iterations, before vortex shedding set in and further modifications to the stirrers' cross-section yielded improved mixedness. For this reason a non-circular initial geometry will be chosen, in an attempt to introduce vortex shedding already in the first iteration and on which the direct-adjoint optimisation will further improve. A more expedited convergence is thus expected. More particularly, we initialise the stirrers as four-pointed hypocycloids (astroids), examples of which can be seen in figure \ref{Fig:ShapeGeometriesPic}. While vortex shedding from the cusps will commence immediately, initially the vortices will be trapped in the astroids' cavities and show little interaction. This situation is then the starting point for our direct-adjoint shape-optimisation algorithm. 

The geometric arrangements and rotational speed ($\omega = \frac{2\pi}{8}$) of the star-shaped stirrers have been chosen identical to our previous study \citep{Eggl2018} to allow a qualitative comparison. The Reynolds and P\'eclet number have also been taken from the same reference. In addition, the control space for the stirrer shapes has been limited to the five largest wave numbers, i.e.,
\begin{align}
    \{a_i, b_i, c_i, d_i \} \text{ for } i=1,\ldots,5.
\end{align}
We curtail the optimisation horizon to one single rotation, i.e., to $T=8$. This choice focuses the optimisation effort on short-term features, rather than on residual diffusive phenomena. In effect, we force the optimisation to achieve efficient mixing with severe limitations in time.

\subsection{Case 1: Shape optimisation of one stirrer}
\begin{figure}
\centering
 \begin{adjustwidth}{-1cm}{-1cm}
\begin{tabular}{cc}
\includegraphics[trim=3cm 0 3cm 0,width=0.5\textwidth]{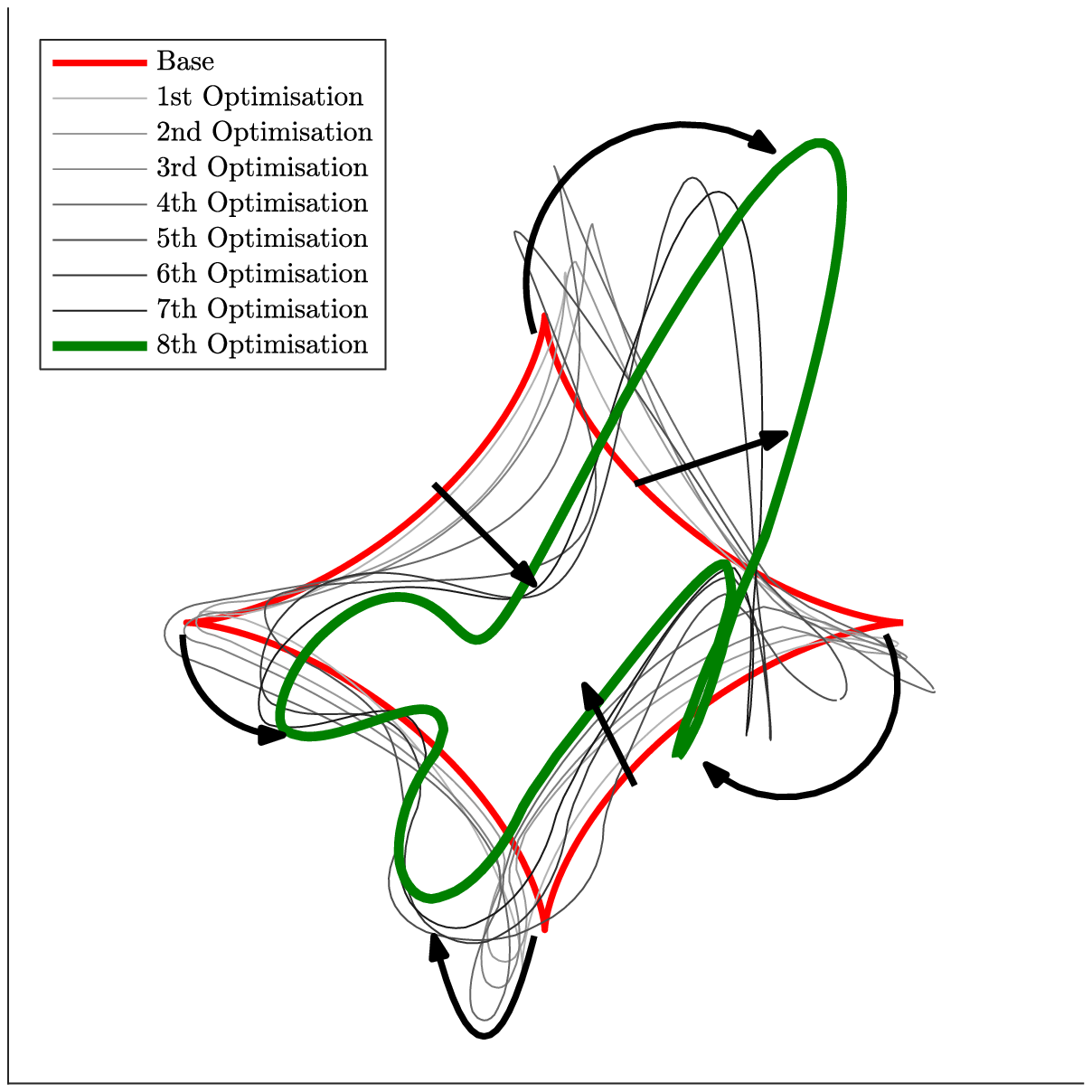} &
\includegraphics[trim=3cm 0 3cm 0,width=0.5\textwidth]{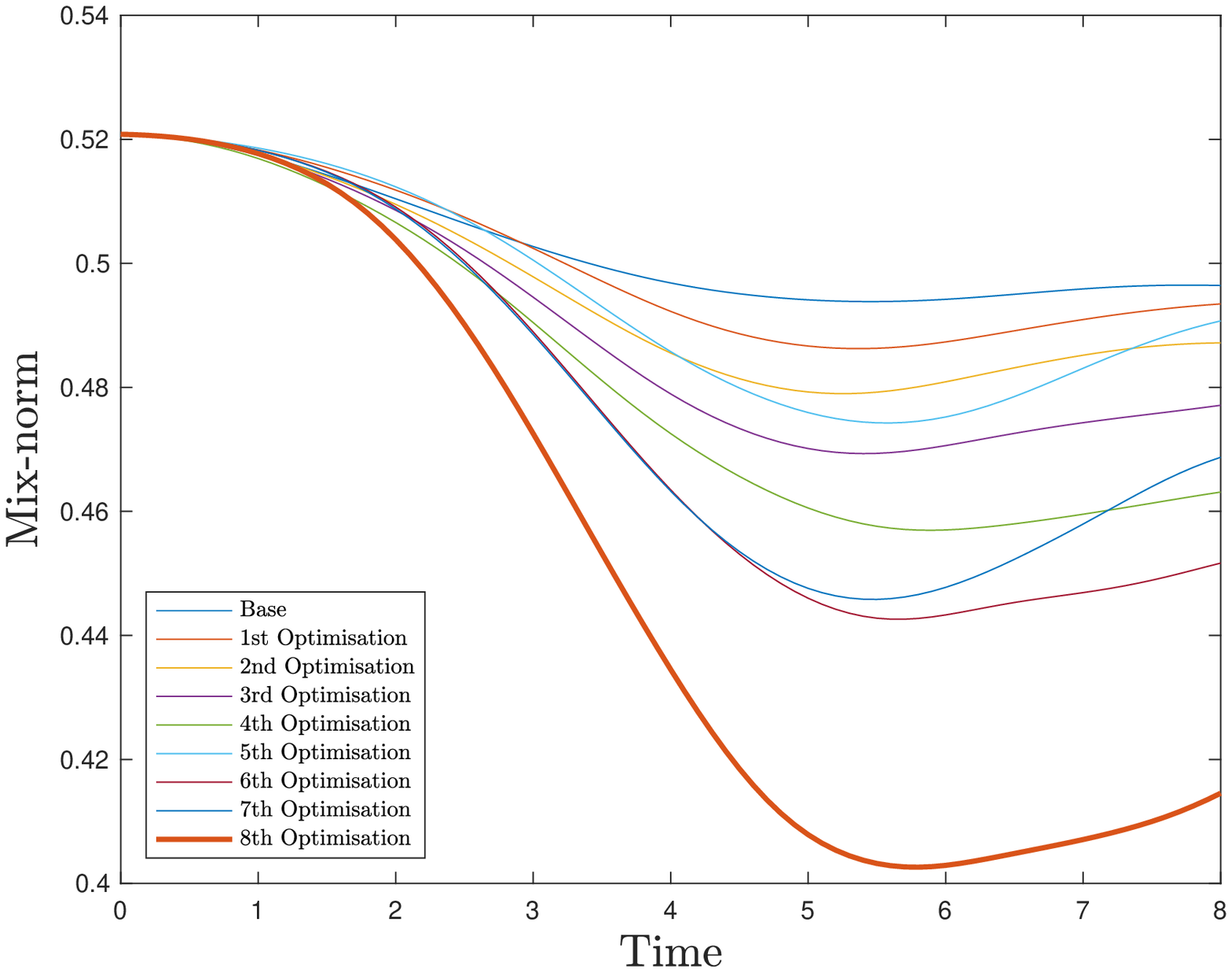}
\end{tabular}
\end{adjustwidth}
\includegraphics[trim=3cm 0 3cm 0,width=0.5\textwidth]{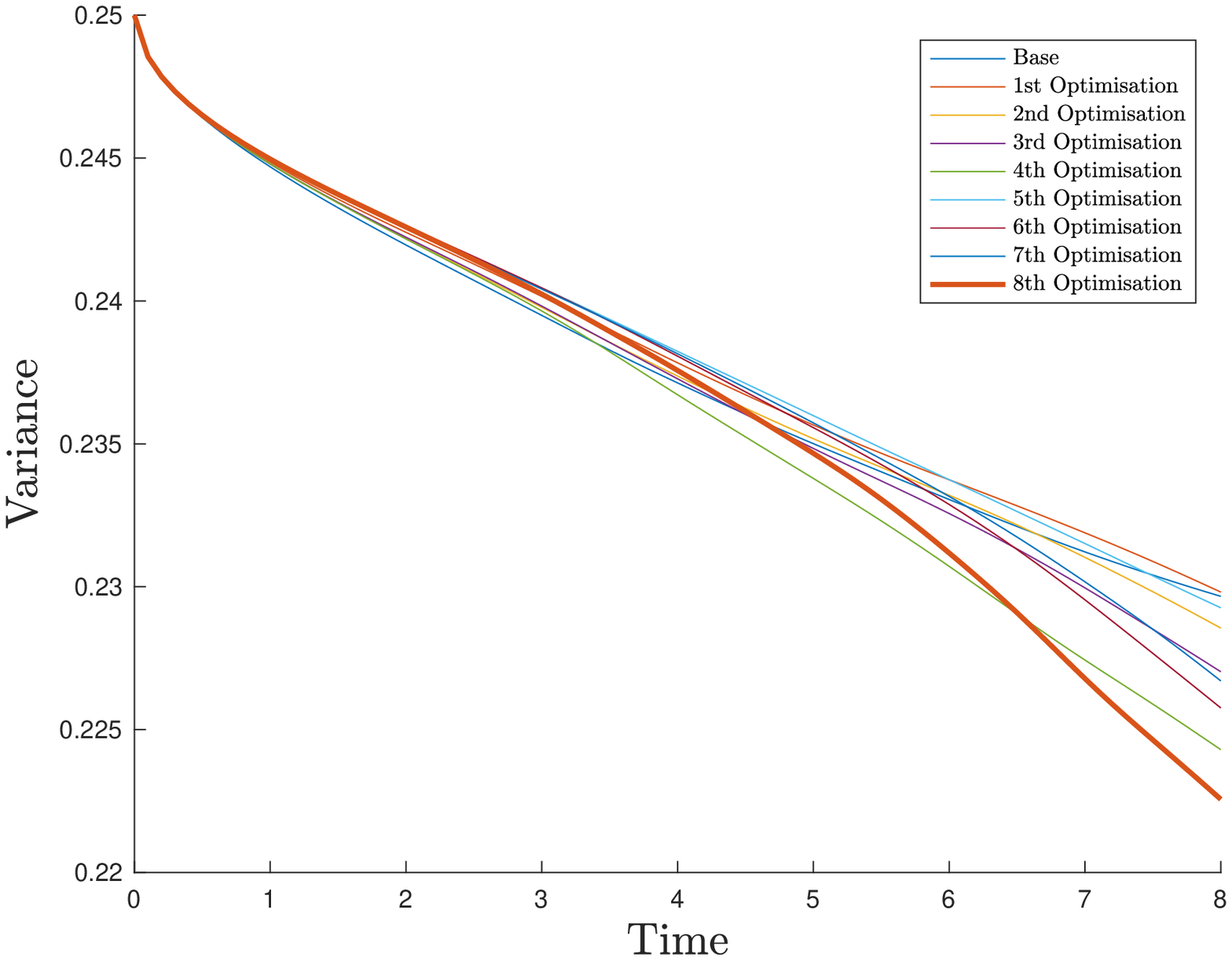}
\caption{Case 1: mixing optimisation using one
    fixed rotating stirrer. No untwisting or thickening was applied. (a) Evolution of the shape over the optimisation cycle. The red line is the initial configuration, the green line the final shape. Arrows have been added in black to illustrate the changes in shape. (b) Mix-norm of the passive scalar versus time
    $t \in [0,\ T].$ {(c) Variance of the passive scalar versus time
    $t \in [0,\ T].$}}
\label{Fig:Shape1}
\end{figure}
\begin{figure}
  \centering
  \begin{tabular}{cc}
    \includegraphics[trim = 1.5in 0.6in 1.3in 0.4in, clip,width=0.35\textwidth]{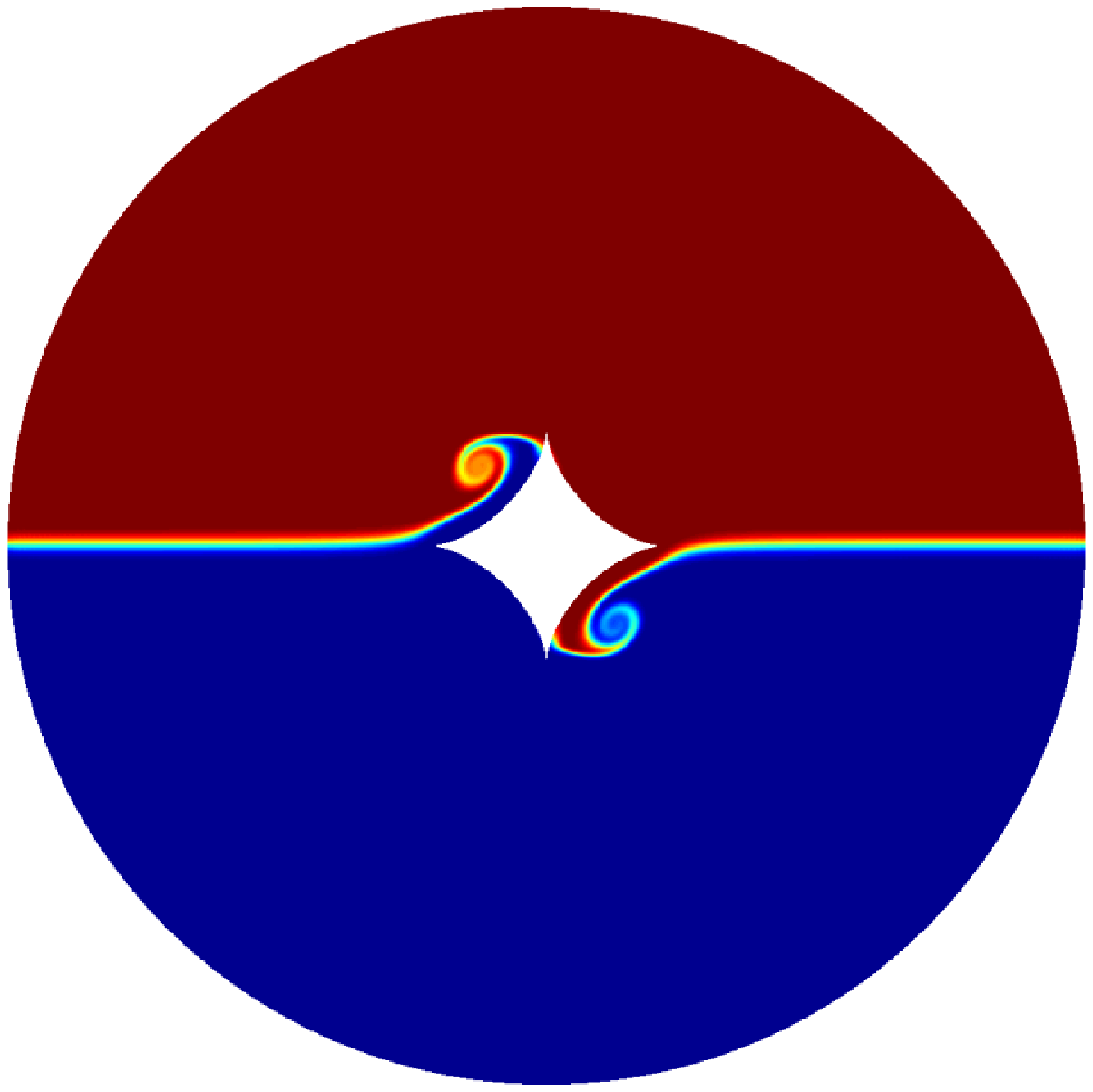} & \hspace{1truecm}
    \includegraphics[trim = 1.5in 0.6in 1.3in 0.4in, clip,width=0.35\textwidth]{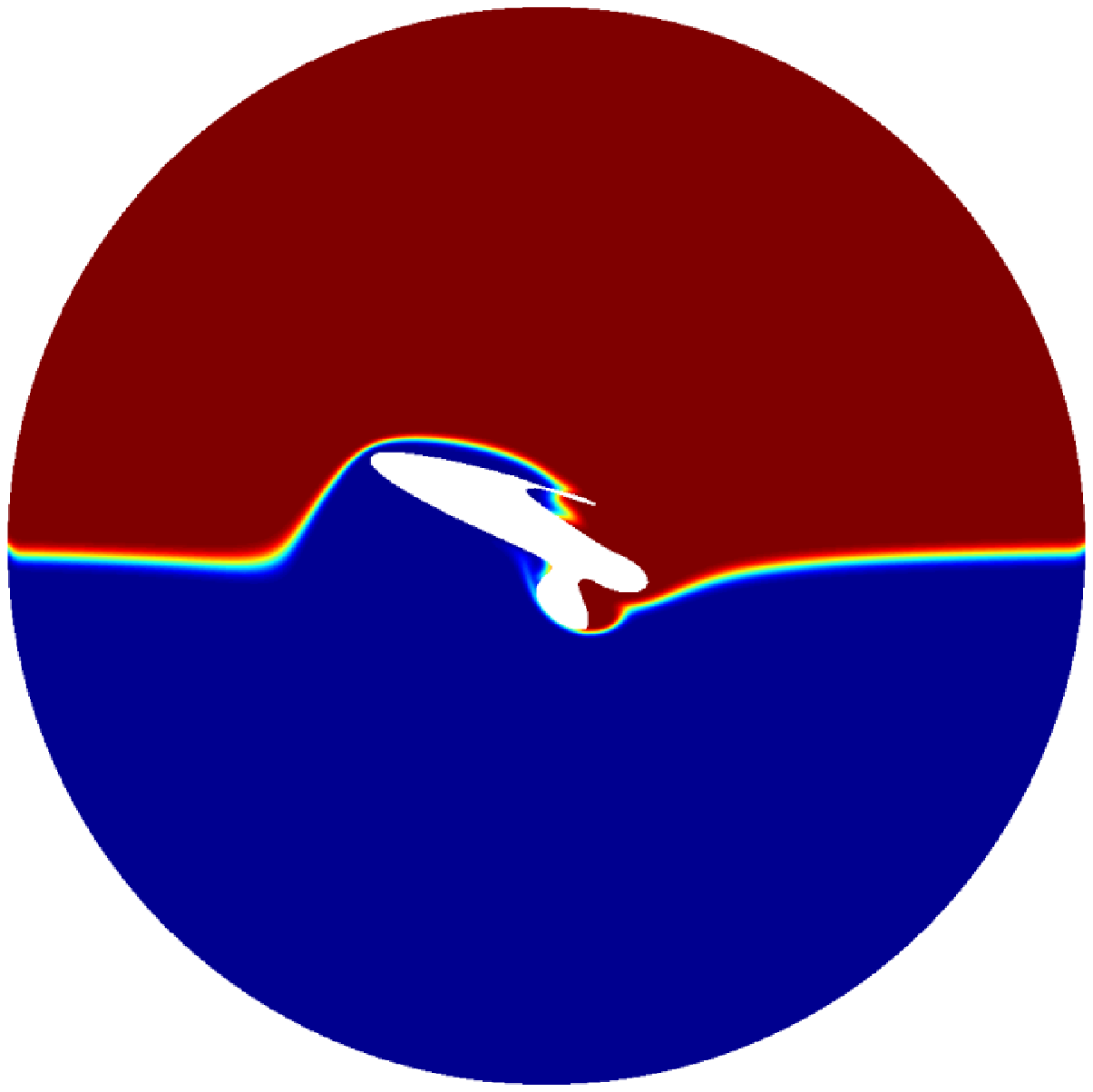} \\
    \includegraphics[trim = 1.5in 0.6in 1.3in 0.4in, clip,width=0.35\textwidth]{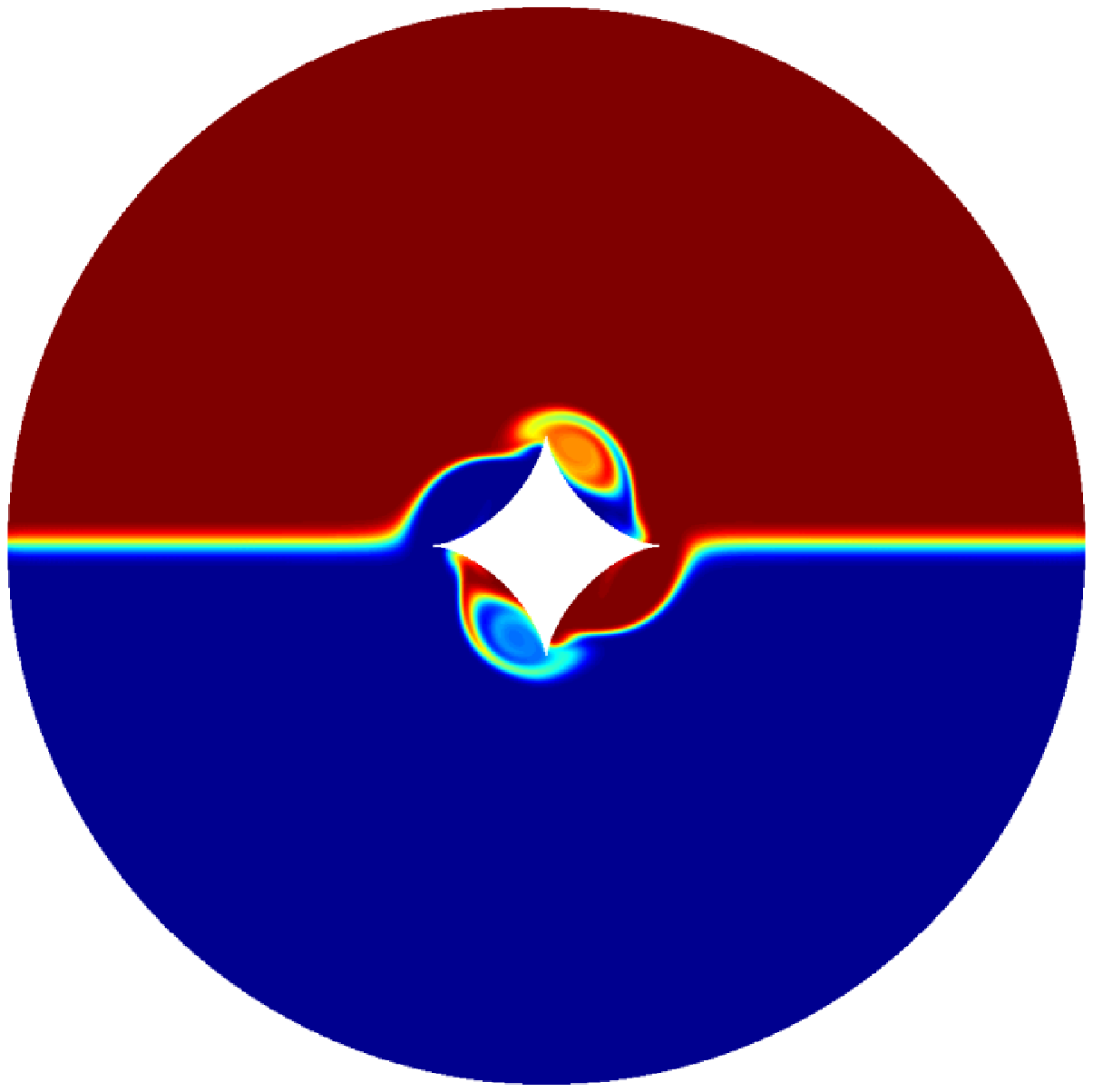} & \hspace{1truecm}
    \includegraphics[trim = 1.5in 0.6in 1.3in 0.4in, clip,width=0.35\textwidth]{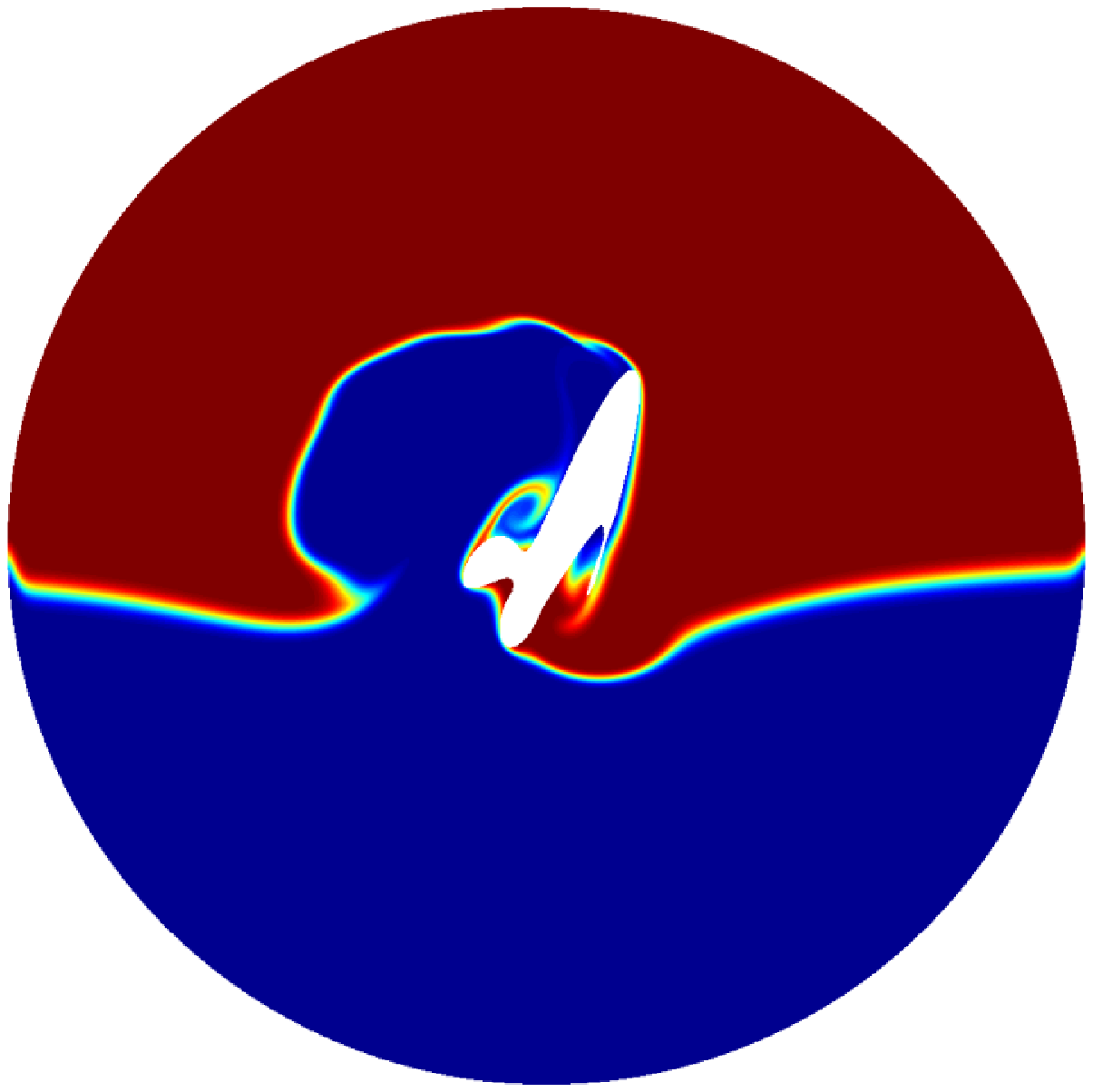} \\
    \includegraphics[trim = 1.5in 0.6in 1.3in 0.4in, clip,width=0.35\textwidth]{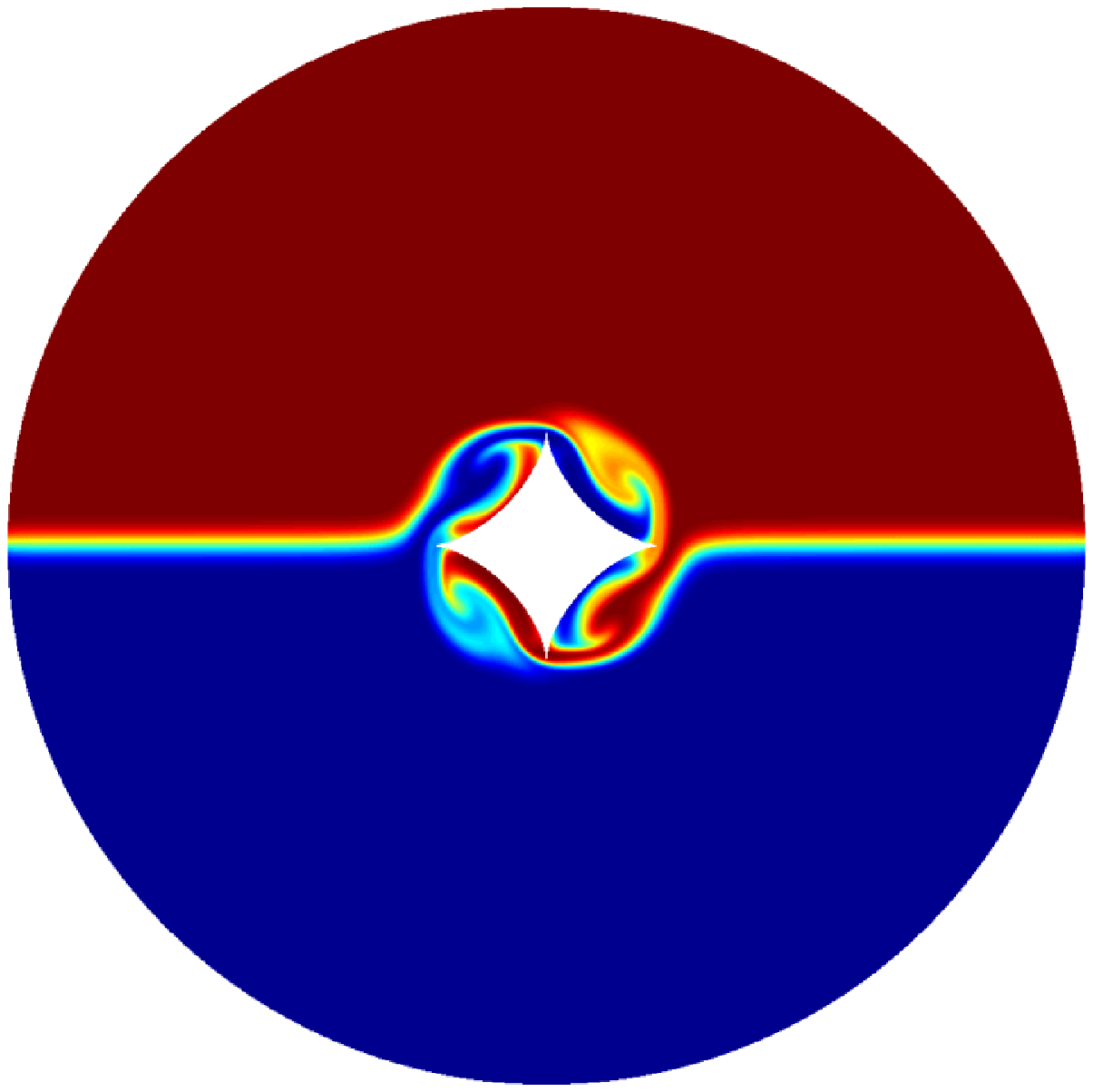} & \hspace{1truecm}
    \includegraphics[trim = 1.5in 0.6in 1.3in 0.4in, clip,width=0.35\textwidth]{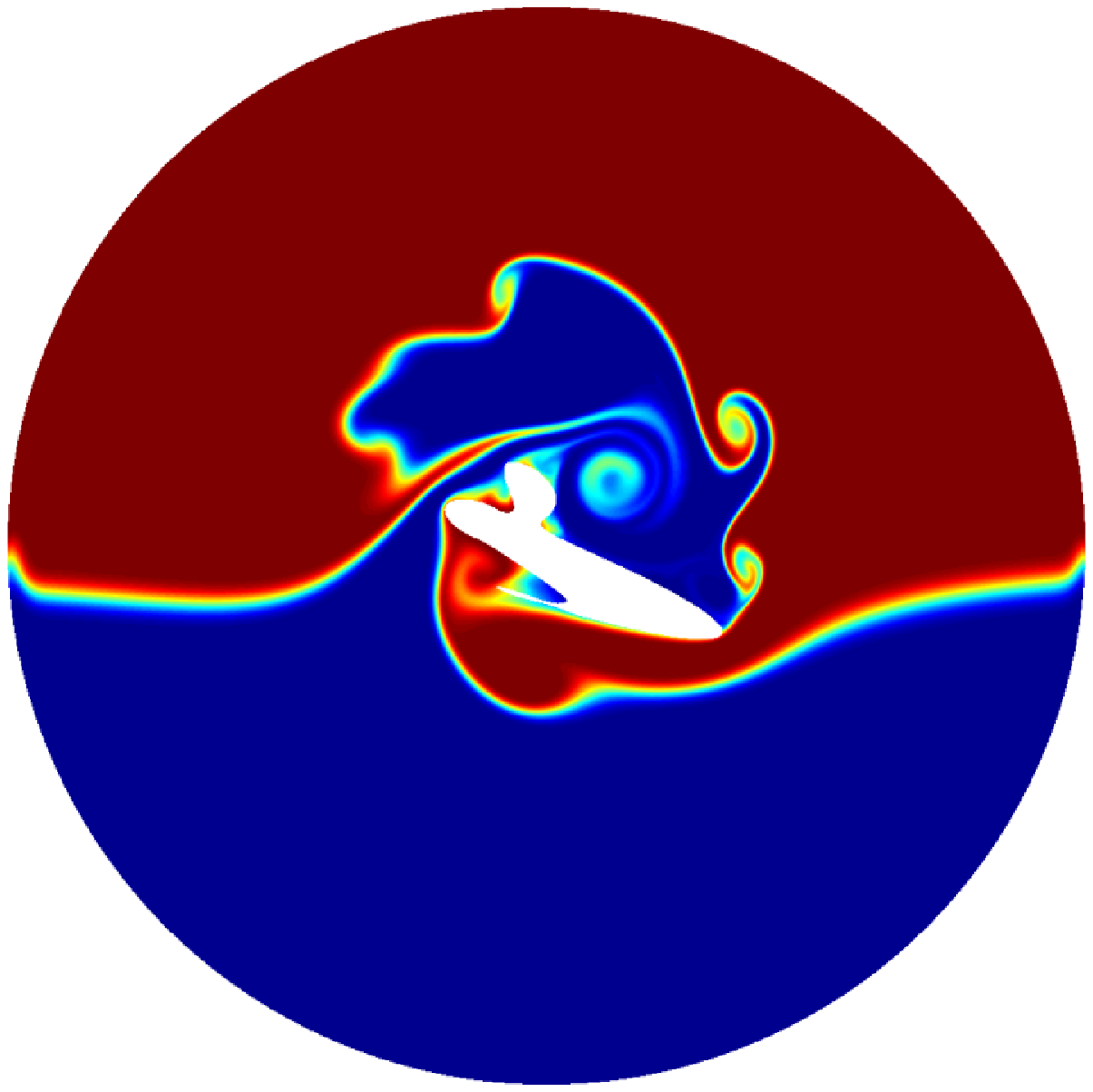} \\
    \includegraphics[trim = 1.5in 0.6in 1.3in 0.4in, clip,width=0.35\textwidth]{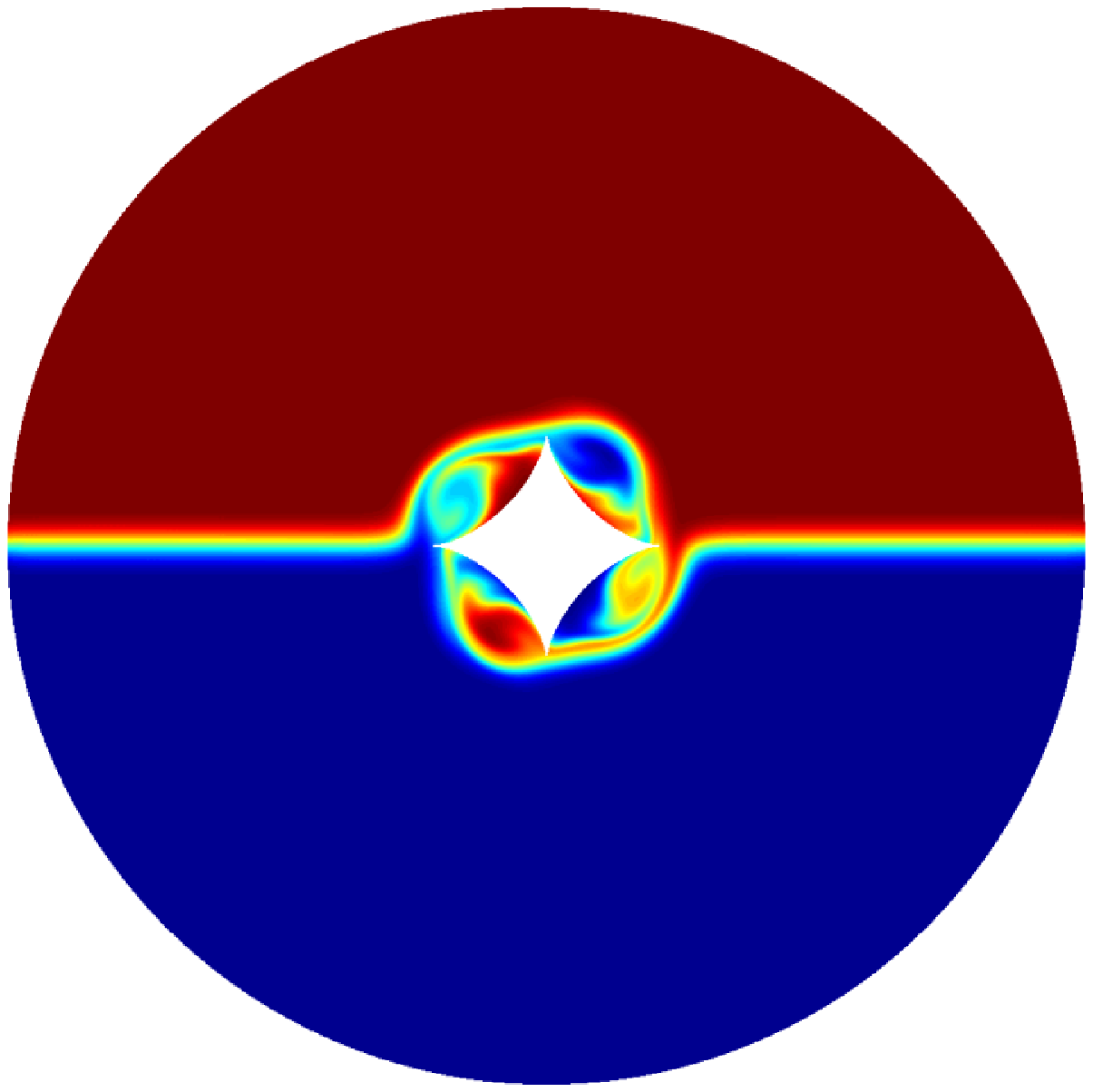} & \hspace{1truecm}
    \includegraphics[trim = 1.5in 0.6in 1.3in 0.4in, clip,width=0.35\textwidth]{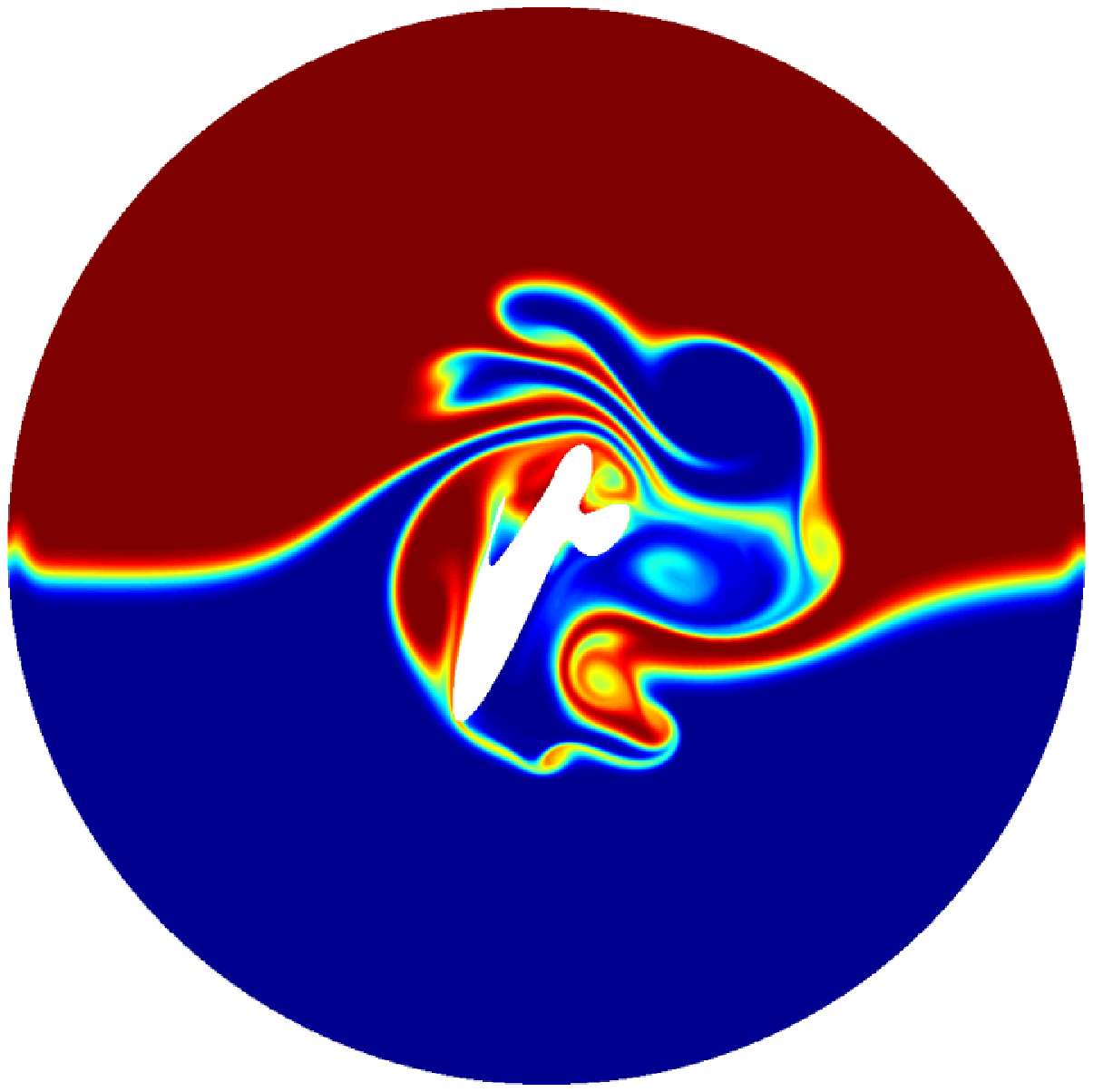}
  \end{tabular}
  \caption{\label{Fig:Shape1Pic} Case 1: mixing optimisation using one
    rotating stirrer. Left column: unoptimised
    configuration, with snapshots at $t = 2, 4, 6, 8$
    (top to bottom). Right column: after eight direct-adjoint
    optimisations, with snapshots at $t = 2, 4, 6, 8$
    (top to bottom). For videos of these scenarios please refer to
    {\tt{Shape1NoOpt.mp4}} and {\tt{Shape1Opt.mp4}} for the left and
    right column, respectively.}
\end{figure}

We begin by considering a single, centrally placed stirrer. The effect on the scalar field by the action of the unoptimised (astroidal) stirrer and the fully optimised stirrer is displayed in figure \ref{Fig:Shape1Pic}. We observe that the shape generated by the optimisation is drastically more asymmetric than the initial (symmetric) configuration. This complex shape induces the shedding of vortices and their mutual interaction. In the absence of a second stirrer, a fully cooperative mixing strategy is not possible; rather, the single stirrer must achieve a self-collaborative strategy among its shed vortices. One strategy is to elongate the stirrer's shape: with a constant rotational speed, a longer shape causes a higher velocity at the trailing edge and a more vigorous shedding of vortices. This principal tendency is further augmented by an accompanying alignment of the three remaining cusps along the axis of the principal cusp (see figure \ref{Fig:Shape1} for the gradual evolution into a final shape, with black arrows guiding the eye). We recall the conservation of the initial cross-sectional area of the stirrer, causing the observed stretching and thinning of the solid body. 

The converged shape, after our final iteration, leads to the formation of a trailing-line vortex street which can be seen at $t=6$ in the right column of figure \ref{Fig:Shape1Pic}. Furthermore, one of the protruding cusps near the centre of the stirrer generates a patch of high circulation which entrains unmixed passive scalar and, later in the rotation, launches vorticity towards the trailing-line vortex street ($t=6$ to $t=8$). We urge the reader to consult the animations in the supplemental material to fully appreciate this dynamic process.

One item of particular interest in this optimisation is the capability of the direct-adjoint looping, coupled with shape optimisation, to explore competing minima beyond a current local solution. This manifests itself in figure \ref{Fig:Shape1}b where the final values of the mix-norm after the fifth and seventh iteration are significantly higher, indicating worse mixing, than during the previous iterations. Further analysis of this observation reveals that the shapes (shown in figure \ref{Fig:Shape1}a) go through drastic changes, i.e., there is a determined shift from an initial, presumably optimal configuration to a more efficient one. Throughout the initial stages of the optimisation, the first attempts at a better shape reinforce a particular configuration. During this reinforcement, and owing to the global nature of the Fourier-based shape representation, the optimisation scheme becomes aware of a more optimal shape at a different point in design space. Though in the pursuit of this new configuration, the optimisation must pass through a regime of sub-optimality. Once this traversal is completed, we ultimately obtain a stirrer geometry with significantly enhanced mixing efficiency. Aiding in this behaviour is a strong and global link between all points of the Fourier-represented curves, and thus changes at a particular point are communicated effectively to the remaining curve. This strong global link allows the optimisation to probe multiple local minima. However, it must be called to mind that, due to the non-convexity of the problem, convergence to a global minimum cannot be guaranteed.

After eight optimisations we terminate the direct-adjoint loop, even though better mixing efficiency could be achieved in subsequent iterations, as the resulting shapes increasingly become structurally delicate and physically unfeasible. During these eight optimisations, no post-processing such as thickening or untwisting has become necessary; the resulting optimal shapes can thus conceivably be implemented in an industrial setting.

As a final observation, we note the rise of the mix-norm towards the end of the simulation. Similar behaviour has been observed in \cite{vermach2018}. This may seem counter-intuitive and raise the question why the optimisation did not terminate when the mix-norm is lowest. We remind the reader that our cost-functional focuses on the end-time evaluation of the mix-norm and, therefore, any influence prior to the final time horizon does not directly enter the optimisation. Parenthetically, the variance continues to decrease beyond the optimisation horizon, which implies that the fluid continues to mix.

\subsection{Case 2: Shape optimisation of two stirrers}

We complicate the geometry and optimisation scheme by introducing a second stirrer which we also place at the interface between the two fluids (see figure \ref{Fig:ShapeGeometriesPic}). Both stirrers are initially taken as a four-pointed hypocycloid and are placed sufficiently far from each other such that communication between them is weak. The interest in this configuration is two-fold: on one hand, we intend to assess the proximity-effects of the outer wall, on the other hand, we wish to investigate a possible collaborative mixing strategy between the two stirrers.

\begin{figure}
  \centering
  \begin{tabular}{cc}
    \includegraphics[trim = 1.5in 0.6in 1.3in 0.4in, clip,width=0.35\textwidth]{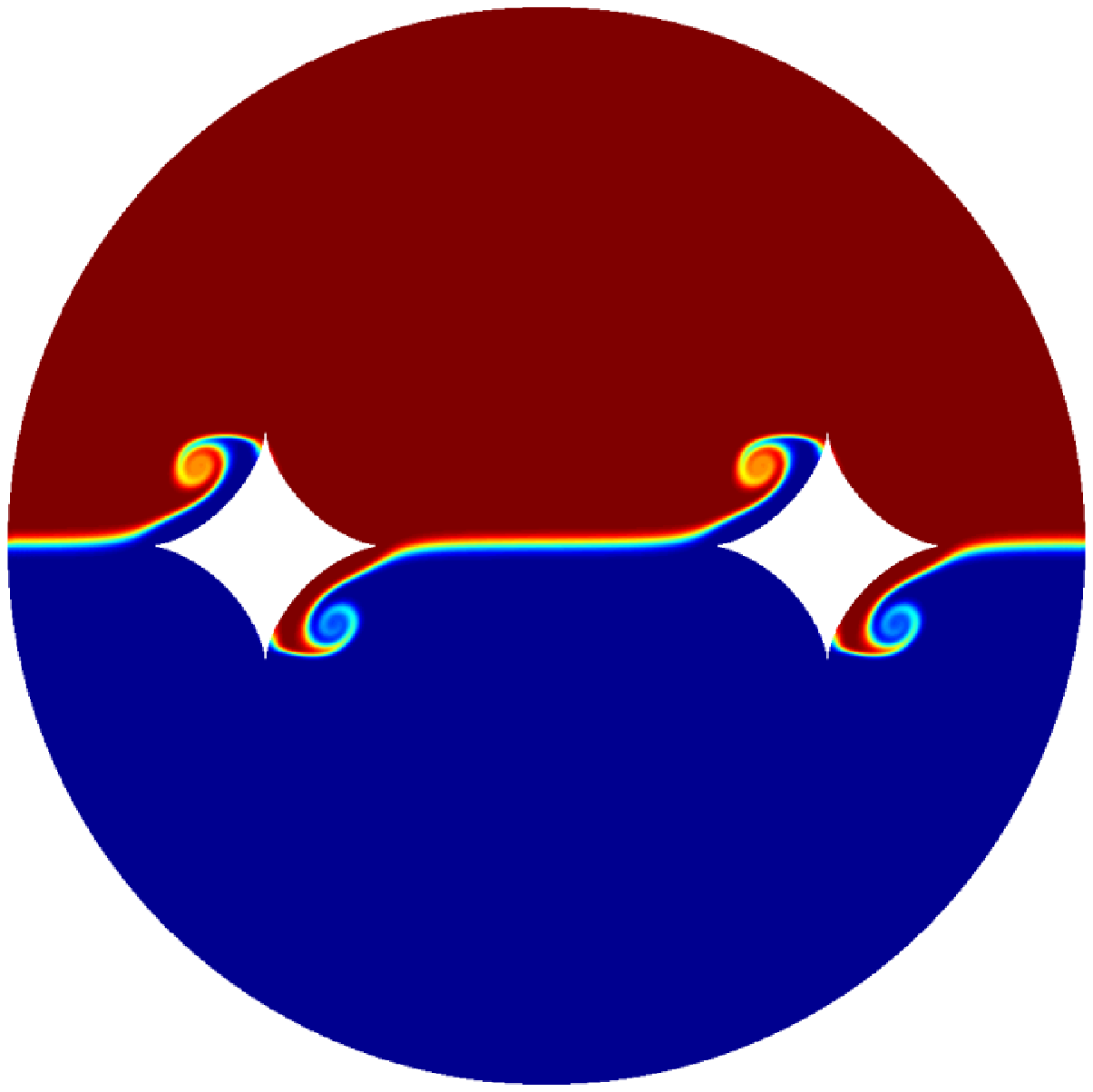} & \hspace{1truecm}
    \includegraphics[trim = 1.5in 0.6in 1.3in 0.4in, clip,width=0.35\textwidth]{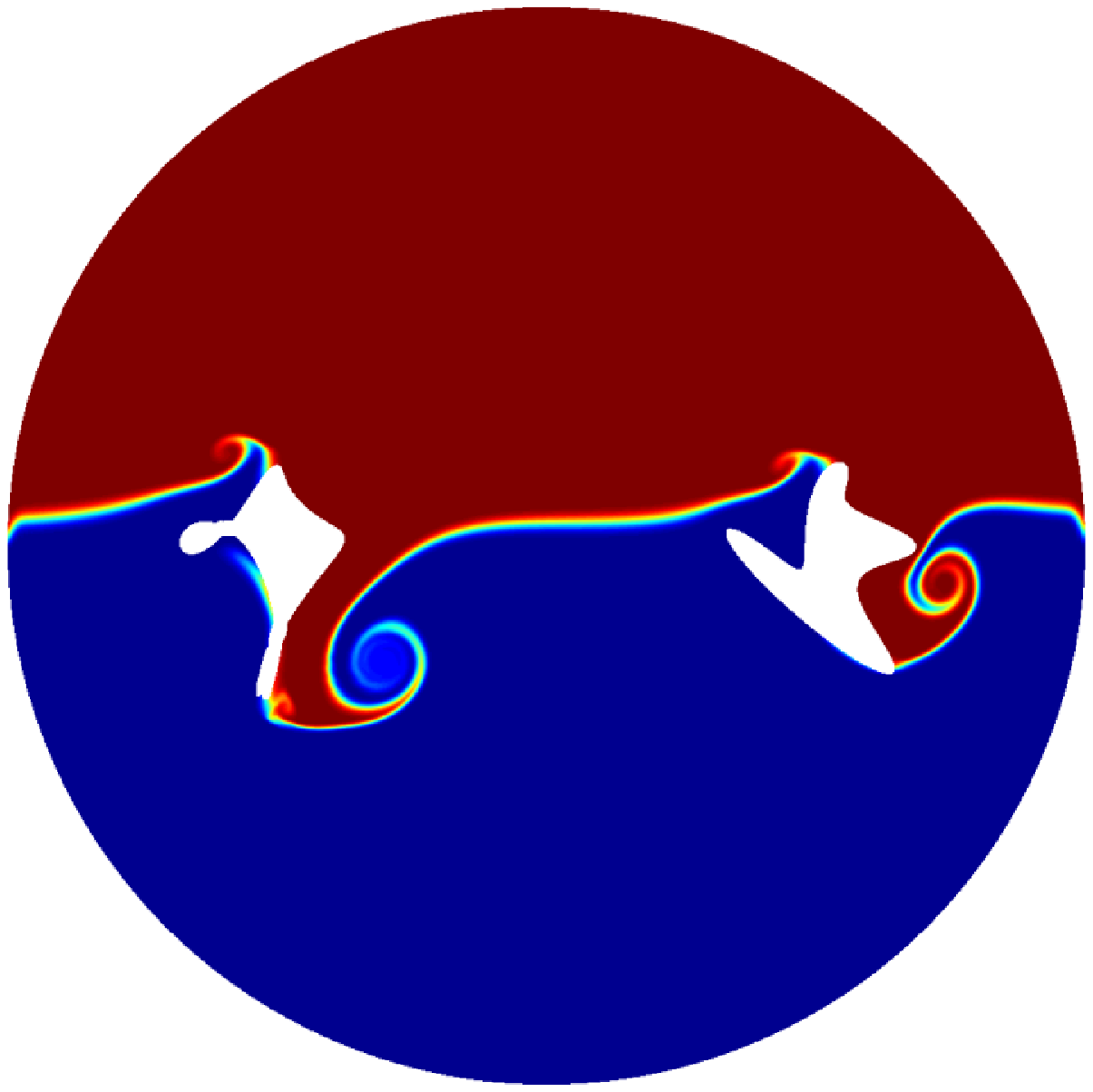} \\
    \includegraphics[trim = 1.5in 0.6in 1.3in 0.4in, clip,width=0.35\textwidth]{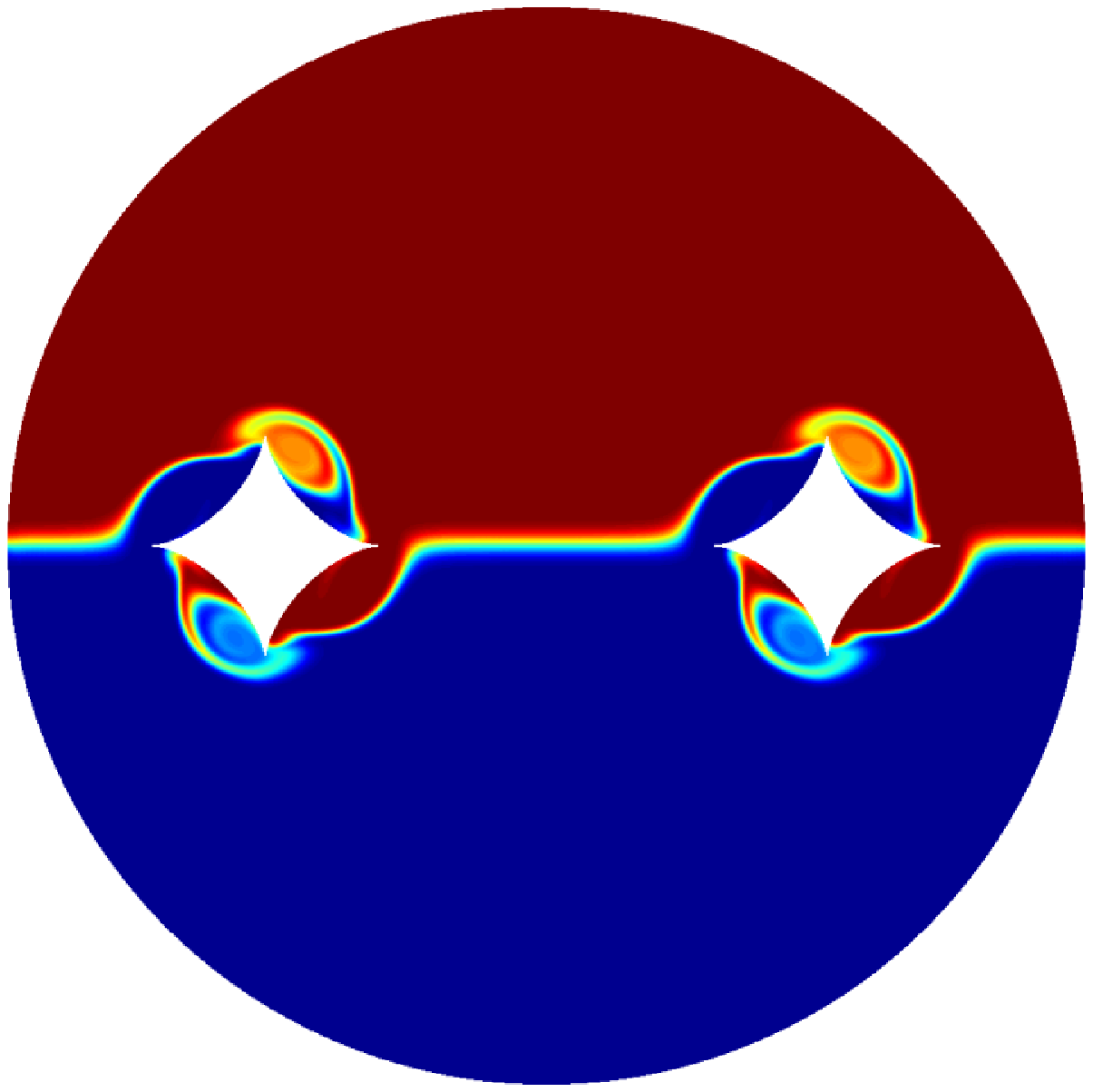} & \hspace{1truecm}
    \includegraphics[trim = 1.5in 0.6in 1.3in 0.4in, clip,width=0.35\textwidth]{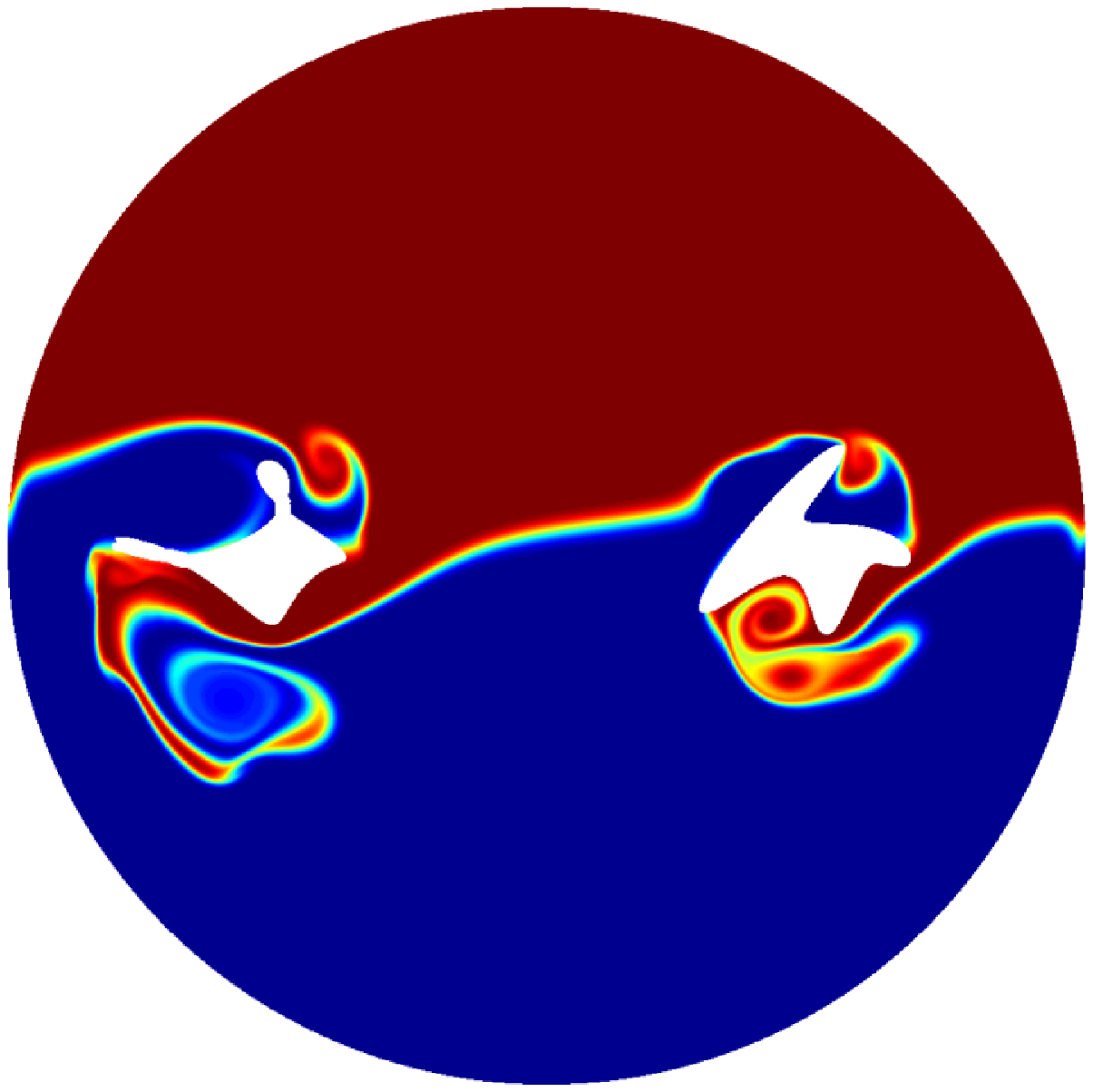} \\
    \includegraphics[trim = 1.5in 0.6in 1.3in 0.4in, clip,width=0.35\textwidth]{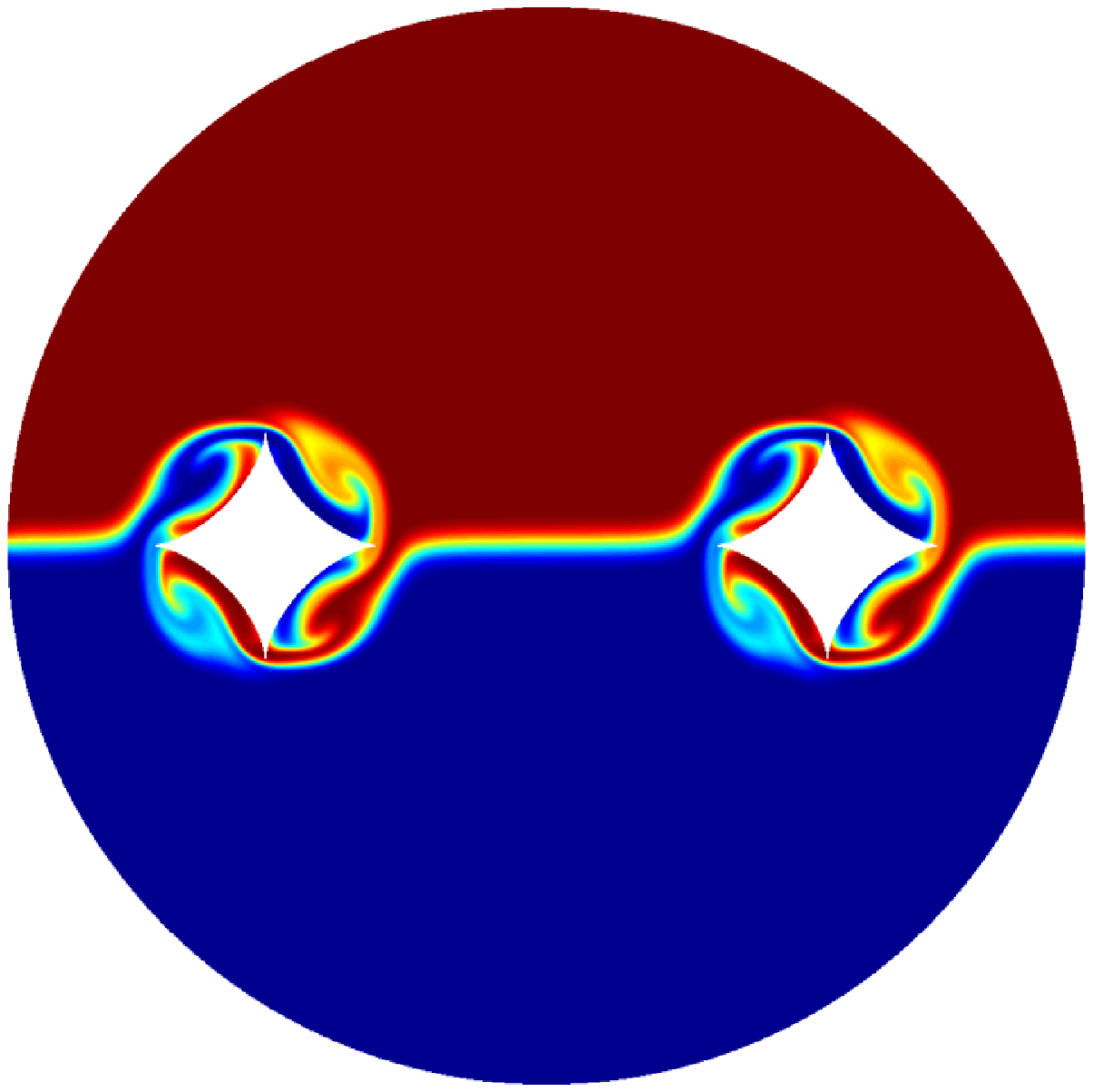} & \hspace{1truecm}
    \includegraphics[trim = 1.5in 0.6in 1.3in 0.4in, clip,width=0.35\textwidth]{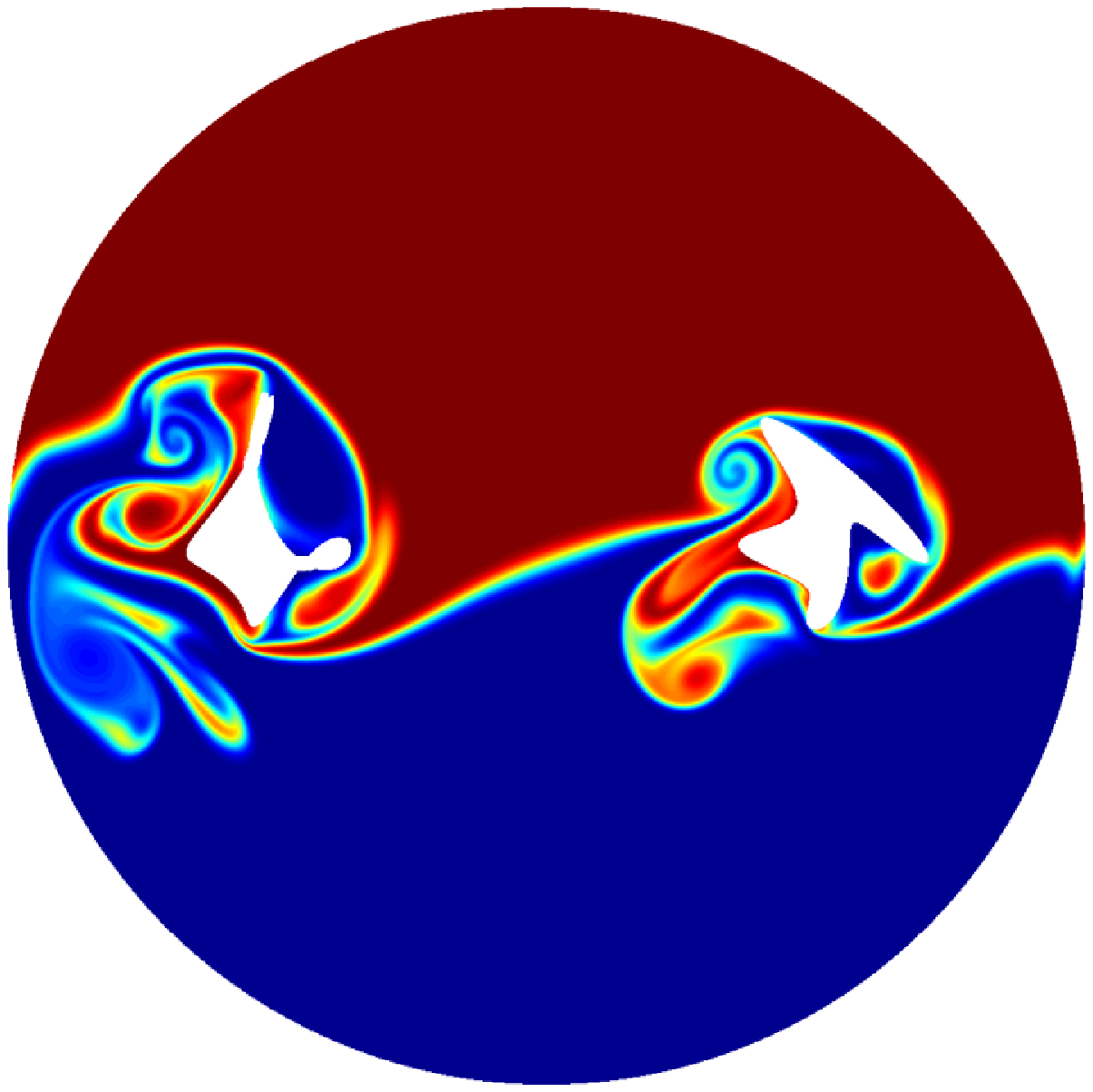} \\
    \includegraphics[trim = 1.5in 0.6in 1.3in 0.4in, clip,width=0.35\textwidth]{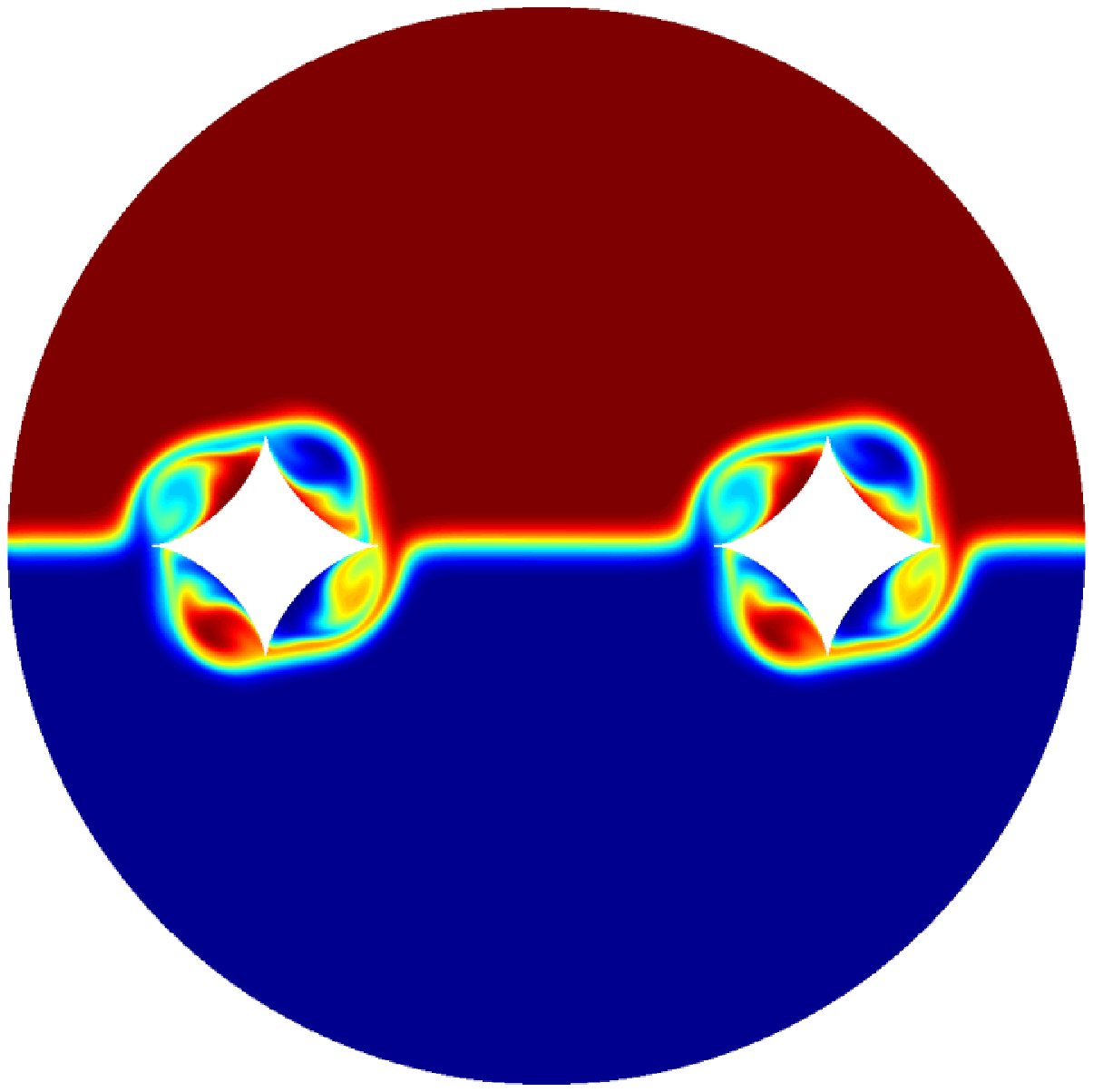} & \hspace{1truecm}
    \includegraphics[trim = 1.5in 0.6in 1.3in 0.4in, clip,width=0.35\textwidth]{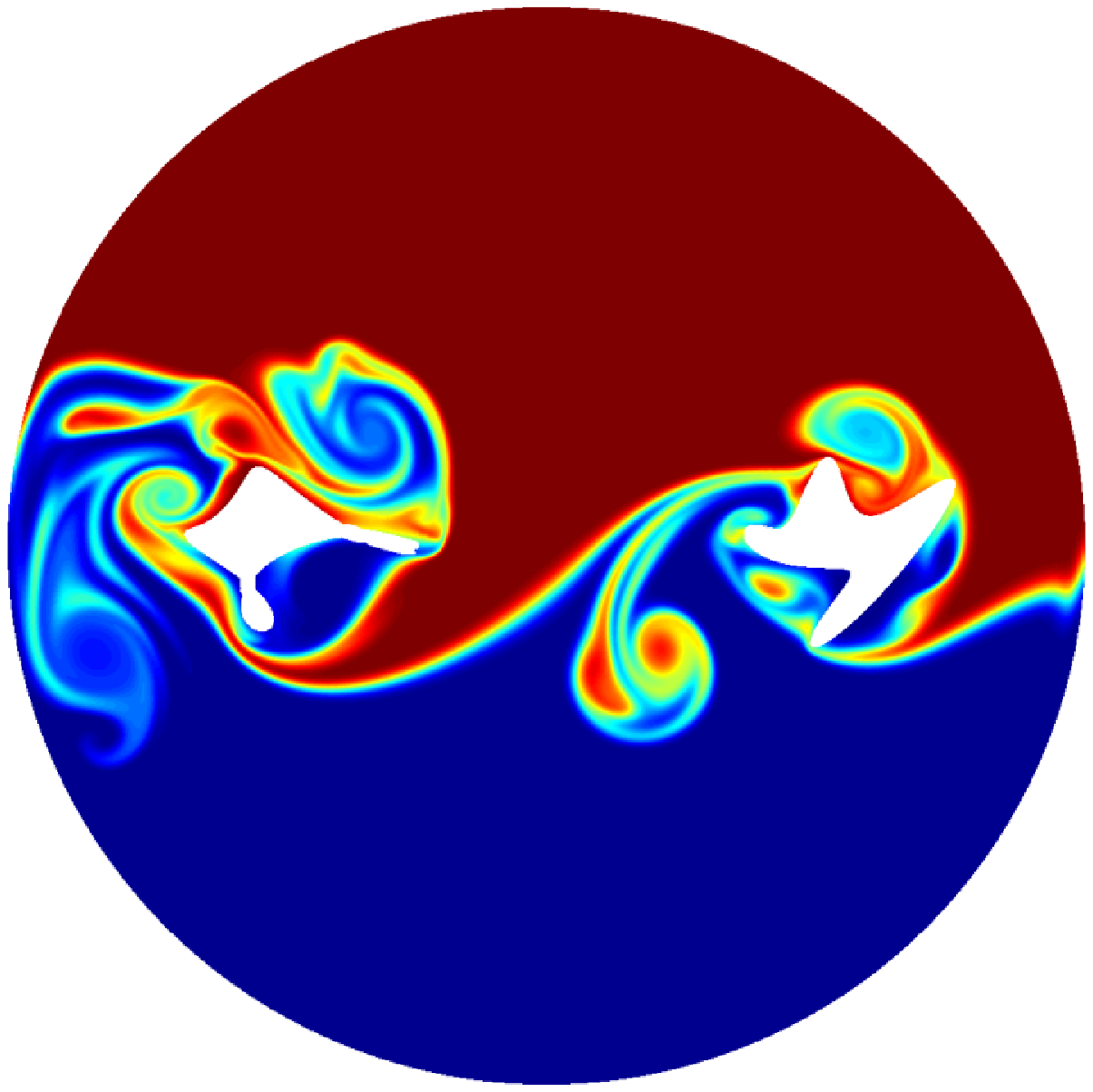}
  \end{tabular}
  \caption{\label{Fig:Shape2Pic} Case 2: mixing optimisation using two rotating stirrers. Left column: unoptimised
    configuration, with snapshots at $t = 2, 4, 6, 8$
    (top to bottom). Right column: after nine direct-adjoint
    optimisations, with snapshots at $t = 2, 4, 6, 8$
    (top to bottom). For videos of these scenarios, please refer to
    {\tt{Shape2NoOpt.mp4}} and {\tt{Shape2Opt.mp4}} for the left and
    right column, respectively.}
\end{figure}

As can be seen from the left column of figure \ref{Fig:Shape2Pic}, showing the unoptimised state, there is no visible interaction between the stirrers and the outer wall; the shed vortices appear unaware of any other solid in their vicinity. In contrast, the optimised results illustrate an influence of the outer wall, in particular for the left stirrer, while the right stirrer tends to concentrate its mixing efforts on the local vortex dynamics. This break in symmetry can be attributed to the identical rotational direction of the stirrers. Firstly, we observe an elongation in the horizontal axis for both stirrers. This echoes the behaviour of the single cylinder optimisation, where elongation has been used to generate a vortex street. However, one fundamental difference prevails: whereas the only elongation for the single-stirrer optimisation was tilted at an angle to the interface to take advantage of a maximum plunging action, the elongation for the left stirrer in the present case is in the horizontal direction. The principal reason can be found in the presence of the wall, which enables and encourages effective vortex-wall collisions. The right stirrer, sensing far less influence of the wall, engages more in a plunging action and therefore develops a secondary elongation at an angle to the interface, similar to the one-stirrer case.

In general, we observe geometric features similar to the one-stirrer case, i.e., elongation in a principal direction and formation of concavities and protuberances, which generate and induce interacting vortices. While the left stirrer takes advantage of the outer wall in its vortex dynamics, the right stirrer follows the previous strategy of more local vortex interplays.

Overall, only a weak level of collaboration between the two stirrers can be discerned. Minor and final changes in the right stirrer geometry can nonetheless be attributed to an exchange of information via the the velocity field; an example of this can be seen in form of the centre vortex at $t=8$, which forms the start of a joint and collaborative mixing strategy.
 
\begin{figure}
  \centering
    \begin{adjustwidth}{-0.5cm}{-0.5cm}
  \begin{tabular}{cc}
    \includegraphics[trim=3cm 0 2.5cm 0,width=0.5\textwidth]{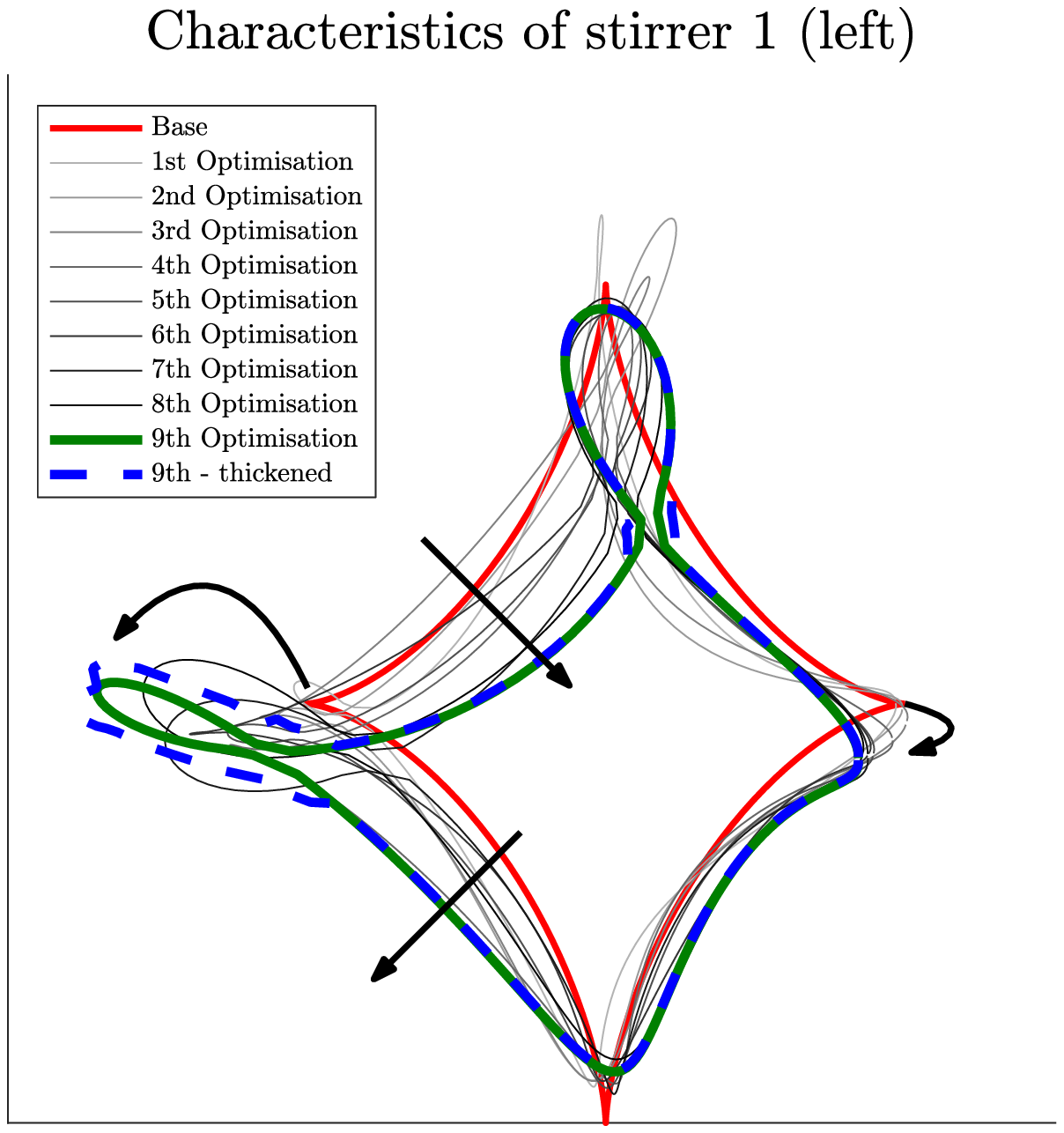} &
    \includegraphics[trim=3cm 0 2.5cm 0,width=0.5\textwidth]{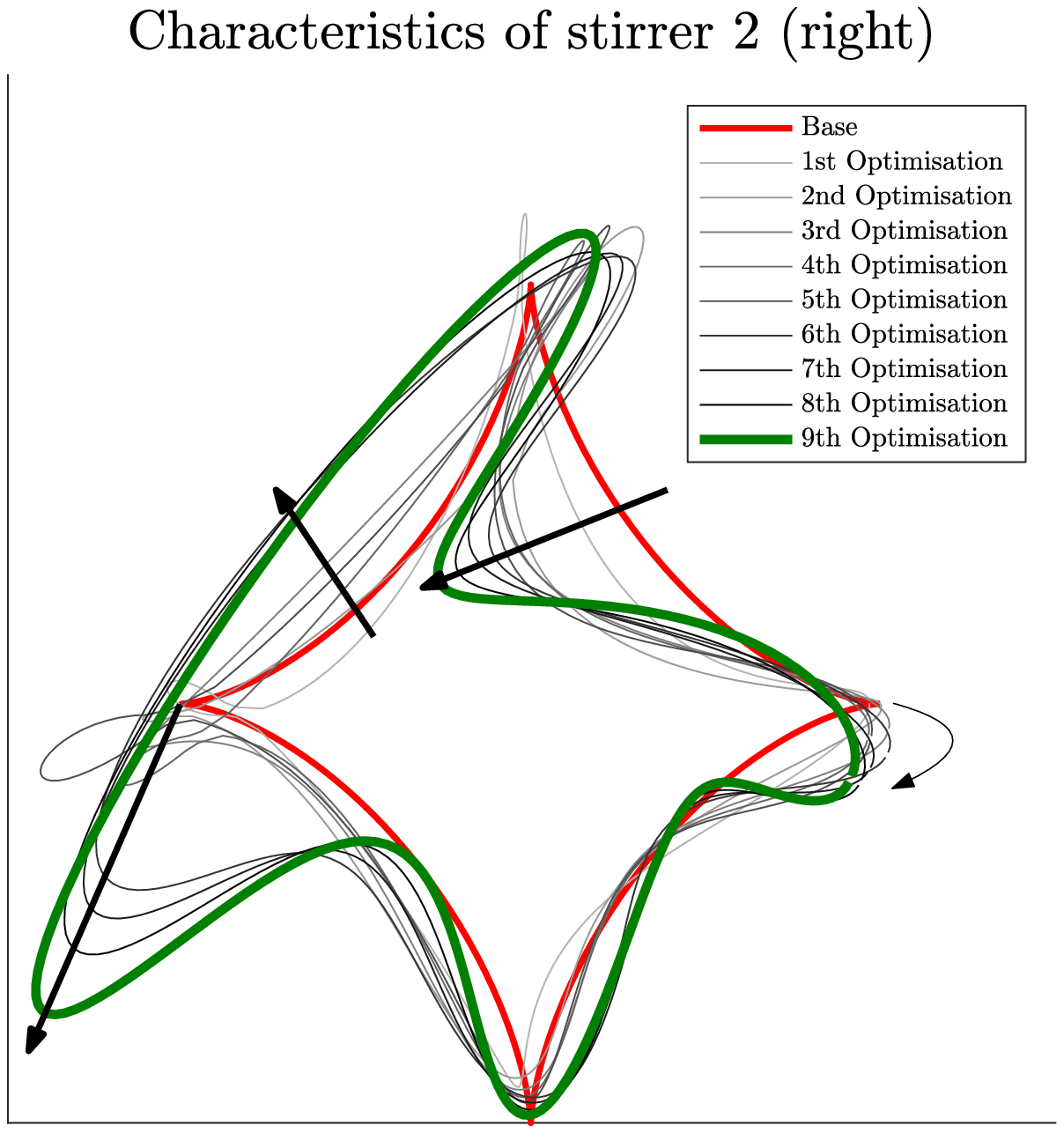} \\
      \includegraphics[trim=0cm 0cm 0cm 0cm,width=0.5\textwidth]{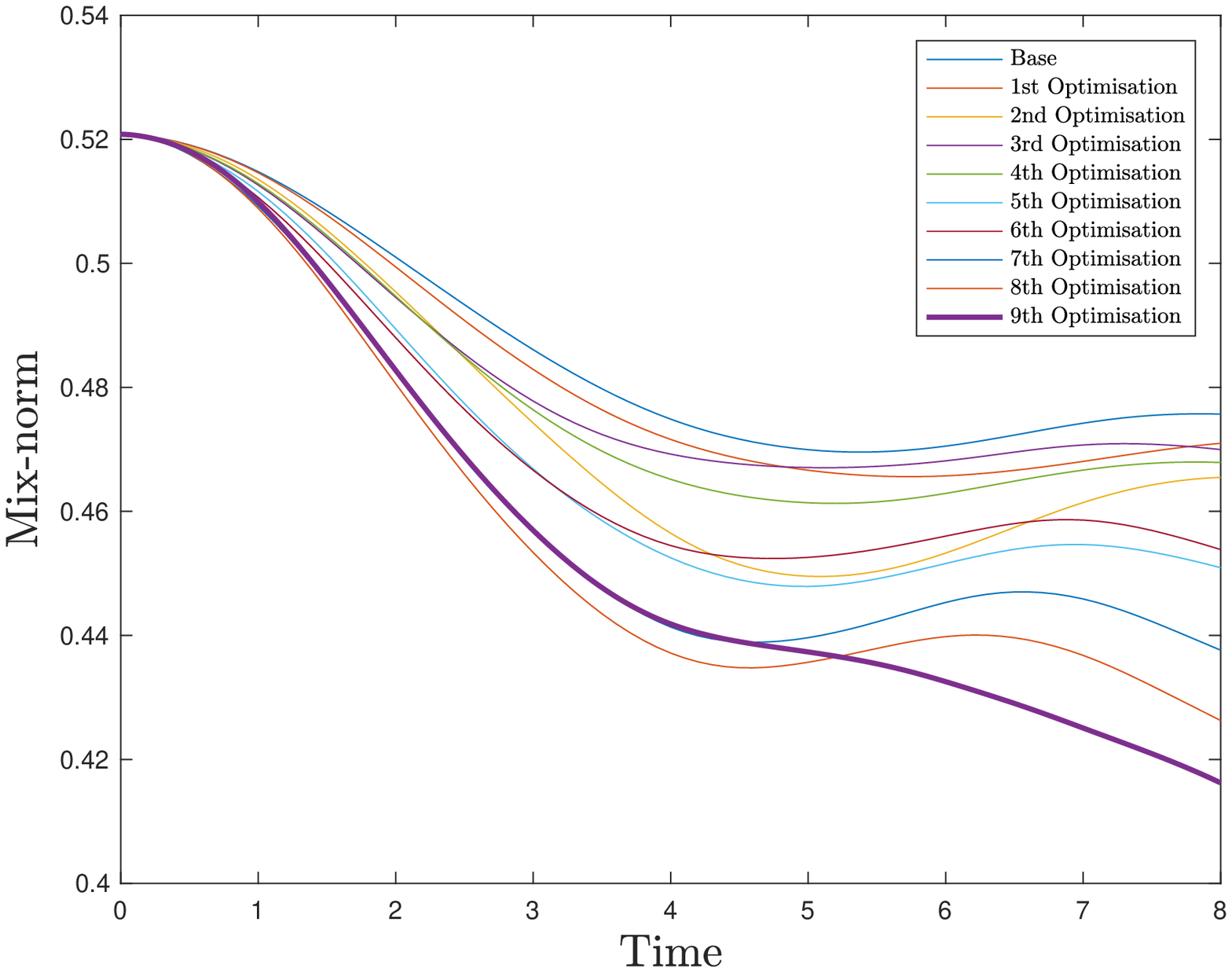} &
        \includegraphics[trim=0cm 0cm 0cm 0cm,width=0.5\textwidth]{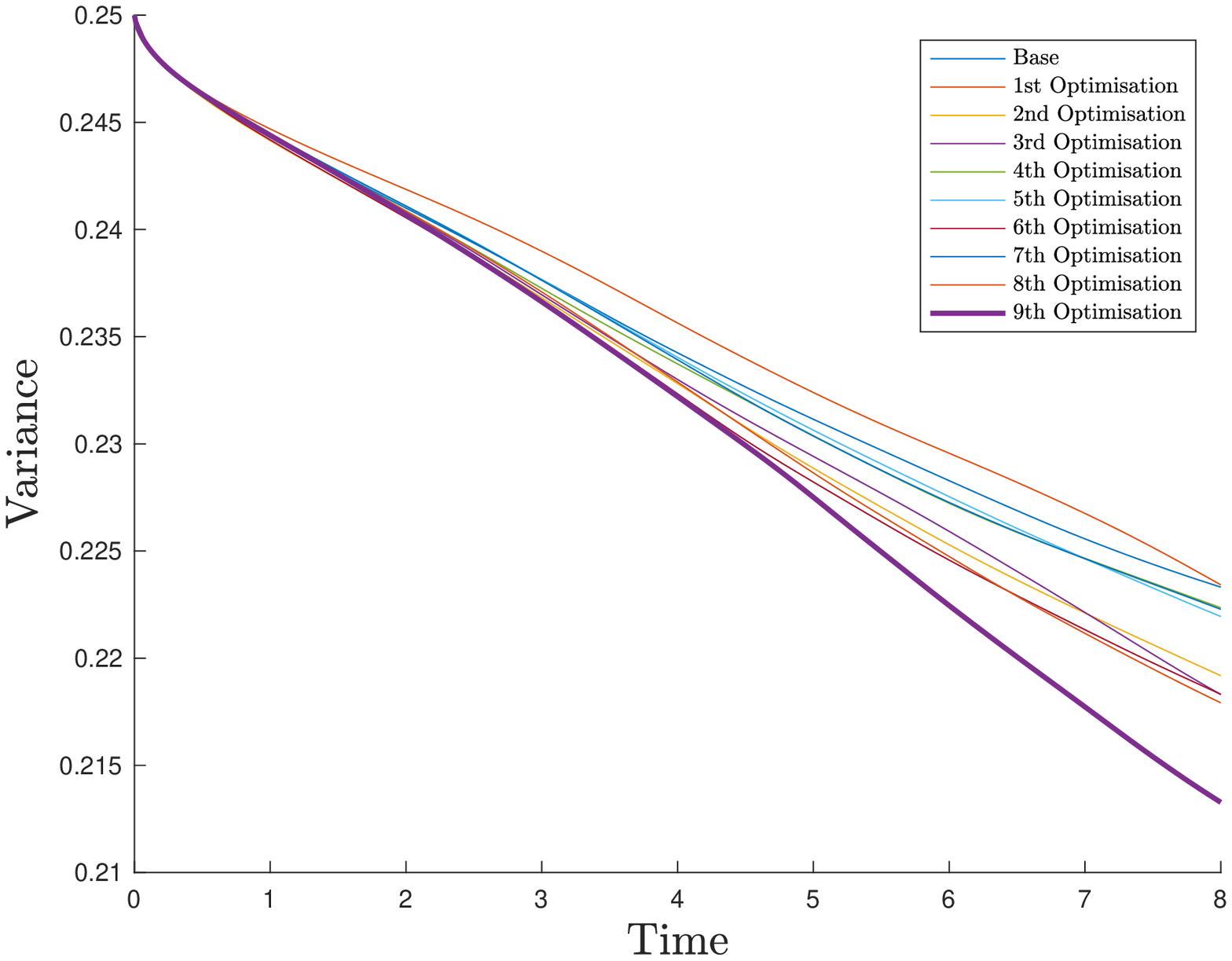}
  \end{tabular}
      \end{adjustwidth}

  \caption{\label{Fig:2Shape}Case 2: mixing optimisation using two
    fixed rotating stirrers. (a) Evolution of the shape of the left stirrer throughout the optimisation steps. The red line is the initial configuration, green the final shape, and the blue dashed line refers to the thickening routine applied to the final optimised geometry. Arrows have been added in black to illustrate the changes in shape. (b) Evolution of the shape of the right stirrer throughout the optimisation steps. (c) Mix-norm of the passive scalar versus time
    $t \in [0,\ T].$ {(d) Variance of the passive scalar versus time
    $t \in [0,\ T].$}}
\end{figure}

Lastly, constraints on the geometry (untwisting and thickening) had to be employed to the left stirrer, ensuring a singly connected stirrer cross-section. The enforcement of these constraints mildly compromised the optimality suggested by the adjoint system. In other words, we traded a minor loss in optimality for physical feasibility. Despite this loss, a significant enhancement in mixing efficiency could be achieved again (see figure \ref{Fig:2Shape}c).

\subsection{Case 3: Shape optimisation of five stirrers}

Encouraged by the advancement of a collaborative mixing strategy for multiple stirrers, we further facilitate this process by embedding additional stirrers into the fluid. Three stirrers will be placed on the interface, with two additional stirrers at an off-set to the interface; see figure \ref{Fig:ShapeGeometriesPic} for a sketch of the configuration. The close proximity of the three stirrers is expected to result in a coordinated strategy between them and, possibly, an interaction with the off-set elements. Intuitively we may identify the off-set cylinders as non-critical to the overall mixing process. With no imposed energy constraints, however, this may not necessarily be the case for the optimisation algorithm.

\begin{figure}
  \centering
   \begin{adjustwidth}{-0.75cm}{-0.5cm}
  \begin{tabular}{cc}
    \includegraphics[trim=3cm 0 2cm 0,width=0.5\textwidth]{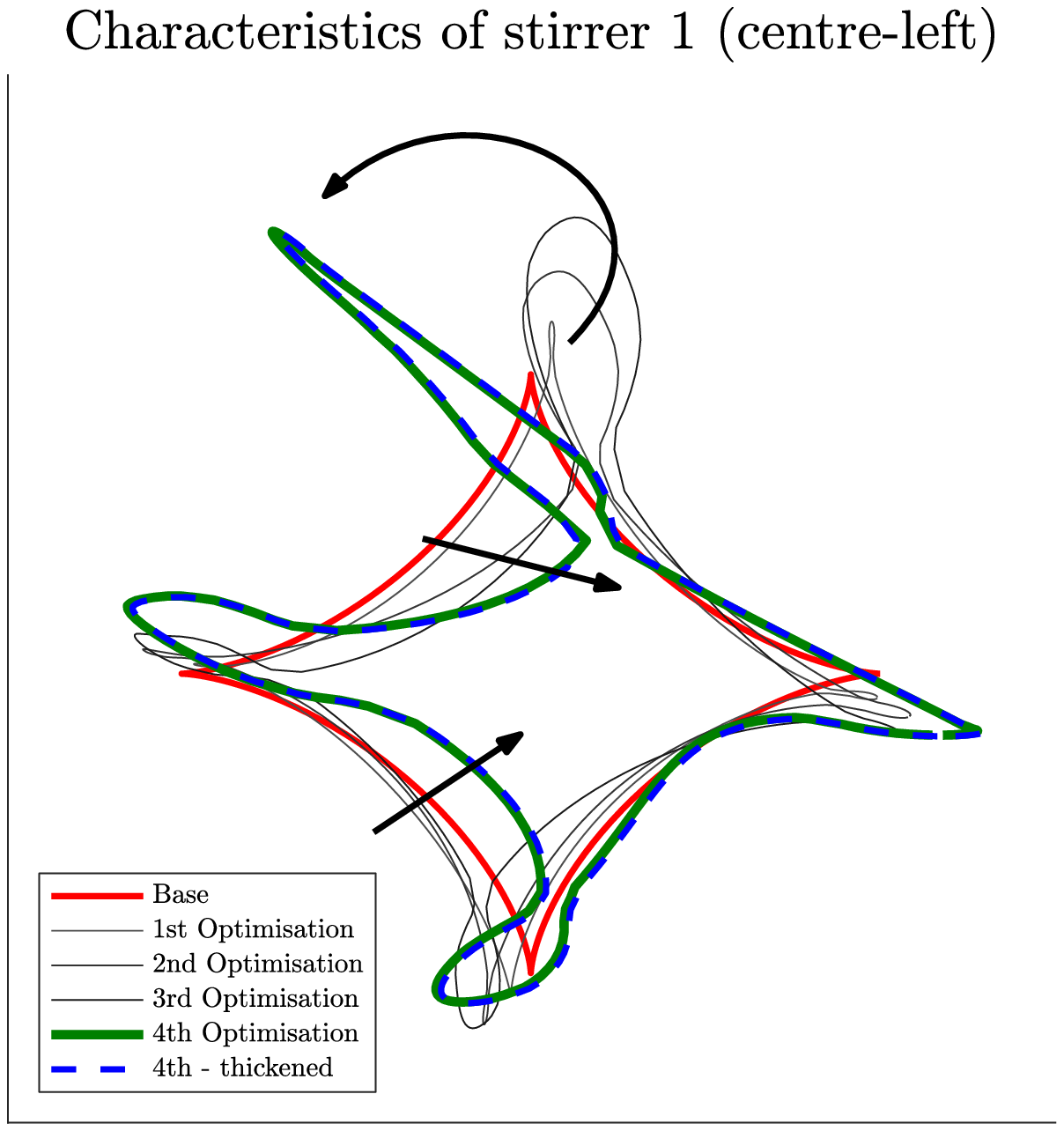} &
    \includegraphics[trim=3cm 0 2cm 0, width=0.5\textwidth]{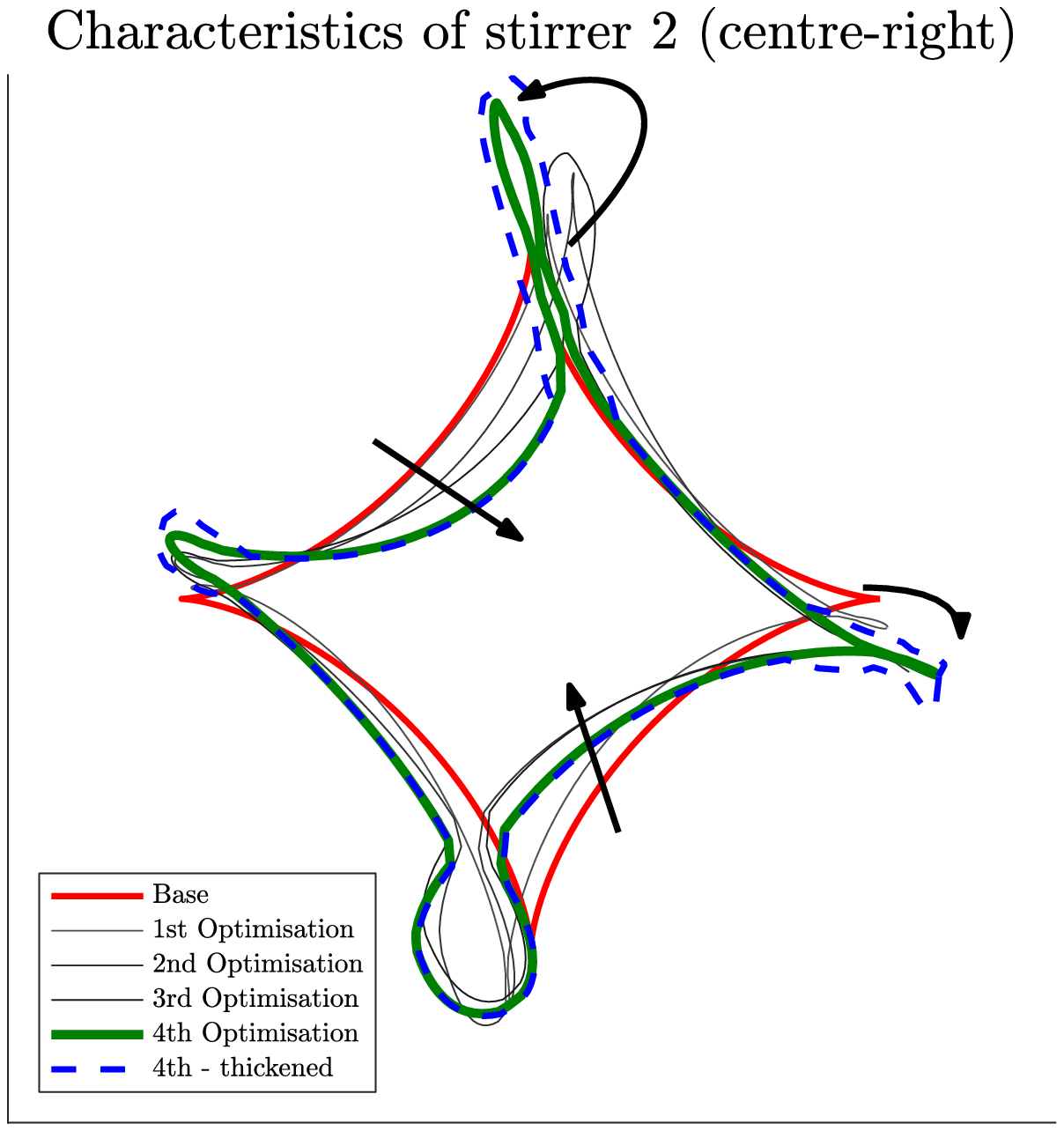} \\
  \end{tabular}
  \end{adjustwidth}
      \includegraphics[trim=3cm 0 2cm 0, width=0.5\textwidth]{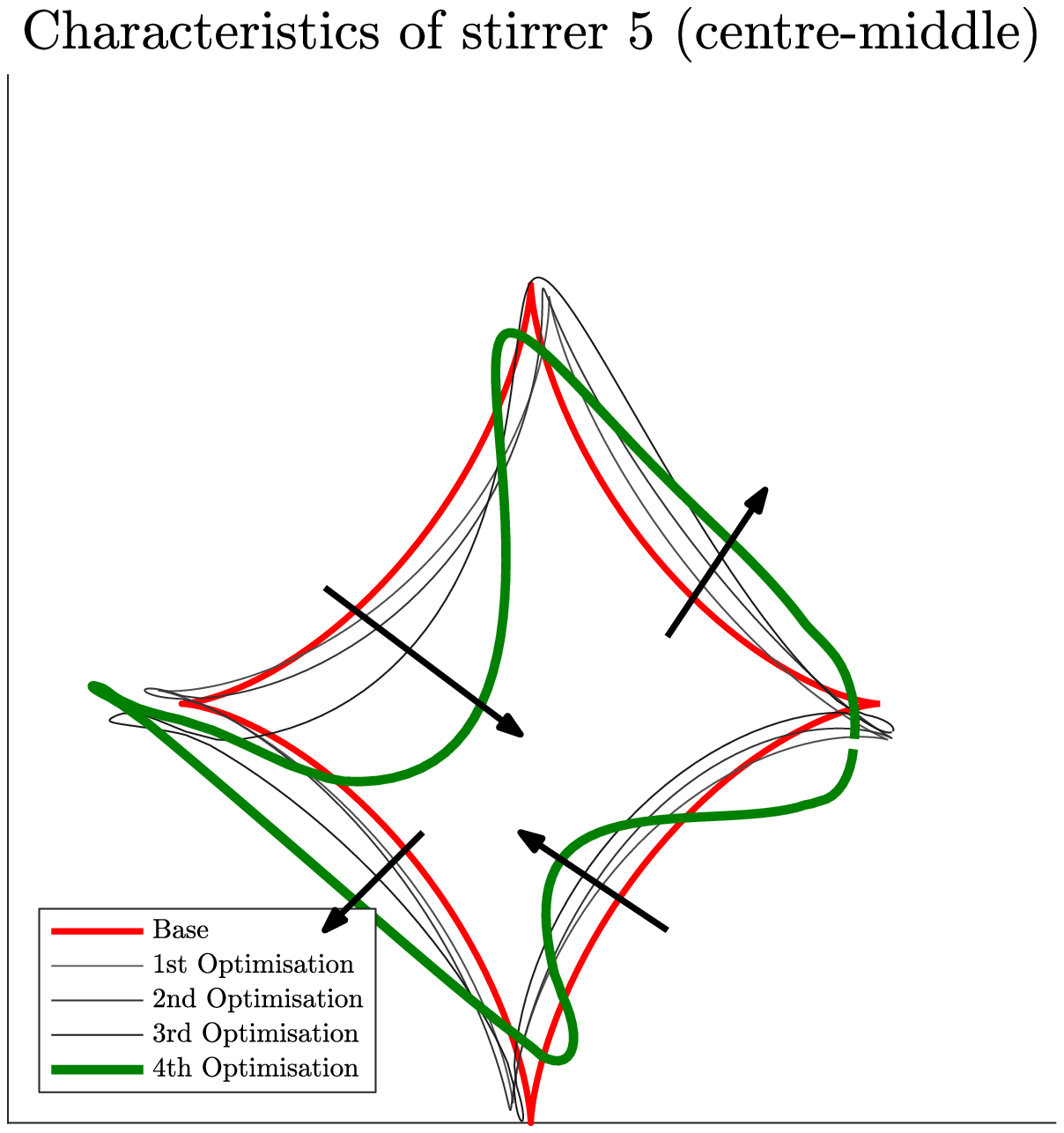}
  \caption{\label{Fig:5Shapes} Case 3: mixing optimisation using five
    stationary rotating stirrers. (a) Evolution of the shape of stirrer 1 throughout the optimisation steps. The red line is the initial configuration, green the final one and the blue dashed line refers to the thickening routine applied to the final optimised shape. Arrows have been added in black to illustrate the changes in the shape. (b) Evolution of the shape of stirrer 2 throughout the optimisation steps. (c) Evolution of the shape of stirrer 5 throughout the optimisation steps.}
\end{figure}

\begin{figure}
  \centering
  \begin{tabular}{cc}
    \includegraphics[trim = 1.5in 0.6in 1.3in 0.4in, clip,width=0.35\textwidth]{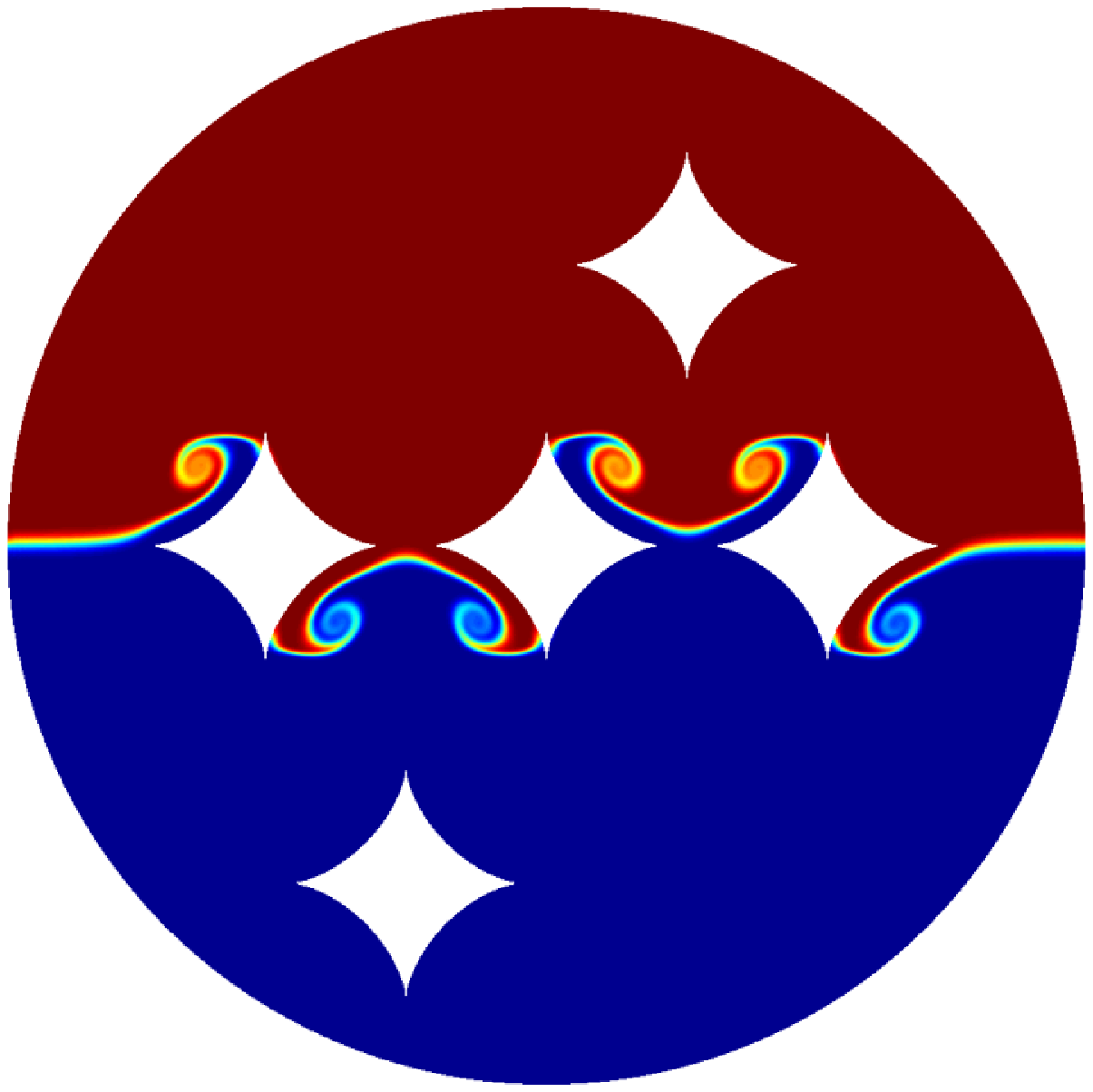} & \hspace{1truecm}
    \includegraphics[trim = 1.5in 0.6in 1.3in 0.4in, clip,width=0.35\textwidth]{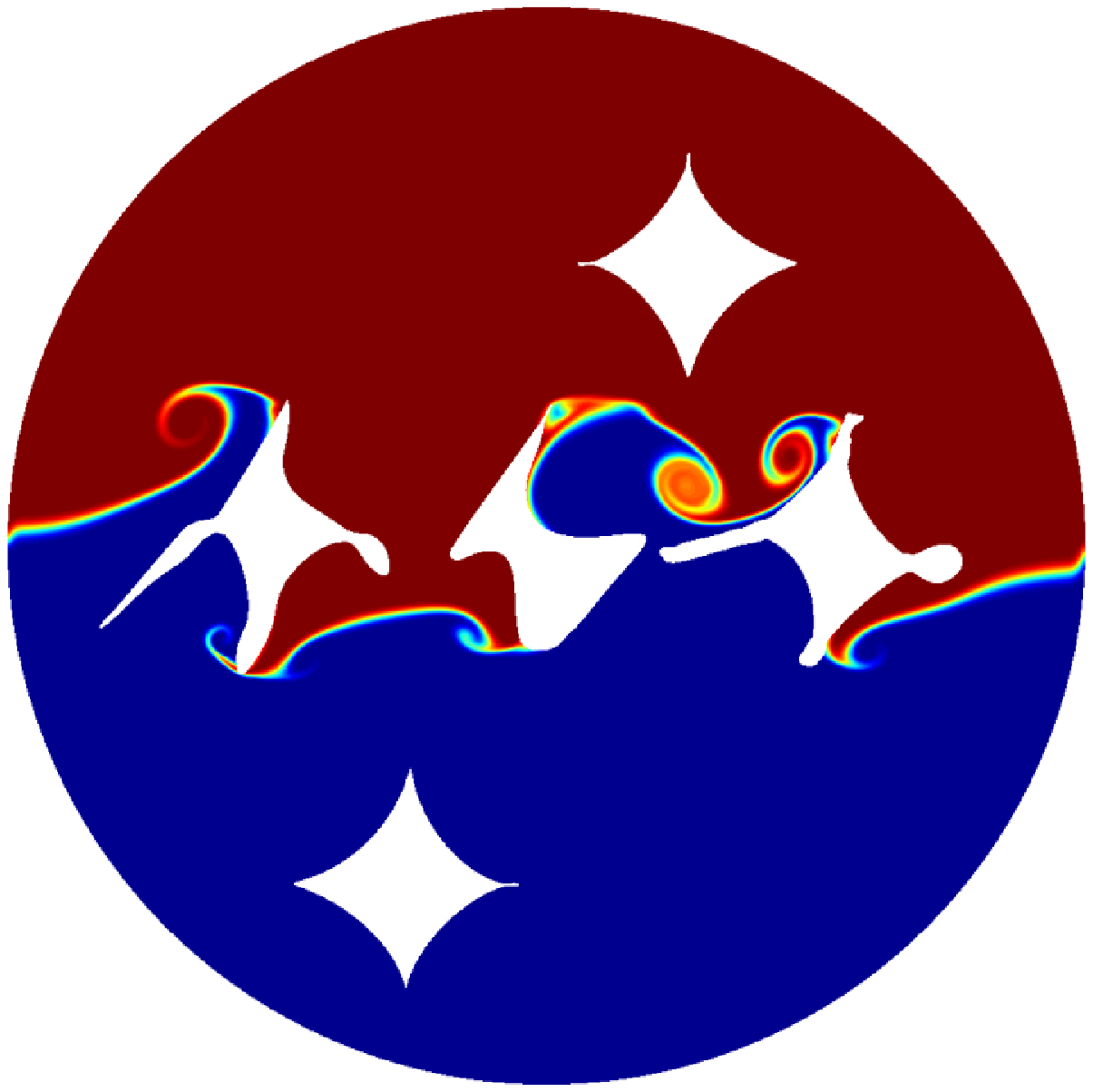} \\
    \includegraphics[trim = 1.5in 0.6in 1.3in 0.4in, clip,width=0.35\textwidth]{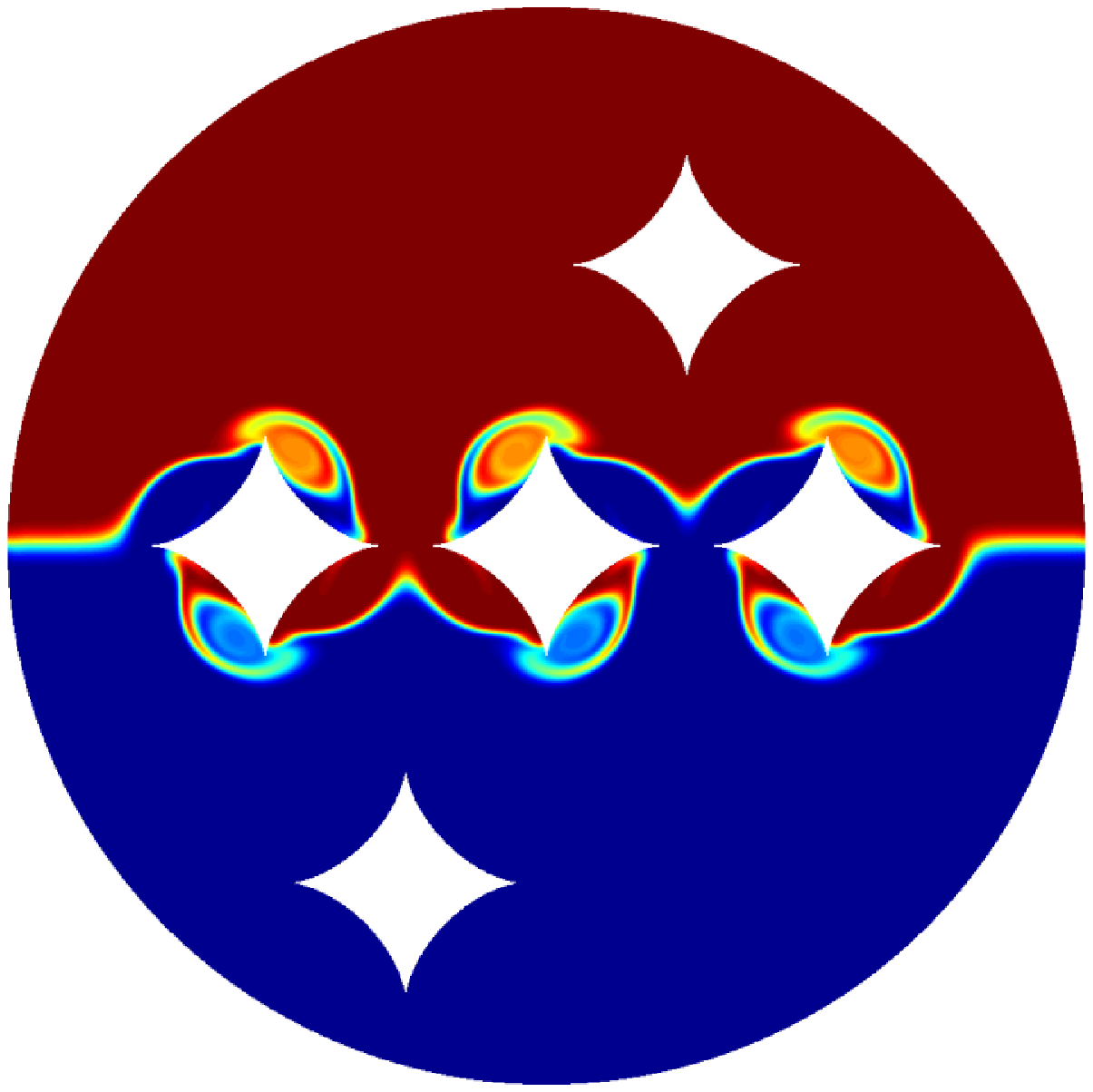} & \hspace{1truecm}
    \includegraphics[trim = 1.5in 0.6in 1.3in 0.4in, clip,width=0.35\textwidth]{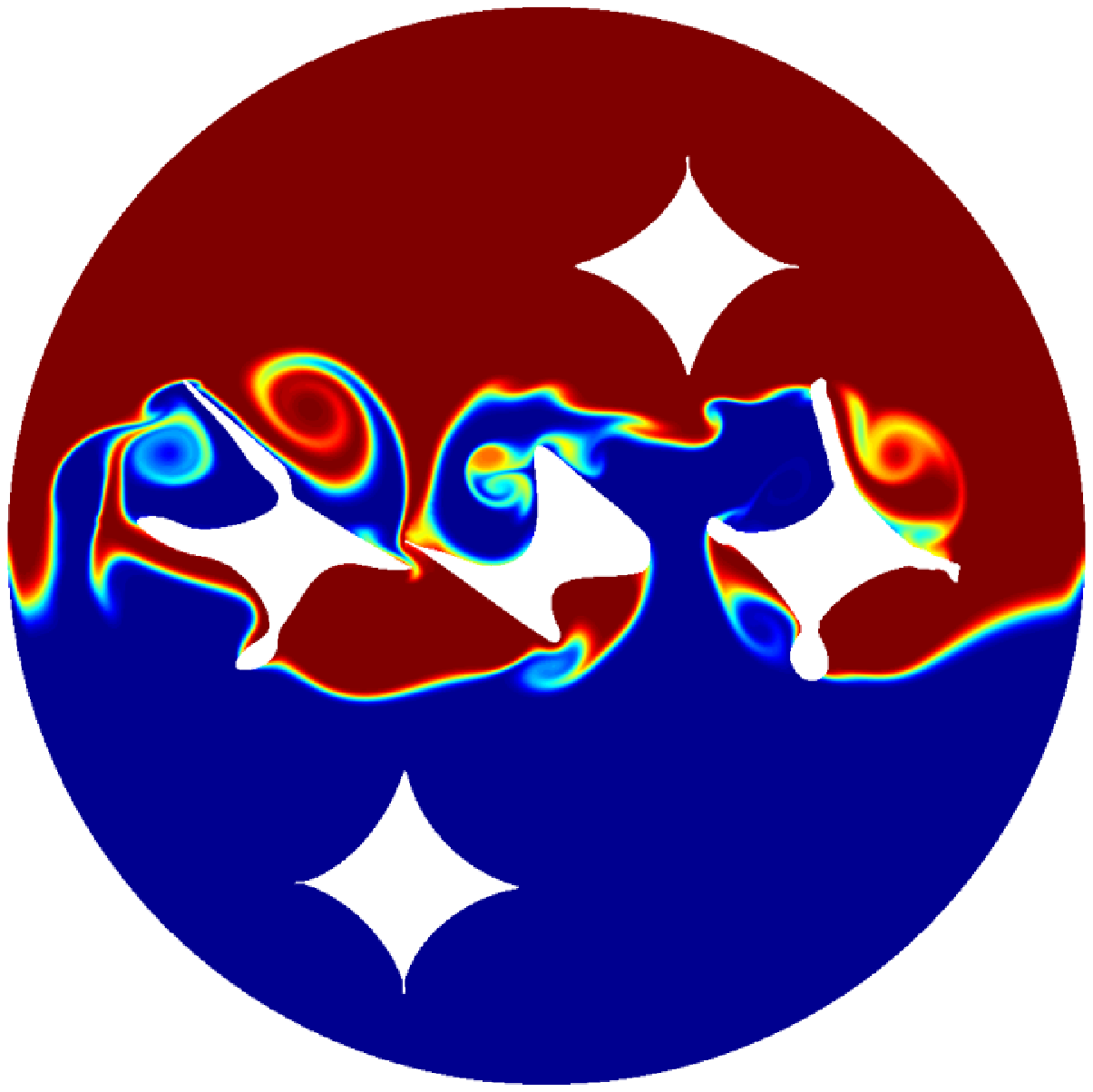} \\
    \includegraphics[trim = 1.5in 0.6in 1.3in 0.4in, clip,width=0.35\textwidth]{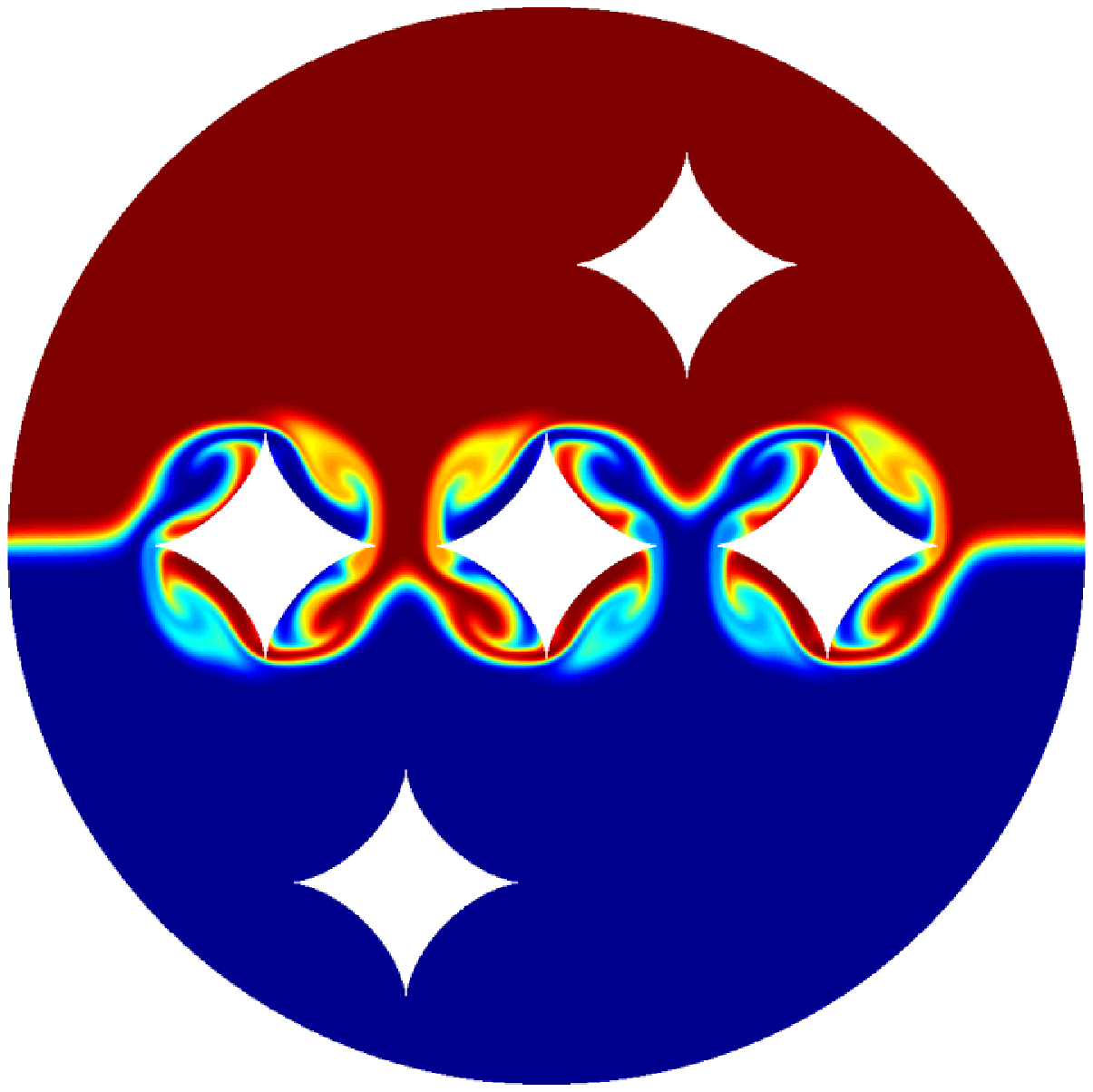} & \hspace{1truecm}
    \includegraphics[trim = 1.5in 0.6in 1.3in 0.4in, clip,width=0.35\textwidth]{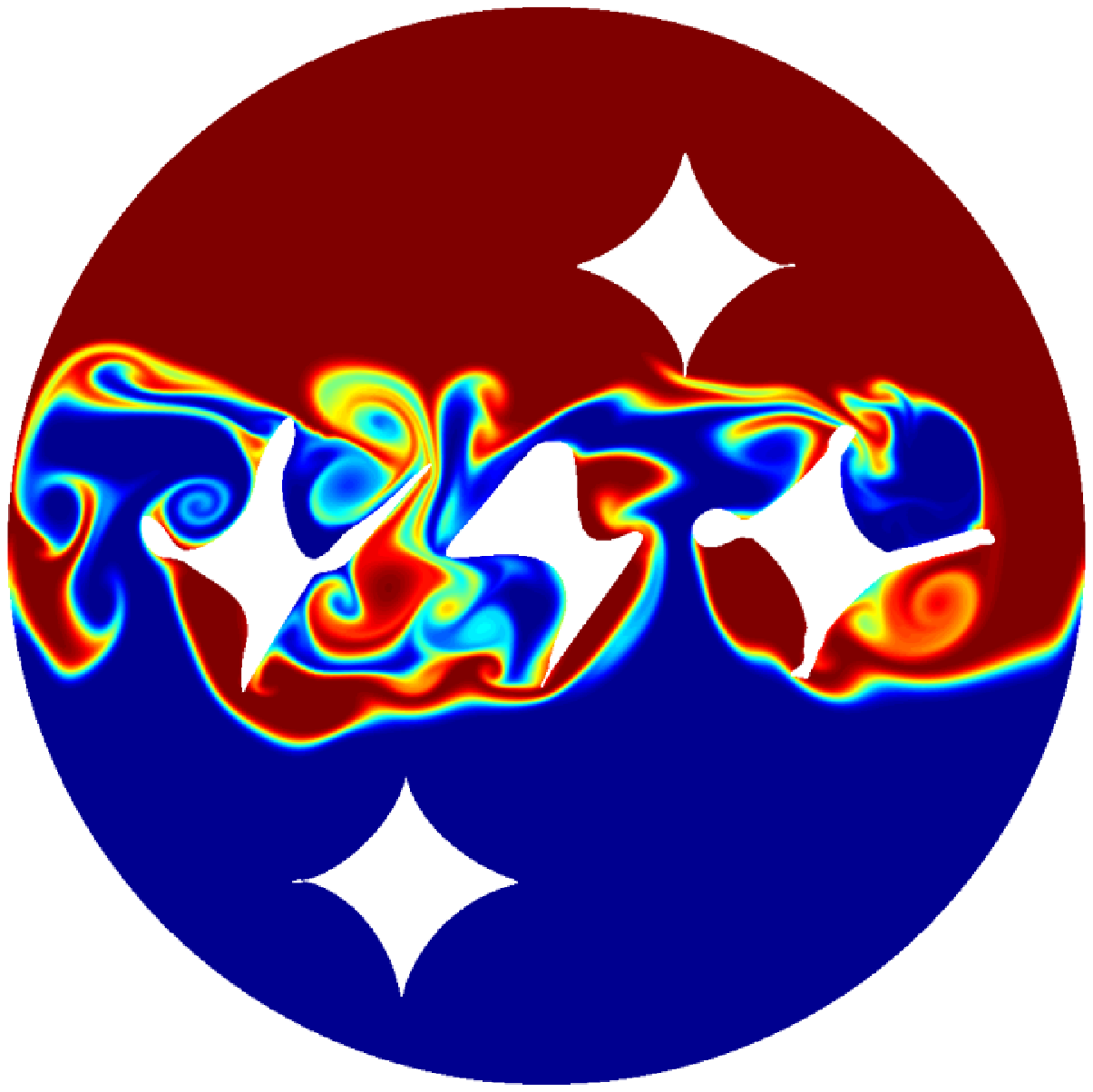} \\
    \includegraphics[trim = 1.5in 0.6in 1.3in 0.4in, clip,width=0.35\textwidth]{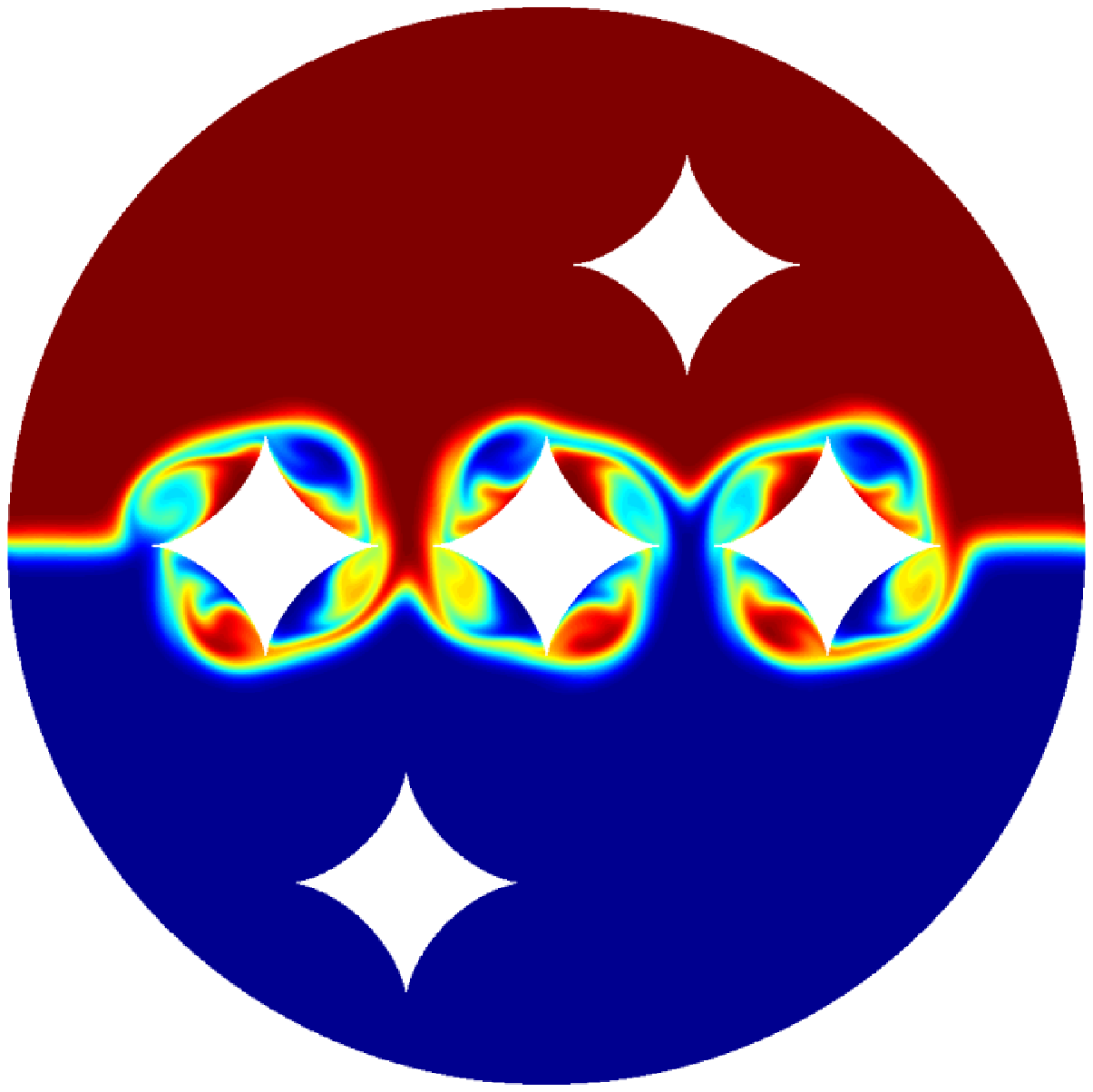} & \hspace{1truecm}
    \includegraphics[trim = 1.5in 0.6in 1.3in 0.4in, clip,width=0.35\textwidth]{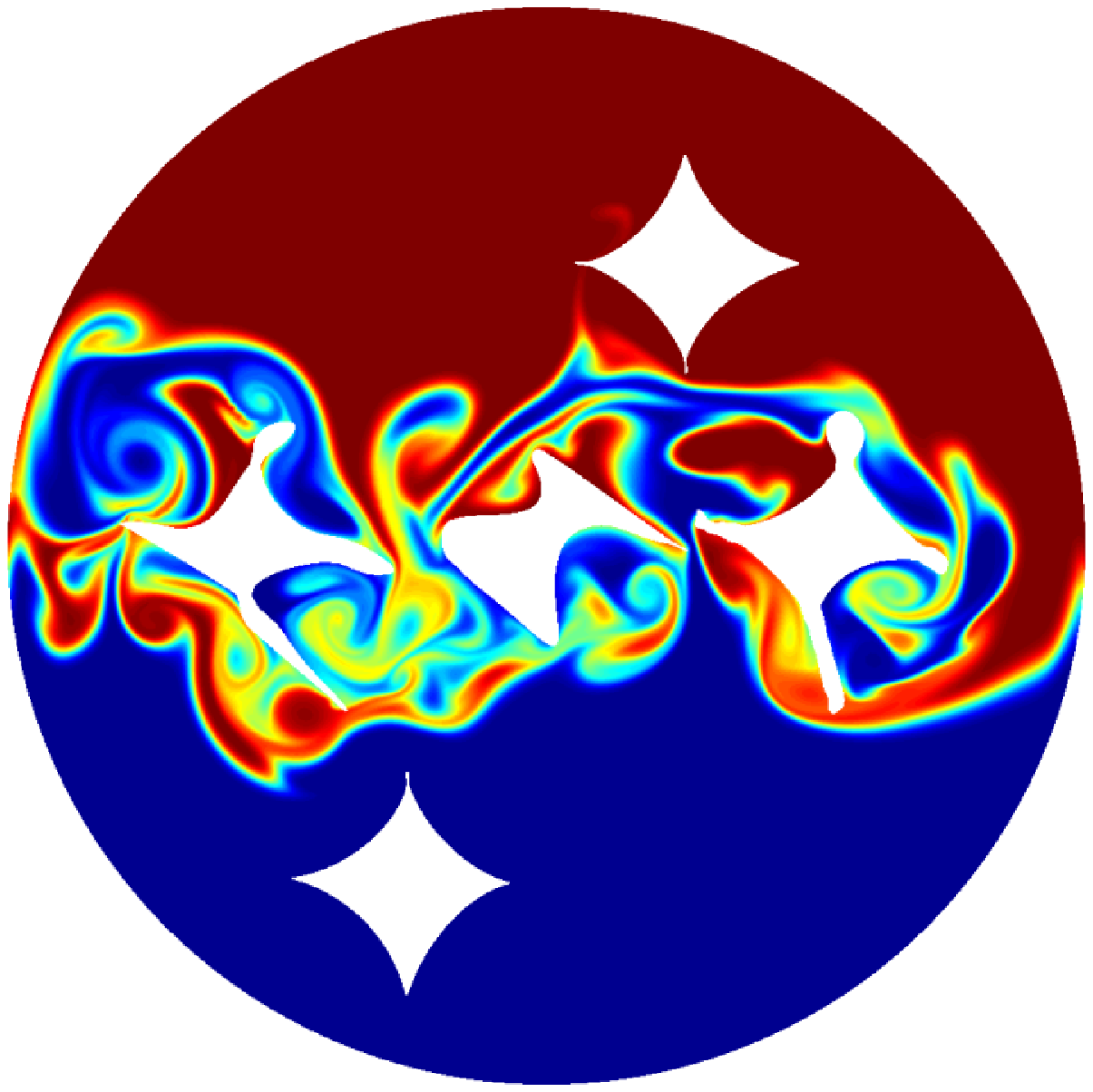}
  \end{tabular}
  \caption{\label{Fig:Shape5Pic} Case 3: mixing optimisation using
    five rotating {stirrers}. Left column: unoptimised
    configuration, with snapshots at $t = 2,4, 6, 8$
    (top to bottom). Right column: after four direct-adjoint
    optimisations, with snapshots at $t = 2, 4, 6, 8$
    (top to bottom). For videos of these scenarios please refer to
    {\tt{Shape5NoOpt.mp4}} and {\tt{Shape5Opt.mp4}} for the left and
    right column, respectively.}
\end{figure}

Considering the unoptimised case (see the left column of figure \ref{Fig:Shape5Pic}), we note that, despite the close proximity of the central stirrers, little to no interaction is observed between the generated vortices. After only four iterations (right column), an entirely different picture arises. Stirrers one and two have been modified regarding their relative angles between the cusps and by the introduction of areas of high concavity (see figures \ref{Fig:5Shapes}a,b; the black arrows illustrate the evolution of the stirrer shapes). These modifications, when combined with the changes that the middle stirrer (five) has undergone, produce an interlocking gear mechanism. In the narrow gaps between the central stirrers, a great many filaments in the passive scalar are produced and thoroughly mixed. Less emphasis is placed on the interaction with the outer wall, and more focus is directed towards the collaboration between the stirrers, resulting in a highly distorted interface and the generation of small-scale structures. Vortices generated by one stirrer collide with a neighbouring stirrer. The middle stirrer acts as a mediator between the two outer and more pointed stirrers, enabling a full and effective communication between the three central cylinders.  We stress that the interlocking geometry resulting in the gear-like motion of the rotating stirrers was produced by the adjoint optimisation without explicitly enforcing or encouraging it.

The outer off-set cylinders undergo far less change, but nevertheless contribute in a minor fashion to the overall mixing process. In contrast to the analogous case studied in \cite{Eggl2018}, where energy considerations ($L^2$-norm of $\bm{u}_s$) played an important role to de-emphasise the outer cylinders, in the present case (with no energy constraints) the off-set stirrers eventually participate in decreasing the mix-norm. Specifically the upper off-set stirrer has been altered, so that two of its cusps have been extended to reach the mixed fluid. This can be observed at $t=8$ in figure \ref{Fig:Shape5Pic}, where minute filaments of fluid are transported into the upper half of the vessel as a consequence.

One of the more remarkable aspects of this result consists of the fact that it was achieved in only four iterations. The proximity of the stirrers, and the information they communicate to each other, is an effective incentive for the optimisation algorithm and ensures rapid convergence to a collaborative strategy. We emphasise once again that the initial rotational velocity of each stirrer has been kept constant throughout, and solely shape optimisation has been performed. In view of these restrictions, and despite a rather limited time horizon of one rotation, a significant decrease in the mix-norm has been achieved.
\begin{figure}
  \centering
   \begin{adjustwidth}{-0.5cm}{0cm}
  \begin{tabular}{cc}
    \includegraphics[trim=1cm 0 1cm 0,width=0.5\textwidth]{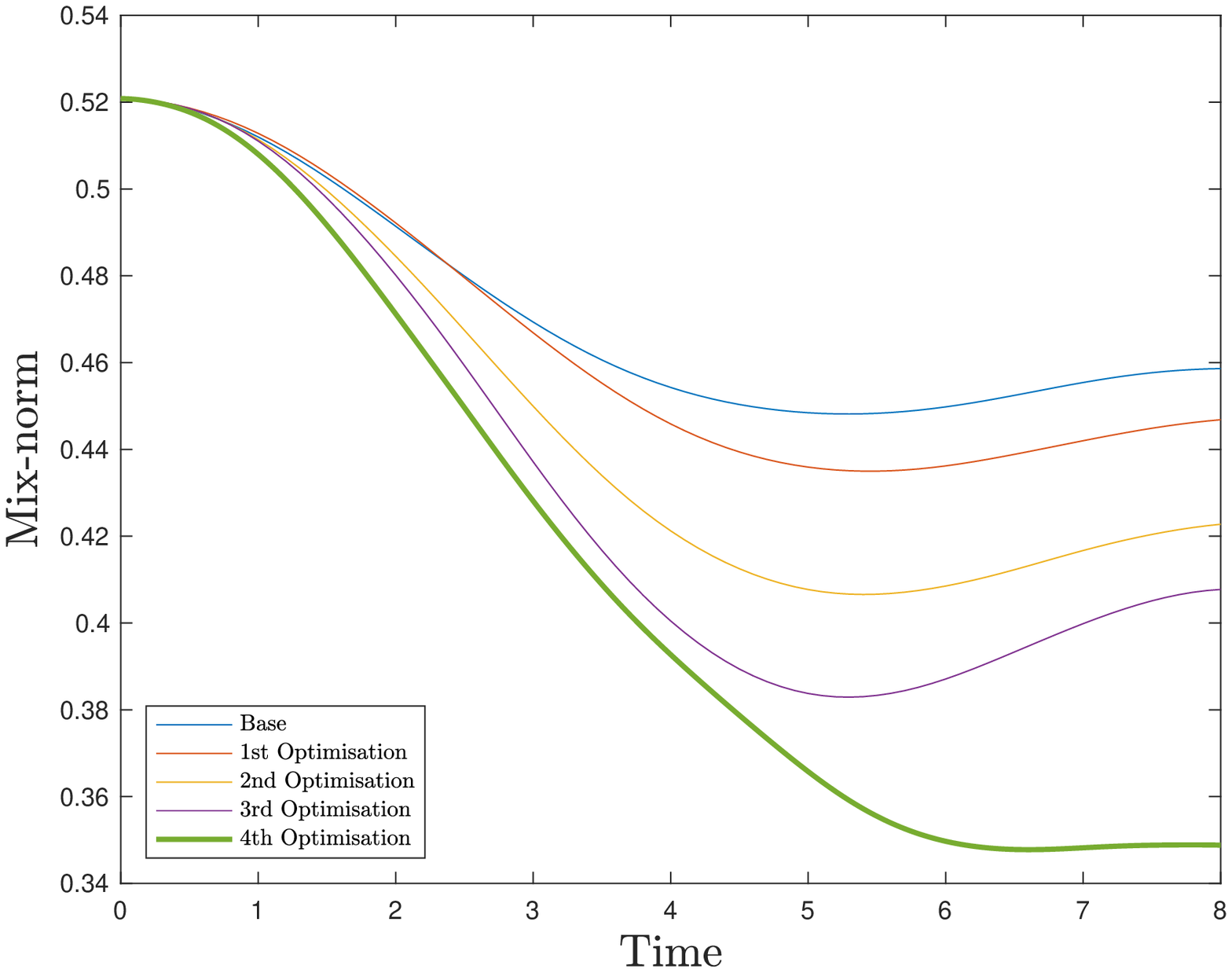} &
    \includegraphics[trim=1cm 0 1cm 0,width=0.5\textwidth]{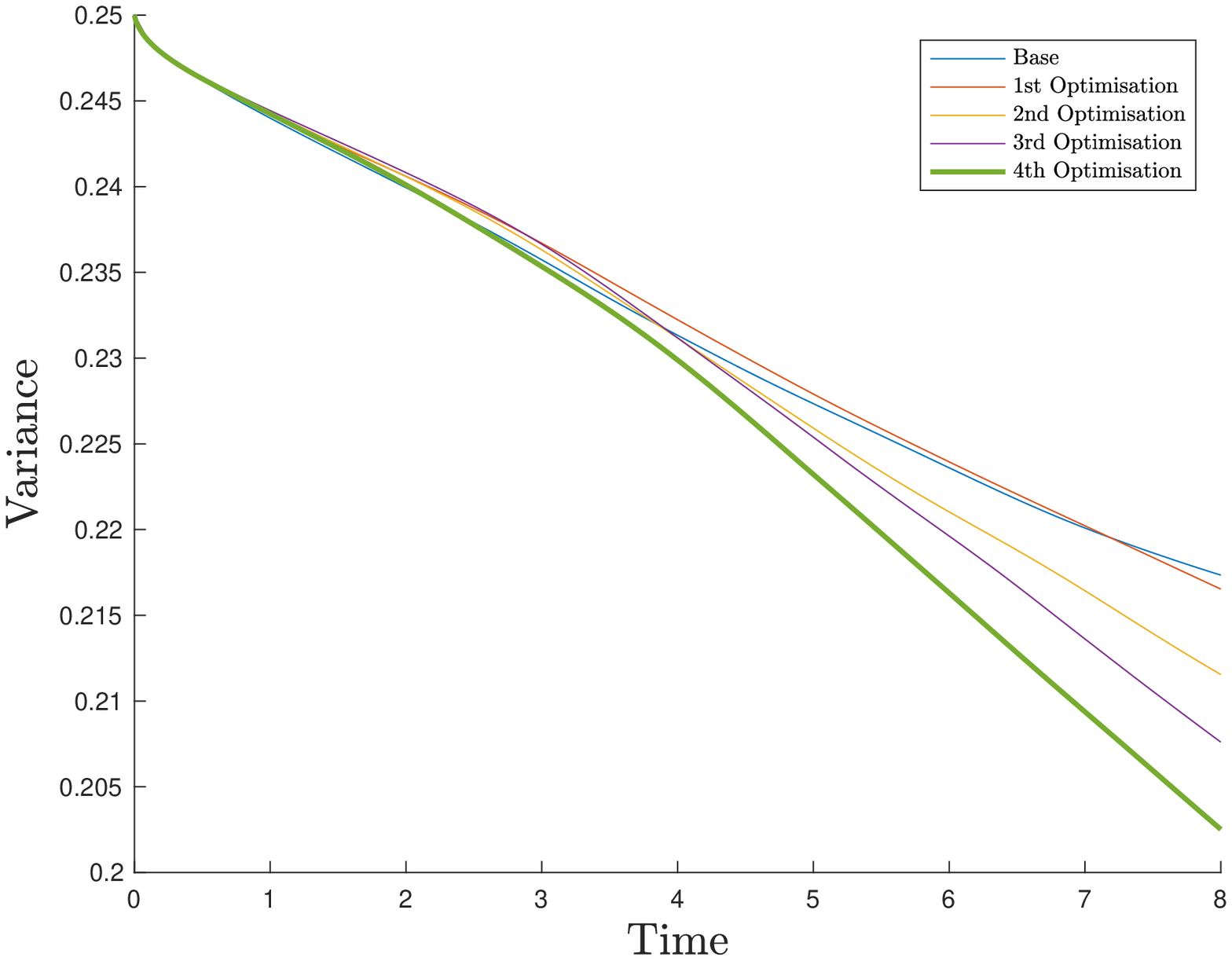}
  \end{tabular}
  \end{adjustwidth}
  \caption{\label{Fig:5Norms}{Case 3: mixing optimisation using five
    stationary rotating stirrers. (a) Mix-norm of the passive scalar versus time
    $t \in [0,\ T].$ (b) Variance of the passive scalar versus time
    $t \in [0,\ T].$}}
\end{figure}

\newpage

\section{Summary and conclusions}
\label{Sec:conclusion}

We have introduced a mathematical and computational framework, based on direct-adjoint iterative optimisation, for treating the mixing enhancement in binary fluids by topological shape modification of the embedded stirrers, which are represented by a Fourier-based parametric curve. 
The derivation and application of this shape parameterisation is surprisingly simple, and the resulting design space, while rather low-dimensional, allows for a large number of possible stirrer configurations.

Three test cases of increasing complexity have been presented to demonstrate the ease of use and power of the computational methodology: (i) the optimisation of one central stirrer,
(ii) a double configuration along the initial interface, and (iii) a layout with five stirrers.
In all cases significant improvement in mixing efficiency could be accomplished,
and the optimisation algorithm showed notable robustness in finding a more optimal solution. These optimal solutions often include non-intuitive shapes that, when implemented in an industrial setting, could lead to substantial savings in expended effort or to higher-quality products. 

A key observation of the obtained solutions is the generation of vortex filaments by the moving stirrers. Their mutual interactions dominates the mixing process. In contrast to the spatially fixed stirrers, these vortical elements are able to move freely within the mixing vessel and therefore can affect areas of the fluid that are otherwise inaccessible to the stirrers but that are highly beneficial to mixing enhancement. This effect could be observed by the convergence of the geometries towards protuberances and concavities for each of the stirrers. These two geometric components, combined with the rotation of the stirrers, generated areas of fluid recirculation and lead to the creation and shedding of vortices. The vortices then interacted and collided with previously and subsequently generated vortices, further distorting the interface and creating small scales for their later diffusion.
 
A further observation that lead to significant mixing enhancement (also reported in \cite{Eggl2018}) was the evolution of collaborative mixing strategies when multiple stirrers are optimised. The interlocking gear design of the optimised shapes of the five stirrer configuration is an example of this cooperation. In this case, the cusp of a rotating stirrer is placed directly in the path of a vortex being shed by another solid. This joint strategy requires little extra expended kinetic energy, but can yield a substantial increase in mixing efficiency. Further mixing studies may take advantage of a multi-stirrer configuration and exploit this collaboration of stirrers to additionally enhance mixing.

Despite these promising results, a few challenges remain. As is the
case for any gradient-based optimisation applied to a non-convex
problem, only a local minimum can be guaranteed by our algorithm. The Fourier-based shape parameterisation somewhat alleviates this problem, as the global nature of the representation allows the exploration of multiple local minima. Nonetheless, a global minimum is unattainable with our approach. A possible solution could lie in coupling our framework with an annealing-type stochastic algorithm \cite{kirkpatrick1983}, but the excessive computational cost associated with these methods prohibits its use in our case. Despite this issue, local optima should not be dismissed {\it{ab initio}}, as even small improvements in efficiency can equate into substantial gains and profits at an industrial scale.

A further challenge arises from the application of the Fourier-based shape parameterisation. The thickening and untwisting procedures have been  embedded via a projection approach, after the optimisation step has taken place. While this approach achieved a significant increase in mixing efficiency, the resulting shapes were not strictly optimal. Mathematically, the geometric constraints should be incorporated into the cost functional; the details of this operation, however, remain an open challenge.

While keeping these challenges in mind, the investigations presented in this article point towards several areas where our optimisation framework can excel and achieve enhanced mixing efficiency.

A natural extension consists in a cooperative combination of the Fourier-based shape optimisation and the optimisation of time-dependent quantities, such as the velocity of stirrers along a given path. 
Exploring this {\it{combined}} design space will also give insight into the underlying mechanisms that drive mixing, e.g., flapping, plunging, heaving or rotation.

Extending the Fourier representation of the shape, the equivalent mathematical representation based on parametric curves can also be applied to optimise the trajectory of moving stirrers. The Fourier-based parameterisation for the stirrers' paths opens new areas of the design space. For example, proposing parameterised Lissajous curves \citep{cundy1961} for each stirrer could yield pertinent and interesting results, with potentially novel and non-intuitive mixing strategies.
This approach has the added benefit that for path optimisation, the constraints of untwisting and thickening of the geometry can be safely ignored, since any continuous, and even self-intersecting, trajectory represents a viable mixing strategy. However, care must be taken to avoid collisions among stirrers and/or the wall. 

All the above extensions follow the two-dimensional nature of the test cases studied in this article and ignore three-dimensional effects. The computational framework is highly parallelised and scales well to many cores on parallel computer architectures; the Fourier-based parameterisation can also naturally be extended to three dimensions. Scalable performance for large-scale shape optimisations of binary mixing problems is thus expected. 

In view of the wide scope of possibilities offered by the direct-adjoint optimisation approach and its robustness in finding enhanced mixing solutions, promising results can be expected in future studies.

\bibliographystyle{imamat}
\bibliography{ImaPaper}

\providecommand{\noopsort}[1]{}\providecommand{\singleletter}[1]{#1}%
\begin{thebibliography}{}

\bibitem[Aamo \& Krstic, 2013]{aamo2013}
Aamo, O.~M. {\&} Krstic, M. (2013) {\em Flow control by feedback: stabilization
  and mixing}.
Springer Science \& Business Media.

\bibitem[Alexandersen et~al., 2014]{alexandersen2014}
Alexandersen, J., Aage, N., Andreasen, C.~S. {\&} Sigmund, O. (2014)  Topology
  optimisation for natural convection problems. {\em Int. J. Num. Meths.
  Fluids}, \textbf{76}(10), 699--721.

\bibitem[Angot et~al., 1999]{Angot1999}
Angot, P., Bruneau, C.-H. {\&} Fabrie, P. (1999)  A penalization method to take
  into account obstacles in incompressible viscous flows. {\em Num. Math.},
  \textbf{81}(4), 497--520.

\bibitem[Annaswamy \& Ghoniem, 1995]{Annaswamy1995}
Annaswamy, A. {\&} Ghoniem, A. (1995)  Active control in combustion systems.
  {\em IEEE Contr. Sys.}, \textbf{15}(6), 49--63.

\bibitem[Aref, 1984]{Aref1984}
Aref, H. (1984)  Stirring by chaotic advection. {\em J. Fluid Mech.},
  \textbf{143}(1), 1--21.

\bibitem[Arquis \& Caltagirone, 1984]{arquis1984}
Arquis, E. {\&} Caltagirone, J. (1984)  Sur les conditions hydrodynamiques au
  voisinage d'une interface milieu fluide-milieu poreux: applicationa la
  convection naturelle. {\em C.R. Acad. Sci. Paris II}, \textbf{299}, 1--4.

\bibitem[Batchelor, 1952]{batchelor1952}
Batchelor, G.~K. (1952)  The effect of homogeneous turbulence on material lines
  and surfaces. {\em Proc. Roy. Soc. Lond. A}, \textbf{213}(1114), 349--366.

\bibitem[Boiron et~al., 2009]{Boiron2009}
Boiron, O., Chiavassa, G. {\&} Donat, R. (2009)  A high-resolution penalization
  method for large {M}ach number flows in the presence of obstacles. {\em Comp.
  \& Fluids}, \textbf{38}(3), 703--714.

\bibitem[Bruneau et~al., 2013]{Bruneau2013}
Bruneau, C.-H., Chantalat, F., Iollo, A., Jordi, B. {\&} Mortazavi, I. (2013)
  Modelling and shape optimization of an actuator. {\em Struct. Multidisc.
  Optim.}, \textbf{48}(6), 1143--1151.

\bibitem[{Buckingham}, 1938]{Buckingham1938}
{Buckingham}, R.~A. (1938)  The Classical Equation of State of Gaseous Helium,
  Neon and Argon. {\em Proc. Roy. Soc.Lond. A}, \textbf{168}, 264--283.

\bibitem[Chantalat et~al., 2009]{Chantalat2009}
Chantalat, F., Bruneau, C.-H., Galusinski, C. {\&} Iollo, A. (2009)  Level-set,
  penalization and {C}artesian meshes: {A} paradigm for inverse problems and
  optimal design. {\em J. Comp. Phys.}, \textbf{228}(17), 6291--6315.

\bibitem[Chien et~al., 1986]{chien1986}
Chien, W.-L., Rising, H. {\&} Ottino, J. (1986)  Laminar mixing and chaotic
  mixing in several cavity flows. {\em J. Fluid Mech.}, \textbf{170}, 355--377.

\bibitem[Corrsin, 1957]{corrsin1957}
Corrsin, S. (1957)  Simple theory of an idealized turbulent mixer. {\em AIChE
  J.}, \textbf{3}(3), 329--330.

\bibitem[Corrsin \& Karweit, 1969]{corrsin1969}
Corrsin, S. {\&} Karweit, M. (1969)  Fluid line growth in grid-generated
  isotropic turbulence. {\em J. Fluid Mech.}, \textbf{39}(1), 87--96.

\bibitem[{Crimmins}, 1982]{Crimmins1982}
{Crimmins}, T.~R. (1982)  A Complete Set of {F}ourier Descriptors for
  Two-Dimensional Shapes. {\em IEEE Trans. Sys., Man, and Cybern.},
  \textbf{12}(6), 848--855.

\bibitem[Cundy \& Rollett, 1961]{cundy1961}
Cundy, H.~M. {\&} Rollett, A.~P. (1961) {\em Mathematical models}.
Clarendon Press Oxford.

\bibitem[D'Alessandro et~al., 1999]{DAlessandro1999}
D'Alessandro, D., Dahleh, M. {\&} Mezic, I. (1999)  Control of mixing in fluid
  flow: a maximum entropy approach. {\em IEEE Trans. Autom. Contr.},
  \textbf{44}(10), 1852--1863.

\bibitem[De~Boor et~al., 1987]{deboor1987}
De~Boor, C., H{\"o}llig, K. {\&} Sabin, M. (1987)  High accuracy geometric
  {H}ermite interpolation. {\em Comp. Aided Geom. Design}, \textbf{4}(4),
  269--278.

\bibitem[Dimotakis, 2000]{dimotakis2000}
Dimotakis, P.~E. (2000)  The mixing transition in turbulent flows. {\em J.
  Fluid Mech.}, \textbf{409}, 69--98.

\bibitem[Dimotakis \& Catrakis, 1999]{dimotakis1999}
Dimotakis, P.~E. {\&} Catrakis, H.~J. (1999)  Turbulence, fractals, and mixing.
  In {\em Mixing}, pages 59--143. Springer.

\bibitem[Eggl \& Schmid, 2018]{Eggl2018}
Eggl, M. {\&} Schmid, P. (2018)  A gradient-based framework for maximizing
  mixing in binary fluids. {\em J. Comp. Phys.}, \textbf{368}, 131 -- 153.

\bibitem[Engels et~al., 2015]{Engels2015}
Engels, T., Kolomenskiy, D., Schneider, K. {\&} Sesterhenn, J. (2015)  {FluSI}:
  A novel parallel simulation tool for flapping insect flight using a {F}ourier
  method with volume penalization. {\em SIAM J. Sci. Comp.}, pages 1--21.

\bibitem[Foures et~al., 2014]{Foures2014}
Foures, D., Caulfield, C. {\&} Schmid, P. (2014)  Optimal mixing in
  two-dimensional plane {P}oiseuille flow at finite {P}{\'{e}}clet number. {\em
  J. Fluid Mech.}, \textbf{748}, 241--277.

\bibitem[Glowinski et~al., 1996]{Glowinski1996}
Glowinski, M., Pan, T., Wells~Jr., R. {\&} Zhou, X. (1996)  Wavelet and finite
  element solutions for the {N}eumann problem using fictitious domains. {\em J.
  Comp. Phys.}, \textbf{126}(1), 40--51.

\bibitem[Han et~al., 2009]{han2009}
Han, X.-A., Ma, Y. {\&} Huang, X. (2009)  The cubic trigonometric {B}{\'e}zier
  curve with two shape parameters. {\em Appl. Math. Letts.}, \textbf{22}(2),
  226--231.

\bibitem[Hejlesen et~al., 2015]{hejlesen2015}
Hejlesen, M.~M., Koumoutsakos, P., Leonard, A. {\&} Walther, J.~H. (2015)
  Iterative Brinkman penalization for remeshed vortex methods. {\em J. Comp.
  Phys.}, \textbf{280}, 547--562.

\bibitem[Hemrajani \& Tatterson, 2004]{Handbook6}
Hemrajani, R. {\&} Tatterson, G. (2004) {\em Mechanically Stirred Vessels},
  pages 345--390.
John Wiley \& Sons, Inc.

\bibitem[Hessel et~al., 2005]{Hessel2005}
Hessel, V., L{\"{o}}we, H. {\&} Sch{\"{o}}nfeld, F. (2005)  Micromixers -- A
  review on passive and active mixing principles. In {\em Chemical Engineering
  Sciences}, volume~60, pages 2479--2501.

\bibitem[Horn \& Johnson, 2012]{Horn2012}
Horn, R. {\&} Johnson, C. (2012) {\em {Matrix Analysis}}.
Cambridge University Press, New York, NY, USA, 2nd edition.

\bibitem[Hunt \& Linden, 1999]{Hunt1999}
Hunt, G. {\&} Linden, P. (1999)  The fluid mechanics of natural ventilation --
  displacement ventilation by buoyancy-driven flows assisted by wind. {\em
  Buildg. and Environm.}, \textbf{34}(6), 707--720.

\bibitem[Jalali et~al., 2015]{jalali2015}
Jalali, M.~A., Khoshnood, A. {\&} Alam, M.-R. (2015)  Microswimmer-induced
  chaotic mixing. {\em J. Fluid Mech.}, \textbf{779}, 669--683.

\bibitem[Juniper, 2010]{Juniper}
Juniper, M. (2010)  Optimization with nonlinear adjoint looping. In {\em Int.
  Workshop on Nonnormal and Nonlinear Effects in Aero- and Thermoacoustics},
  Munich, Germany.

\bibitem[Kadoch et~al., 2012]{Kadoch2012}
Kadoch, B., Kolomenskiy, D., Angot, P. {\&} Schneider, K. (2012)  {A volume
  penalization method for incompressible flows and scalar advection-diffusion
  with moving obstacles}. {\em J. Comp. Phys.}, \textbf{231}(12), 4365--4383.

\bibitem[Kevlahan \& Ghidaglia, 2001]{Kevlahan2001}
Kevlahan, N. {\&} Ghidaglia, J.-M. (2001)  Computation of turbulent flow past
  an array of cylinders using a spectral method with {B}rinkman penalization.
  {\em Eur. J. Mech. B/Fluids}, \textbf{20}(3), 333--350.

\bibitem[Kirkpatrick et~al., 1983]{kirkpatrick1983}
Kirkpatrick, S., Gelatt, C.~D. {\&} Vecchi, M.~P. (1983)  Optimization by
  simulated annealing. {\em Science}, \textbf{220}(4598), 671--680.

\bibitem[Lekien et~al., 2005]{Lekien2005}
Lekien, F., Coulliette, C., Mariano, A., Ryan, E., Shay, L., Haller, G. {\&}
  Marsden, J. (2005)  Pollution release tied to invariant manifolds: A case
  study for the coast of {Florida}. {\em Physica D}, \textbf{210}(1-2), 1--20.

\bibitem[Lin \& Chellappa, 1987]{lin1987}
Lin, C. {\&} Chellappa, R. (1987)  Classification of partial 2-{D} shapes using
  {F}ourier descriptors. {\em IEEE Trans. Pat. Anal. and Mach. Intel.}, pages
  686--690.

\bibitem[Lin et~al., 2011]{Lin2011}
Lin, Z., Thiffeault, J.-L. {\&} Doering, C. (2011)  Optimal stirring strategies
  for passive scalar mixing. {\em J. Fluid Mech.}, \textbf{675}, 465--476.

\bibitem[Linden, 1979]{Linden1979}
Linden, P. (1979)  Mixing in stratified fluids. {\em Geophys. \& Astrophys.
  Fluid Dyn.}, \textbf{13}(1), 3--23.

\bibitem[Linden, 1999]{Linden1999}
Linden, P. (1999)  The fluid mechanics of natural ventilation. {\em Annu. Rev.
  Fluid Mech.}, \textbf{31}(1), 201--238.

\bibitem[Liu \& Vasilyev, 2007]{Liu2007}
Liu, Q. {\&} Vasilyev, O. (2007)  A {B}rinkman penalization method for
  compressible flows in complex geometries. {\em J. Comp. Phys.},
  \textbf{227}(2), 946--966.

\bibitem[Liu, 2008]{liu2008}
Liu, W. (2008)  Mixing enhancement by optimal flow advection. {\em SIAM J.
  Contr. Opt.}, \textbf{47}(2), 624--638.

\bibitem[Long, 1978]{long1978}
Long, R.~R. (1978)  A theory of mixing in a stably stratified fluid. {\em J.
  Fluid Mech.}, \textbf{84}(1), 113–124.

\bibitem[Madsen et~al., 2000]{madsen2000}
Madsen, J.~I., Shyy, W. {\&} Haftka, R.~T. (2000)  Response surface techniques
  for diffuser shape optimization. {\em AIAA J.}, \textbf{38}(9), 1512--1518.

\bibitem[Marcotte \& Caulfield, 2018]{marcotte2018}
Marcotte, F. {\&} Caulfield, C. (2018)  Optimal mixing in two-dimensional
  stratified plane {P}oiseuille flow at finite {P}\'eclet and {R}ichardson
  numbers. {\em J. Fluid Mech.}, \textbf{853}, 359–385.

\bibitem[Mathew et~al., 2007]{Mathew2007}
Mathew, G., Mezic, I., Grivopoulos, S., Vaidya, U. {\&} Petzold, L. (2007)
  Optimal control of mixing in {S}tokes fluid flows. {\em J. Fluid Mech.},
  \textbf{580}, 261--281.

\bibitem[Mattiussi, 2000]{Mattiussi2000}
Mattiussi, C. (2000)  The finite volume, finite element, and finite difference
  methods as numerical methods for physical field problems. {\em Adv. Imag. and
  Elec. Phys.}, \textbf{113}, 1--146.

\bibitem[Mohr et~al., 1957]{mohr1957}
Mohr, W., Saxton, R. {\&} Jepson, C. (1957)  Mixing in laminar-flow systems.
  {\em Ind. \& Eng. Chemistry}, \textbf{49}(11), 1855--1856.

\bibitem[Nguyen \& Wu, 2005]{Nguyen2005}
Nguyen, N.-T. {\&} Wu, Z. (2005)  Micromixers: a review. {\em J. Micromech.
  Microeng.}, \textbf{15}(2), R1--R16.

\bibitem[Ottino, 1990]{Ottino1990}
Ottino, J. (1990)  Mixing, chaotic advection, and turbulence. {\em Annu. Rev.
  Fluid Mech.}, \textbf{22}(1), 207--254.

\bibitem[Ottino et~al., 1988]{ottino1988}
Ottino, J., Leong, C., Rising, H. {\&} Swanson, P. (1988)  Morphological
  structures produced by mixing in chaotic flows. {\em Nature},
  \textbf{333}(6172), 419.

\bibitem[Ottino, 1989]{ottino1989}
Ottino, J.~M. (1989) {\em The kinematics of mixing: stretching, chaos, and
  transport}, volume~3.
Cambridge University Press.

\bibitem[Palacz et~al., 2016]{palacz2016}
Palacz, M., Smolka, J., Kus, W., Fic, A., Bulinski, Z., Nowak, A.~J., Banasiak,
  K. {\&} Hafner, A. (2016)  CFD-based shape optimisation of a CO2 two-phase
  ejector mixing section. {\em Appl. Thermal Eng.}, \textbf{95}, 62--69.

\bibitem[Paul et~al., 2004]{paul2004}
Paul, E.~L., Atiemo-Obeng, V.~A. {\&} Kresta, S.~M. (2004) {\em Handbook of
  industrial mixing: science and practice}.
John Wiley \& Sons.

\bibitem[Peltier \& Caulfield, 2003]{peltier2003}
Peltier, W. {\&} Caulfield, C. (2003)  Mixing efficiency in stratified shear
  flows. {\em Annu. Rev. Fluid Mech.}, \textbf{35}(1), 135--167.

\bibitem[Pingen et~al., 2007]{pingen2007}
Pingen, G., Evgrafov, A. {\&} Maute, K. (2007)  Topology optimization of flow
  domains using the lattice Boltzmann method. {\em Struct. Multidisc. Optim.},
  \textbf{34}(6), 507--524.

\bibitem[Saglietti et~al., 2017]{saglietti2017}
Saglietti, C., Schlatter, P., Monokrousos, A. {\&} Henningson, D.~S. (2017)
  Adjoint optimization of natural convection problems: differentially heated
  cavity. {\em Theor. Comput. Fluid Dyn.}, \textbf{31}(5-6), 537--553.

\bibitem[Saglietti et~al., 2018]{saglietti2018}
Saglietti, C., Schlatter, P., Wadbro, E., Berggren, M. {\&} Henningson, D.~S.
  (2018)  Topology optimization of heat sinks in a square differentially heated
  cavity. {\em Int. J. Heat \& Fluid Flow}, \textbf{74}, 36--52.

\bibitem[Smolka, 2013]{smolka2013}
Smolka, J. (2013)  Genetic algorithm shape optimisation of a natural air
  circulation heating oven based on an experimentally validated 3-D CFD model.
  {\em Int. J. Thermal Sci.}, \textbf{71}, 128--139.

\bibitem[Spencer \& Wiley, 1951]{spencer1951}
Spencer, R. {\&} Wiley, R. (1951)  The mixing of very viscous liquids. {\em J.
  Coll. Sci.}, \textbf{6}(2), 133--145.

\bibitem[Tang et~al., 2009]{Tang2009}
Tang, W., Caulfield, C. {\&} Kerswell, R. (2009)  A prediction for the optimal
  stratification for turbulent mixing. {\em J. Fluid Mech.}, \textbf{634},
  487--497.

\bibitem[Taylor, 1934]{Taylor1934}
Taylor, G.~I. (1934)  The formation of emulsions in definable fields of flow.
  {\em Proc. Roy. Soc. Lond. A}, \textbf{146}(858), 501--523.

\bibitem[Thiffeault, 2012]{Thiffeault2012}
Thiffeault, J.-L. (2012)  Using multiscale norms to quantify mixing and
  transport. {\em Nonlinearity}, \textbf{25}(2), R1.

\bibitem[Uhl, 2012]{uhl2012}
Uhl, V. (2012) {\em Mixing V1: Theory And Practice}.
Elsevier.

\bibitem[Vermach \& Caulfield, 2018]{vermach2018}
Vermach, L. {\&} Caulfield, C. (2018)  Optimal mixing in three-dimensional
  plane {P}oiseuille flow at high {P}\'eclet number. {\em J. Fluid Mech.},
  \textbf{850}, 875--923.

\end{thebibliography}

\end{document}